\documentclass{aa}  

\usepackage{multirow}  
\usepackage{xcolor}
\usepackage{lscape}
\usepackage{textcomp}
\usepackage{graphicx}
\usepackage{txfonts}
\usepackage{amsmath,bm}
\usepackage{color}
\usepackage{mathtools}
\usepackage{hyperref}
\usepackage[normalem]{ulem}

\newcommand{\visibility}[2]{ \Gamma_{#1,#2} }
\newcommand{\rmstar}{\star}
\newcommand{\rmplanet}{\medbullet} 

\newcommand{\noise}{\sigma}
\newcommand\norm[1]{\left\lVert#1\right\rVert}
\newcommand\scalar[2]{\langle#1\,|\,#2\rangle}
\newcommand\dscalar[2]{\langle\langle#1\,|\,#2\rangle\rangle}

\DeclareMathOperator{\EX}{\mathbb{E}}

\begin{document} 

   \title{Exoplanets in reflected starlight with dual-field interferometry}
   \subtitle{A case for shorter wavelengths and a fifth Unit Telescope at VLTI/Paranal}

   \titlerunning{Exoplanets in reflected starlight with dual-field interferometry}
   
   \author{
      S.~Lacour\inst{\ref{lesia},\ref{esog}}
      \and
      Ó.~Carrión-González\inst{\ref{lesia}}
      \and
      M.~Nowak\inst{\ref{kavli},\ref{ioa}}
   }

   \institute{
      LESIA, Observatoire de Paris, PSL, CNRS, Sorbonne Universit\'e, Universit\'e de Paris, 5 place Janssen, 92195 Meudon, France
      \label{lesia}      
      \and
      European Southern Observatory, Karl-Schwarzschild-Stra\ss{}e 2, 85748 Garching, Germany
      \label{esog}
      \and
      Kavli Institute for Cosmology, University of Cambridge, Madingley Road, Cambridge CB3~0HA, United-Kingdom
      \label{kavli}
      \and
      Institute of Astronomy, University of Cambridge, Madingley Road, Cambridge CB3~0HA, United-Kingdom
      \label{ioa}
   }
    
   \date{Received September 15, 1996; accepted March 16, 1997}

 
         \abstract
         {The direct observation of cold and temperate planets within 1 to 10 AU would be extremely valuable for uncovering their atmospheric compositions but remains a formidable challenge with current astronomical methods. Ground-based optical interferometry, capable of high angular-resolution imaging, offers a promising avenue for studying these exoplanets.}
         {Our objective is to explore the fundamental limits of dual-field interferometry and assess its potential for characterising exoplanets in reflected light using the Very Large Telescope Interferometer (VLTI).}
         {We developed analytical expressions to describe the performance of dual-field interferometry and integrated these with simulations of atmospheric wavefronts corrected by extreme adaptive optics. An analytical solution for optimal phase apodization was formulated to enhance starlight rejection when injected into a single-mode fibre. This framework was applied to determine the detectability of known exoplanets in reflected light across various wavelength bands for both the current VLTI and a proposed extended version.}
         {Our results indicate that employing shorter wavelengths improves detectability, enabling at least seven Jupiter-mass exoplanets to be observed in the J band with current VLTI's baselines. Adding new baselines with lengths beyond 200 meters significantly enhances VLTI's capabilities, increasing the number of detectable exoplanets and revealing potential habitable zone candidates such as $\tau$ Ceti e and Proxima Centauri b.}
         {To substantially improve the VLTI's exoplanet characterisation capabilities, we recommend developing instrumentation at wavelengths shorter than 1$\,\mu$m, and increasing the baselines length by the addition of a fifth Unit Telescope (UT5).}


   \maketitle

%
\section{Introduction}

Over the past two decades, direct imaging has become a key method for the characterisation of exoplanets. Its development is now a priority  within the community, as illustrated by key community reports and space agencies (Astro 2020 Decadal Survey, and ESA’s Voyage 2050 Senior Committee report, for example). Direct imaging has a particular role to play as it can overcome some of the main limitations of other methods. For example, observing molecular (bio)signatures via transit spectroscopy can be made extremely challenging by the fact that these molecules are often concentrated in lower atmospheric layers \citep{Lustig-Yaeger2019}. Consequently, several direct imaging space observatories have been proposed, and are under study. The Coronagraph Instrument on the Nancy Grace Roman Space Telescope \citep{spergeletal2015} is poised to be the first to directly image exoplanets in reflected light from space. Future flagship missions, such as the Habitable Worlds Observatory, are being conceptualised based on the LUVOIR \citep{luvoirteam2018} and HabEx \citep{mennessonetal2016, gaudietal2018} studies. In Europe, momentum is building for the LIFE project \citep{2022A&A...664A..21Q}, a mid-infrared optical interferometer targeting thermal emission.

Compared to their space-based counterparts, ground-based observatories offer distinct advantages, with usually much larger apertures and higher angular resolution, but also specific challenges, such as an important thermal background and atmospheric aberrations. Several young massive exoplanets have already been successfully discovered and directly imaged from the ground. These are usually hot planets, such as $\beta$~Pictoris~b (approximately 1700\,K) and 51 Eri\,b (around 800\,K) which are still cooling off by gradually dissipating the gravitational energy accumulated during their formation. Their thermal emission is comparatively bright against their host stars, with contrast ratios on the order of a few $10^{-5}$. But the vast majority of planets are older and colder than these young giants, with a temperature set by the radiative equilibrium with their stellar hosts. Such bodies primarily emit in the mid-infrared wavelength range. To be detected in the near-infrared, these objects must be observed in reflected light, typically at a contrast on the order of $10^{-7}$ or lower, depending on their radius, semi-major axis, phase angle and albedo.

Given the challenges associated with achieving such contrast levels, an important question is whether exoplanets can be detected and characterised in reflected light from ground-based observatories. \cite{guyonLimitsAdaptiveOptics2005} provides a comprehensive overview of the challenges involved. The first major challenge is atmospheric perturbations, which can be mitigated in single-dish telescopes using adaptive optics (AO) systems, such as the Spectro-Polarimetric High-contrast Exoplanet REsearch (SPHERE) extreme AO \citep{Beuzit2019}. The second challenge involves the diffraction pattern, which generates photon noise at the location of the planet; this issue can be addressed with the use of a coronagraph \citep{Rouan2000}. The third, and perhaps most significant challenge, involves non-common path quasi-static aberrations that produce slowly moving speckles, difficult to calibrate. These speckles, with intensities around $10^{-5}$, require stabilisation to within 1\%, necessitating a wavefront stability of 4\,pm \citep{guyonLimitsAdaptiveOptics2005}. 

These quasi-static speckles can be subtracted using advanced post-processing techniques such as Spectral Differential Imaging \citep[SDI][]{racineSpeckleNoiseDetection1999} and Angular Differential Imaging \citep[ADI][]{maroisAngularDifferentialImaging2006}, or in real time through focal plane wavefront sensing and correction methods similar to the dark-hole method \citep{potierIncreasingRawContrast2022}. However, these methods require that the wavefront remains stable throughout the observation period; otherwise, shifting speckles cannot be accurately removed. This stringent requirement for temporal stability has spurred the development of new techniques that combines high contrast imaging with high resolution spectroscopy \citep{2013ApJ...764..182S,2015A&A...576A..59S,2017A&A...599A..16L}. 
   These methods rely on the fact that the planet's spectrum is different from the star's, and that spectral signatures detected by cross-correlation techniques can be used to separate the two signals. Several instruments already employ this technique on 8-metre class telescopes, such as the VLT \citep[HiRISE][]{2024A&A...682A..16V} and the Keck Telescope \citep[KPIC][]{2021JATIS...7c5006D}. More instruments are also under development, such as RISTRETTO \citep{2022SPIE12184E..1QL}, but it is anticipated that these techniques will be fully exploited with the advent of the Extremely Large Telescope (ELT). Almost all first-generation ELT instruments, including MICADO, HARMONI, and ANDES, will employ this technique to push the contrast range to its limit by leveraging the substantial photon counts captured by the ELT \citep{2023arXiv231117075P}.

\begin{figure*}
   \centering
   \includegraphics[width=15cm]{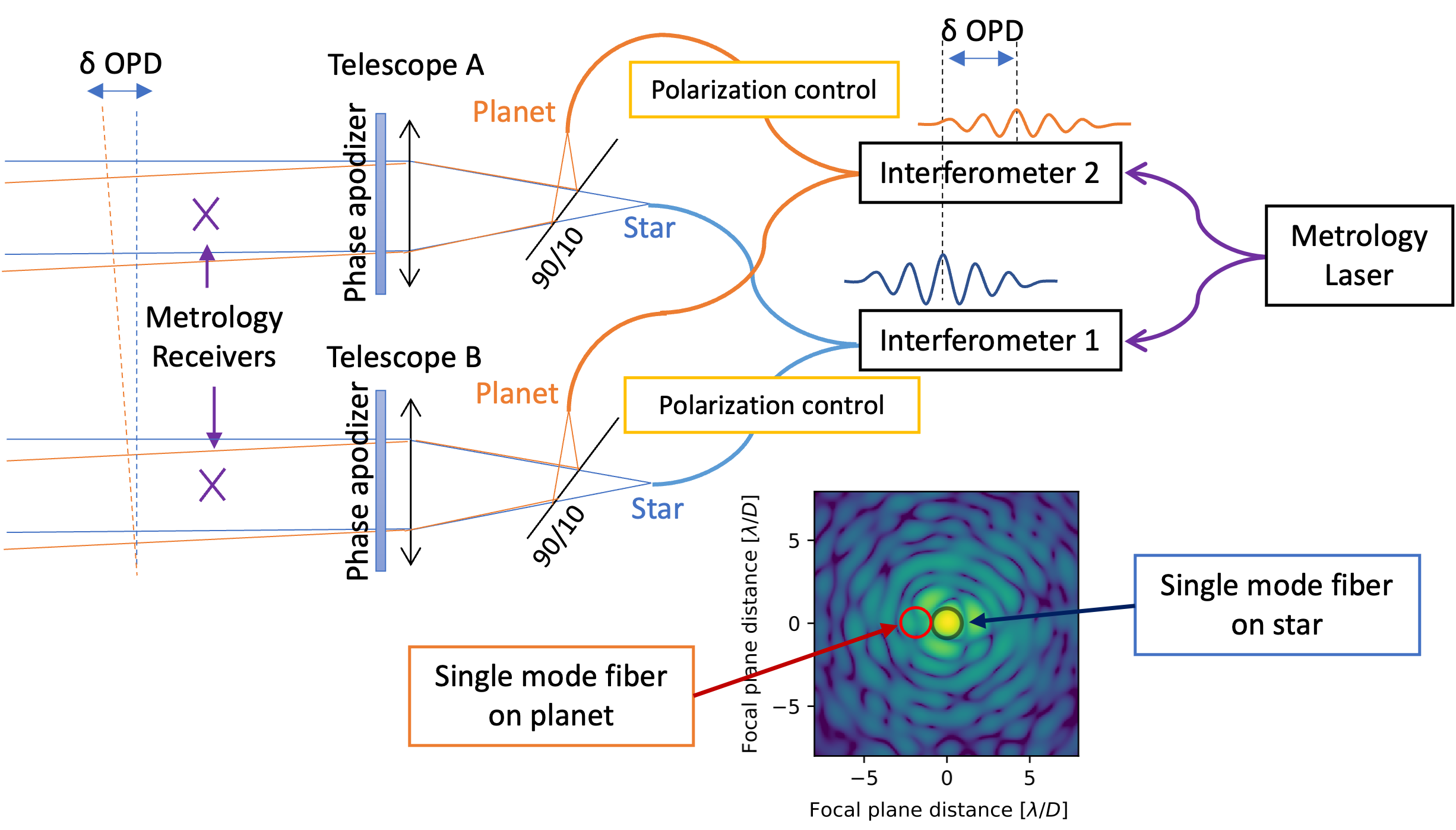}
   \caption{Schematic diagram of a dual field optical interferometer using two telescopes. A beam splitter creates two images of the focal plane, where two fibres are positioned—one on the star to feed Interferometer 1 (serving as the phase reference) and the other on the exoplanet, feeding Interferometer 2. The exoplanet is detected as coherent flux in Interferometer 2, with a phase offset relative to the stellar coherent flux due to an optical path delay $\delta \text{OPD}$, corresponding to the scalar product of the narrow angle baseline vector $\bm{B}$ and the angular separation between the star and exoplanet $\bm{\alpha}$: $\delta \text{OPD}=\bm{\alpha} \cdot \bm{B}$.}
   \label{fig:schema}%
\end{figure*}

In this context, optical interferometry offers a complementary alternative, as its post-processing does not rely on any spectral feature. 
Several high-contrast interferometric designs exist, generally classified into two main types: single-field and dual-field interferometers. Single-field interferometry differentiates the coherent signals of stars and exoplanets using techniques such as closure phase \citep{2018ApJ...855....1G}, nulling \citep{bracewellDetectingNonsolarPlanets1978}, or a combination of these methods, such as Kernel nulling \citep{martinacheKernelnullingRobustDirect2018, Chingaipe2023}. Dual-field interferometry, as summarised in Fig.~\ref{fig:schema}, operates by simultaneously observing two narrow fields of view — typically the size of the telescope's point spread function — situated within a larger field of view that can include both a star and an exoplanet. This configuration allows one interferometer to focus on the star while the other targets the exoplanet, minimising the contamination of the exoplanet signal by stellar photon noise due to the narrowness of each telescope's diffraction pattern. 

The GRAVITY instrument \citep{gravitycollaborationFirstLightGRAVITY2017} at the Very Large Telescope Interferometer (VLTI) has pioneered dual-field interferometry, and was incidentally the first interferometer to directly detect an exoplanet \citep{gravitycollaborationFirstDirectDetection2019}. It has also provided the first direct detection of two short-separation radial-velocities exoplanets: $\beta$\,Pictoris\,c \citep{nowakDirectConfirmationRadialvelocity2020} and HD\,206893\,c \citep{hinkleyDirectDiscoveryInner2023}. These two planets have not been directly observed by any other instrument to date, suggesting that dual-field interferometry is performing better than imaging at such small separations. The commissioning of GRAVITY+ will bring extreme AO to all the Unit Telescopes (UTs) which, coupled with dual-field interferometry \citep{2022Msngr.189...17A}, will further increase this capability and optimise the VLTI for high-contrast \citep{2024arXiv240604003P}.

These developments prompt an inquiry into the fundamental limits of dual-field interferometry and its potential to detect exoplanets using reflected light from the ground. In Section~\ref{sec:simulation}, we introduce a detailed analytical model to assess the performance and limitations of dual-field interferometry using the existing VLTI. Because photon noise is a prime limitation, Section~\ref{sec:starlight} presents a novel, semi-analytical method for determining the optimal phase apodizer necessary for starlight suppression when channeling light into a single-mode fibre. 
Section~\ref{sec:NewUT5} suggests a possible upgrade of the VLTI, in which a new 8~m Unit Telescope (UT5) would be added to optimise the VLTI for exoplanet detection, by creating four additional baselines in the 200~m regime to the North-West direction. In Section~\ref{sec:detect_exo}, we estimate the number of known Radial Velocity (RV) exoplanets that could potentially be characterised by the VLTI, including UT5 or not. Finally, Section~\ref{sec:summary} highlights the main results of this paper.

\section{Simulation of contrast range} \label{sec:simulation}

\subsection{The detection principle}

A concise overview of the working principle of a dual-field optical interferometer used for high-contrast exoplanet observations is given in Fig.~\ref{fig:schema}. This setup involves splitting light from two telescopes into four focal planes, in which single-mode fibres are placed. Two fibres, which feed a first interferometer, are centred on the geometrical image of the star, whereas the other two fibres, which feed a second interferometer, are centred on the geometrical image of the planet. A dedicated metrology system monitors the non-common optical-path differences between the two interferometers.

The detection principle relies on the premise that the exoplanet's location is already known, with a precision well within the field of view of the single mode fibres ($\approx 60\,$mas for an 8\,m telescope in the K band). This implies that this method is more suited for characterising exoplanets (for example, measuring their albedo) rather than discovering new ones.

Given the typical planet to star contrast ratio (at least $10^{-4}$, and usually much smaller), the first interferometer $I_1$ is always dominated by photons from the star, while the second interferometer $I_2$ receives both the photons from the planet, and some non-negligible amount of residual starlight. The light from the planet and the light from the star being fully incoherent, this results in two superimposed systems of fringes observed in the second interferometer. 

Separating these two systems of fringes observed on interferometer $I_2$ to recover the planet coherent flux is one of the main concern of high-contrast dual-field interferometry. For this, two strategies are used in combination: (1) minimising the amount of starlight entering the fibres feeding interferometer $I_2$ in the first place (a process thereafter referred to as "apodization"); (2) separating the two fringe systems in post-processing, by taking advantage of the fact that the on-axis starlight and the off-axis planet light creates fringes centred at different Optical Path Difference (OPD) values.

\subsection{Spelling out the electric fields}

In the next sections of this paper, we will use the symbol $\rmstar$ in subscript to denote parameters associated with the star, and we will use $\rmplanet$ for those associated with the planet. We denote $\bm{\alpha}$ the vector representing the angular separation between the star and the exoplanet: $\bm{\alpha} = \bm{\alpha}_{\rmplanet} - \bm{\alpha}_{\rmstar}$, where $\bm{\alpha}_{\rmplanet}$ points towards the planet, and $\bm{\alpha}_{\rmstar}$ points towards the star. 

We model the interferometric setup using two independent 8~m telescopes, with pupil shapes akin to those of a UT of the VLT. We assume that these telescopes are equipped with an AO system which corrects for the atmospheric turbulence, and we denote $\Phi(\bm{u})$ the phase residuals downstream of the AO on the pupil of a given telescope. A term $\Psi(\bm{u})$ will also be added at the beginning of section~\ref{sec:starlight} to include a pupil phase apodizer. The two-dimensional vector $\bm{u}$ is the X-Y coordinate vector within the telescope's pupil. The narrow-angle baseline, the vector between the two telescopes \citep[the vector delimited by the metrology, as outlined by][]{woillezWIDEANGLENARROWANGLEIMAGING2013}, is denoted as $\bm{B}$.
   
With this setup, a monochromatic point source of wavevector $k = 2\pi/\lambda$, located at a vector position $\bm{\alpha}$ on the sky and radiating on the telescope with a flux $F = A^2$, generates an electric field within the pupil of a given telescope which can be described by:
\begin{equation}
   \mathsf{W}(\bm{u}) = Ae^{-ik\bm{\alpha}\cdot{}\bm{u} + i\Phi(\bm{u})}.
   \label{eq:W}
\end{equation}

The flux entering a single-mode fibre is determined by projecting the electric field onto the fibre's fundamental mode. The integral can be calculated either in the focal plane or in the pupil plane if we assume the far-field Fraunhofer condition (the electric field in the pupil plane is then the Fourier transform of the electric field in the focal plane). This can be demonstrated thanks to the Parseval-Plancherel theorem. Using the pupil-plane approach, the electric field from Eq.~\eqref{eq:W} couples into a single-mode fibre as follows:
\begin{equation}
   \label{eq:injection}
   E = \iint_D  Ae^{-ik\bm{\alpha}\cdot{}\bm{u} + i\Phi(\bm{u})} \mathsf{M}^\ast( \bm{u} ) \ d\bm{u},
\end{equation}
where the integral covers the pupil area with diameter \(D\). The fibre's fundamental mode, represented in the pupil plane as \(\mathsf{M}(\bm{u})\), can be approximated by a Gaussian-shaped mode with beam waist \(w\):
\begin{equation}
   \mathsf{M}(\bm{u}) \propto {\displaystyle\exp\left(-\frac{|\bm{u}|^2}{2 w^2}\right)}\ ,
\end{equation}
where a normalisation factor must be added to ensure the conservation of energy. For a circular aperture telescope without central obstruction, an optimal beam waist that achieve maximum injection can be numerically calculated\footnote{Demonstration in Appendix~\ref{sec:optimalI}.}. The
optimal injection is achieved with \(w=\sqrt{0.1}\,D\) and \(\mathsf{M}(\bm{u})\) can then be exactly spelled as:
\begin{equation}
   \mathsf{M}(\bm{u}) = \frac{2}{\displaystyle  \sqrt{0.1} \pi D^2} {\displaystyle \exp\left(-\frac{|\bm{u}|^2}{0.2 D^2}\right)}\ ,
   \label{eq:Mu}
\end{equation}
leading to the optimal throughput in the absence of phase aberrations:
$\norm{\iint_D  \mathsf{M}( \bm u) \ d\bm{u}}^2 = 0.81.$

This expression of \(\mathsf{M}(\bm{u})\) is valid for a fibre located at the focal point of the telescope (i.e. "on-axis"). If the fibre is moved at the location of the geometrical image on an object situated at the on-sky vector position $\bm{\beta}$, the fibre mode $\mathsf{M}$ in the pupil plane becomes:
\begin{equation}
   \mathsf{M}_\beta(\bm{u}) = \mathsf{M}(\bm u) e^{-ik\bm{\beta}\cdot{}\bm{u}},
\end{equation}
\noindent{}and the coupled electric field becomes:
\begin{align}
   \label{eq:Einjected}
   E &= \iint_D Ae^{i\Phi (\bm{u})} \mathsf{M}( \bm u) e^{ik\left(\bm{\beta}-\bm{\alpha}\right)\cdot{}\bm{u}} \ d\bm{u}.
\end{align}

At this point, it is useful to note that since \(\mathsf{M}\) is a positive non-zero function, we can define a scalar product \(\scalar{\cdot}{\cdot}\) using:
\begin{equation}
   \scalar{\mathsf{G}}{\mathsf{H}} = \iint_D \mathsf{G}(\bm{u})\,\mathsf{M}(\bm{u})\,\mathsf{H}^\ast(\bm{u}) \, d\bm{u},
   \label{eq:scalar}
\end{equation}
as this expression has all the required properties (linearity, conjugate symmetry, and definite-positiveness). It is defined such that \(\scalar{1}{1} = \sqrt{0.81}\) in the case where \(\mathsf{M}(\bm{u})\) is given as in Eq.~\eqref{eq:Mu}.

With this notation, the electric field of Eq.~\eqref{eq:Einjected} is simply the scalar product of the pupil-plane electric field and a complex exponential representing the offset between the source and the fibre:
\begin{align}
   E &= A \scalar{e^{i\Phi (\bm{u})}}{e^{ik\left(\bm{\alpha}-\bm{\beta}\right)\cdot\bm{u}}}.
   \label{eq:EAs}
\end{align}
This formulation simplifies the expression of the fields injected into the fibres of the dual-field interferometer.
These fields are summarised in Table~\ref{tb:elec} for the telescopes $\mathbb A$ and $\mathbb B$. The first set of fields considers a star radiating with a flux \( F_{\rmstar} = A_{\rmstar}^2 \) located on the optical axis. The second set involves a planet with a flux \( F_{\rmplanet} = A_{\rmplanet}^2 \) positioned at \( \bm{\alpha} \). For the first interferometer, we assume \( \bm{\beta} = \bm{0} \). For the second interferometer, we assume \( \bm{\beta} = \bm{\alpha} \).

\begin{table}
   \caption{The eight electric fields inside the four fibres of a two-telescope dual-field interferometer.}
   \label{tb:elec}
   \centering          
   \begin{tabular}{l  |  l}
   \hline      
   \multicolumn{1}{c  |}{Interferometer 1} & \multicolumn{1}{c}{Interferometer 2}  \\ 
   \multicolumn{1}{c  |}{(on-axis; $\bm{\beta}=\bm{0}$)} & \multicolumn{1}{c}{(off-axis; $\bm{\beta}=\bm{\alpha}$)} \\
    \hline                    
    \hline                
     $E_{\mathbb A, 1, \rmstar} =  A_\rmstar \scalar{e^{i\Phi_{\mathbb A}(\bm u)}}{1}$ & $E_{\mathbb A, 2, \rmstar} =  A_\rmstar \scalar{e^{i\Phi_{\mathbb A}(\bm u)}}{e^{-ik\bm{\alpha}\cdot{}\bm{u}}}$ \\    
     $E_{\mathbb A, 1, \rmplanet} \approx 0$ & $E_{\mathbb A, 2, \rmplanet} =  A_\rmplanet \scalar{e^{i\Phi_{\mathbb A}(\bm u)}}{1}$ \\     
   \hline                  
   \hline                    
    $E_{\mathbb B, 1, \rmstar} =  A_\rmstar \scalar{e^{i\Phi_{\mathbb B}(\bm u)}}{1}$ & $E_{\mathbb B, 2, \rmstar} =  A_\rmstar \scalar{e^{i\Phi_{\mathbb B}(\bm u)}}{e^{-ik\bm{\alpha}\cdot{}\bm{u}}}$ \\
     $ E_{\mathbb B, 1, \rmplanet} \approx 0 $ & $E_{\mathbb B, 2, \rmplanet} =  A_\rmplanet e^{-ik\bm{\alpha}\cdot{}\bm{B}} \scalar{e^{i\Phi_{\mathbb B}(\bm u)}}{1}$ \\
   \hline                  
   \end{tabular}
   \tablefoot{In this table $\Phi_{\rm A}(\bm u)$ and $\Phi_{\rm B}(\bm u)$ are the phase aberrations of, respectively, telescopes $\mathbb A$ and $\mathbb B$.}
\end{table}

The two quantities of interest which are derived from the observations on the two interferometers are the coherent energies, denoted $\Gamma$, and given by:
\begin{align}
   \Gamma_1 &=  E_{\mathbb A,1,\rmstar}E_{\mathbb B, 1,\rmstar}^\ast ,\\
   \Gamma_2 &=  E_{\mathbb A,2,\rmstar}E_{\mathbb B, 2,\rmstar}^\ast + E_{\mathbb A, 2,\rmplanet}E_{\mathbb B, 2,\rmplanet}^\ast,
\end{align}
where we implicitly neglect the exoplanet light in the first interferometer. The exoplanet signal $\Gamma_{2,\rmplanet} = E_{\mathbb A, 2,\rmplanet}E_{\mathbb B, 2,\rmplanet}^\ast$ is observable in the second interferometer, blended with some amount of residual stellar coherent energy $\Gamma_{2, \rmstar} = E_{\mathbb A, 2,\rmstar}E_{\mathbb B, 2,\rmstar}^\ast$. In the rest of the paper, we usually refer to $\Gamma_{2, \rmstar}$ as "coherent speckle", by analogy with the speckles observed by classical imaging.
 
In practice, though, the coherent energies $\Gamma$ are never observed directly, but are rather derived from measurements of the interference patterns as a function of optical delay $x$, described by:
\begin{small}
\begin{align}
   \label{eq:fringes_1}
   I_1(x)=&|E_{\mathbb A,1,\rmstar}|^2+|E_{\mathbb B,1,\rmstar}|^2+2\Re\left[\Gamma_1 e^{ix}\right],\\
   \label{eq:fringes_2}
   I_2(x)=&|E_{\mathbb A, 2,\rmstar}|^2+|E_{\mathbb B,2,\rmstar}|^2+|E_{\mathbb A,2,\rmplanet}|^2+|E_{\mathbb B, 2,\rmplanet}|^2+2\Re\left[ \Gamma_2 e^{ix}\right].
\end{align}
\end{small}
A common beam combination pattern is the ABCD scheme, for which the detectors of the interferometers record $I_1(x)$ and $I_2(x)$ as a function of four well-calibrated values of $x$: $0, \pi/2, \pi, 3\pi/2$. The coherent energies $\Gamma_1$ and $\Gamma_2$ are then reconstructed by a proper linear combination of these measurements. 

\subsection{Signal to noise ratio}

In the numerator of the S/N, the signal can have multiple definitions:
\begin{eqnarray}
   S_{\rm entrance} &=& \gamma \, F_{\rmplanet} \label{eq:s_raw} \\
   S_{\rm interfero} &=& \gamma \, \Gamma_{2, \rmplanet} = \gamma \, F_{\rmplanet} \, \eta_{\rmplanet}^{\mathrm{inj}} \label{eq:s_coh} \\
   S_{\rm processed} &=& \gamma \, \Gamma_{2, \rmplanet} \, \eta_{\rmplanet}^{\mathrm{proc}} = \gamma \, F_{\rmplanet} \, \eta_{\rmplanet}^{\mathrm{inj}}\, \eta_{\rmplanet}^{\mathrm{proc}} \,. \label{eq:s_proc}
\end{eqnarray}

The definition of $S_{\rm entrance}$ in Eq.~\eqref{eq:s_raw} corresponds to the number of photons of exoplanetary origin entering the interferometer. It depends solely on the flux of the exoplanet $F_{\rmplanet} = A_{\rmplanet}^2$ with a proportionality factor $\gamma$ that encompasses the observation parameters. $S_{\rm entrance}$ represents an upper limit of the signal that we can expect.

The definition of $S_{\rm interfero}$ used in Eq.~\eqref{eq:s_coh} corresponds to the number of coherent exoplanetary photons that we can observe. To simplify the study, we assume a visibility of one, meaning that the exoplanet flux is fully coherent. Compared to $S_{\rm entrance}$, $S_{\rm interfero}$ is therefore reduced by the injection losses of the light into the single-mode fibre. These losses are denoted as $\eta_{\rmplanet}^{\mathrm{inj}}$ and correspond to the product of two scalar products as in Eq.~\eqref{eq:EAs} averaged over time. Its value is always below 1 and is formally discussed in Section~\ref{sec:starlight}.

The definition of $S_{\rm processed}$ in Eq.~\eqref{eq:s_proc} includes an additional term, $\eta_{\rmplanet}^{\mathrm{proc}}$, accounting for the post-processing of the data that reduces the amount of detectable signal ($\eta_{\rmplanet}^{\mathrm{proc}} < 1$). This term depends on the algorithm used for post-processing and is described in Section~\ref{sec:postprocessing}.

The term $\gamma$ is a proportionality factor that converts the flux into the signal in units of photons. This conversion factor incorporates several terms: the total observation duration ($\Delta t$), the telescope's collecting area ($\pi D^2/4$), the spectral bandwidth ($\Delta \lambda / \lambda$), the transmission efficiency ($T_r$), and Planck's constant ($h$):
\begin{equation}
\gamma = \rho \times \frac{1}{h} \times  \frac{\Delta \lambda }{ \lambda} \times \frac{\pi D^2}{4} \times T_r \times \Delta t\,.
\label{eq:gamma}
\end{equation}
The coefficient $\rho$ corresponds to how the flux of a telescope is dispatched between the different interferometric baselines. Under the assumption that all baselines are recombined, it is equal to the number of telescopes divided by the maximum number of baselines:
\begin{equation}
   \rho = N_{\mathrm{tel}} \times \left( \cfrac{N_{\mathrm{tel}} (N_{\mathrm{tel}} - 1)}{2} \right)^{-1} = \frac{2 N_{\mathrm{tel}}}{N_{\mathrm{tel}}^2 - N_{\mathrm{tel}}} \,.
\end{equation}
Thus, we have $\rho = 2$ for a two-telescope interferometer and $\rho = 2/3$ for the recombination of the 4 UTs with the GRAVITY instrument. The signal is then defined for a specific baseline, while the signals from multiple baselines can be combined to obtain the final S/N in Section~\ref{sec:NewUT5}.

In the denominator of the S/N, the noise can be broken into its individual and independent components:
\begin{equation}
   N = \sqrt{\noise_\mathrm{readout}^2 + \noise_\mathrm{thermal}^2 + \noise_\mathrm{shot}^2 + \noise_\mathrm{speckle}^2 },
\end{equation}
where the individual noise elements account for detector readout noise, thermal background photon noise, shot noise, and speckle noise. These elements are discussed in the following subsections. 

Combining all sources of noise, and using $S_{\rm processed}$ for the signal, the S/N ratio takes the form:
\begin{equation}
   \label{eq:snr}
   \mathrm{S/N} = \frac{\gamma F_\rmplanet \, \eta_\rmplanet^\mathrm{inj} \, \eta_\rmplanet^\mathrm{proc}}{\sqrt{\noise_\mathrm{readout}^2 + \noise_\mathrm{thermal}^2 + \noise_\mathrm{shot}^2 + \noise_\mathrm{speckle}^2 }}.
\end{equation}

\subsection{Readout and thermal background noise}

The detector readout noise, \(\noise_{\mathrm{readout}}\), depends on the characteristics of the detector, on the number of frames and on the number of used pixels:
\begin{equation}
   \noise_{\mathrm{readout}} = \sigma_{e^-} \sqrt{4 \times \frac{3 \Delta \lambda}{\delta \lambda} \times \frac{\Delta t}{\delta t}}
\end{equation}
where \(\sigma_{e^-}\) represents the read noise per pixel per frame. For modern Teledyne H2RG detectors using a Fowler-16 detection readout scheme, \(\sigma_{e^-}\) is typically around 5 electrons. The term under the square root quantifies the total number of reads during the entire observation. The factor of 4 accounts for the ABCD recombination scheme. The spectral bandwidth is expressed as \(\Delta\lambda\) and the spectral resolution is denoted by \(\delta\lambda\). With an oversampling factor of 3, the total number of pixels across the band is \(3\Delta\lambda/\delta \lambda\). The number of frames is given by the ratio of the total integration time to the detector integration time, \(\Delta t/\delta t\). 

The thermal background noise, \(\noise_{\mathrm{thermal}}\), can be estimated by assuming that blackbody emission from the laboratory at temperature \(T_{\rm lab}\) is the dominant source of background. This assumption is reasonable given the low throughput (\(T_r < 10\%\)) of the optical chain, meaning that sky background can be neglected\footnote{Airglow lines could also be a concern, but they were not observed in the GRAVITY data and therefore were not included in our simulations.}. 
The background emission can be analytically estimated using the limited acceptance angle of a single-mode fibre (also referred to as the étendue) and Planck's law for blackbody radiation, \(B_\lambda (\rm T_{\rm lab},\lambda)\):
\begin{equation}
   \noise_{\mathrm{thermal}} = \sqrt{\rho \frac{1}{h} \frac{\Delta \lambda}{\lambda} \Delta t 
   \times (1-T_r) \times \lambda^2 \times B_\lambda (\rm T_{\rm lab},\lambda)} \,.
\end{equation}
The term \(\lambda^2\) represents the étendue of the single-mode fibre. The factor \(\rho \frac{1}{h} \frac{\Delta \lambda}{\lambda} \Delta t\), which is also used in Eq.\,\eqref{eq:gamma} for \(\gamma\), is included to convert the flux into a number of detected photons. Finally, the factor \((1-T_r)\) accounts for the VLTI's transparency, such that if $T_r = 0$, the thermal background noise is maximum.
 
\begin{figure}
   \centering
   \includegraphics[width=9cm]{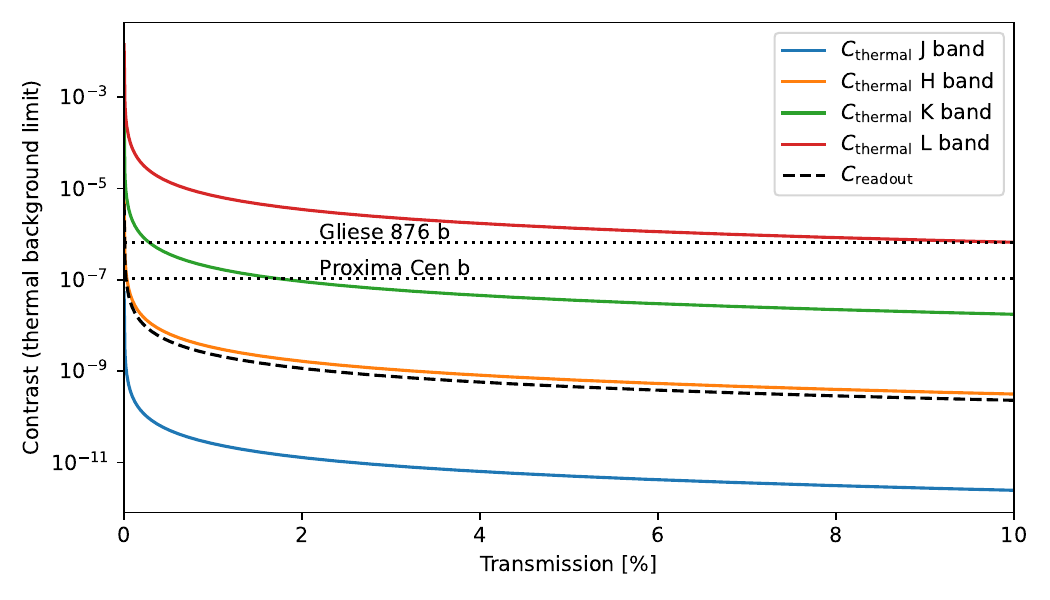}
   \caption{Contrast limit at the VLTI as a function of the transmission efficiency of the interferometer ($T_r$, in percents). The coloured curves represent theoretical contrast limits set by the thermal background of the telescope and interferometric laboratory (300\,K). The dashed curves indicate the contrast limit from detector readout noise. The simulation parameters are detailed in Table~\ref{tb:simulation}.}
   \label{fig:CThermal}
\end{figure}

To better understand the contrasts accessible with these readout and thermal background noise sources, we can convert these noise expressions into first-order contrast limits, by writing the planet raw signal as a function of the stellar flux:
\begin{equation}
   S_{\rm entrance} = \gamma\, C F_{\rmstar}\,,
\end{equation}
where $C$ is the planet to star contrast ratio: $C = F_\rmplanet/F_\rmstar$. 

Using a threshold of $\mathrm{S/N} = 3$, this gives two contrast limits set respectively by the amount of readout and thermal background noise:
\begin{align}
   & C_\mathrm{readout} = 3 \frac{\sigma_{e^-}}{\gamma F_\rmstar} \sqrt{4  \frac{3 \Delta \lambda}{\delta \lambda} \frac{\Delta t}{\delta t}}\,, \label{eq:C_readout}\\
   & C_\mathrm{thermal} = 3 \frac{1}{ \gamma F_\rmstar}  \sqrt{   \rho \frac{1}{h} \frac{\Delta \lambda}{\lambda} \Delta t 
    (1-T_r) \lambda^2  B_\lambda (300\,\mathrm{K},\lambda)} \,. \label{eq:C_thermal}
\end{align}

Both $C_{\text{readout}}$ and $C_{\text{thermal}}$ are depicted in Fig.~\ref{fig:CThermal} for a two telescopes interferometer with parameters as in Table~\ref{tb:simulation}. For comparison, Figure~\ref{fig:CThermal} also shows the contrast ratios in reflected light expected for two known planets: Gliese 876\,b and Proxima Centauri b, assuming an albedo of 0.3. This plot demonstrates the importance of high transmission, especially given the inherent low brightness of exoplanets. At a transmission of 2\%, typical of the VLTI, the readout noise begins to limit detection slightly above $10^{-9}$. The thermal emission equals  readout noise near the H band. At longer wavelengths, near 3 microns, the  300\,K emission of the laboratory strongly limits the achievable contrast and impedes the observation of reflected light from Gliese 876\,b and Proxima Centauri b. 

\begin{table}
   \caption{Simulation parameters}             
   \label{tb:simulation}      
   \centering                          
   \begin{tabular}{c c c c c}        
   \hline\hline                 
        & Fig.~\ref{fig:CThermal} & Fig.~\ref{fig:Astar} & Fig.~\ref{fig:Aprocess} & Fig.~\ref{fig:Cfull} \\    
   \hline                        
   $F_{\rmstar}$ & 10\,Jy & 10\,Jy &  & 10\,Jy \\      
   $\delta t$ & $100$\,s & $100$\,s     & & $100$\,s\\
   $\Delta t$ & 3\,h & 3\,h  &  & 3\,h\\
   $T_r$ & 0-10\% & 5\%  &  & 5\%\\ 
   $\rm T_{\rm lab}$ & 300\,K &300\,K &  & 300\,K \\
   $D$ & 8 & 8  & 8 & 8 \\ 
   $N_{\rm tel}$ & 2 &  2 & 2 & 2 \\
   ${\Delta \lambda}/{\lambda}$  & 0.23 & 0.23    & 0.23 & 0.23\\
   ${\lambda}/{\delta \lambda}$  & 100 & 100  & 100 & 100 \\ 
   $\lambda_B$ & J/H/K/L & I/K  & K & I/J/H/K/L \\ 
   Strehl &  & 98  & 98 & 88/98 \\
   $|\bm B|$ &  &   &  100/200\,m  & 100/200\,m\\ 
   $n_{\rm poly}$ &  &   & 3-7-11  & 11\\ 
   \hline                                   
   \end{tabular}
   \tablefoot{
      $F_{\rmstar}$ is flux of the stellar host. $\delta t$ is the integration time of individual frames while $\Delta t$ the total integration time. $T_r$ is the total transmission of the instrument, $D$ is the diameter of the telescopes and $N_{\rm tel}$ is the number of telescopes. ${\Delta \lambda}/{\lambda}$ the spectral bandwidth and ${\lambda}/{\delta \lambda}$ is the spectral resolution. $\lambda_B$ is the name of the spectral band.  $|\bm B|$ is the projected baseline length, while $n_{\rm poly}$ is the polynomial order used for stellar light post processing.
   }
\end{table}

\subsection{Starlight suppression}
\label{sec:starlight}
   
\subsubsection{Attenuation formulas}

At short wavelengths, thermal background noise and readout noise limit the achievable contrast to values of $\sim 10^{-9}$. Unfortunately, the shot noise caused by stellar speckles can be several orders of magnitude above this limit. Therefore, it is necessary to remove as much of this parasitic light as possible, a process known as starlight suppression.

For high-contrast interferometry, we propose leveraging the efficient spatial filtering provided by single-mode fibres. This filtering can be combined with pupil-plane phase apodization, which can achieve near-perfect rejection of stellar light into the fibre. The effect of the apodization can be modelled with two coupling coefficients: one affecting the stellar light ($\eta_{\rmstar}^\mathrm{inj}$) and the other the planetary light ($\eta_{\rmplanet}^{\rm inj}$). The shot noise caused by the stellar flux then becomes:
\begin{equation}
   \noise_{\rm shot} = \sqrt{\gamma F_{\rmstar} \eta_{\rmstar}^{\rm inj}},
   \label{eq:nshot}
\end{equation}
while the exoplanet signal is given by:
\begin{equation}
   S_{\rm interfero} = \gamma {F_{\rmplanet} \eta_{\rmplanet}^{\rm inj}}.
   \label{eq:s_coh_2}
\end{equation}

Phase apodization works by adding an extra phase term $\Psi(\bm{u})$ in the pupil-plane of the telescope, for example using an offset to the deformable mirror of the AO. With this additional phase term in the pupil-plane, the coupling formulas for the electric field in the second interferometer (see Table~\ref{tb:elec}) are given by:
\begin{align}  
  & E_{2, \rmstar} =  A_\rmstar \scalar{e^{i\Phi(\bm u) + i\Psi(\bm{u})}}{e^{-ik\bm{\alpha}\cdot{}\bm{u}}}\,\\
  & E_{2, \rmplanet} =  A_\rmplanet \scalar{e^{i\Phi(\bm u) + i\Psi(\bm{u})}}{1}\,
\end{align}
and the corresponding apodization coefficients are: 
\begin{align}
   \label{eq:apod_star}
   & \eta_{\rmstar}^{\rm inj} = \frac{ \left|  E_{2, \rmstar} \right|^2}{F_\rmstar} = \left|\scalar{e^{i\Phi(\bm{u}) + i\Psi(\bm{u})}}{e^{-ik\bm{\alpha}\cdot\bm{u}}}\right|^2 \\
   \label{eq:apod_planet}
   & \eta_{\rmplanet}^{\rm inj} = \frac{ \left|  E_{2, \rmplanet}  \right|^2}{F_\rmplanet} = \left|\scalar{e^{i\Phi(\bm{u}) + i\Psi(\bm{u})}}{1}\right|^2.
\end{align}

These expressions are valid for a single value of the wave-vector $k$. For a polychromatic source, these expressions must be integrated over the waveband to obtain appropriate values for the apodization coefficients.

\subsubsection{Designing a phase apodizer}
\label{sec:inj}

The overall objective of a good apodization scheme is of course to minimise the value of $\eta_\rmstar^{\rm inj}$ while maximising the value of $\eta_\rmplanet^{\rm inj}$. There are various strategies for doing so, as multiple solutions exist. State-of-the-art techniques, which include the solutions proposed by \cite{porPhaseapodizedpupilLyotCoronagraphs2020} or \cite{haffertSinglemodeComplexAmplitude2020}, leverage from an iterative optimisation algorithm \citep{Por2017}.

In this study, we adopt a more qualitative approach, which takes advantage of the fact that we are only interested in minimising the flux at a particular position, corresponding to the location of the fibre. Therefore, we know in advance for which value of $\bm{\alpha}$ our apodization function $\Psi_{\bm{\alpha}}$ should be designed. We use a two-step process: the first step involves a slight simplification of the apodization problem to derive a simple analytical solution, which is then used as a starting point for a numerical optimisation to derive the final apodization function. This second step is ultimately intended to be performed in real-time using a dedicated control-loop in the instrument, the details of which are outside the scope of the present study. 

To derive an analytical solution, we first simplify the problem by ignoring the impact of atmospheric fluctuations, setting $\Phi(\bm{u}) = 0$ in Eq.~\eqref{eq:apod_star} and \eqref{eq:apod_planet}. We also simplify the apodization function by enforcing a set of symmetries. Specifically, we look for a solution that is a function of $\bm{\alpha}\cdot{}\bm{u}$ (axially symmetric) and anti-centrosymmetric\footnote{A similar derivation can also be done without the anti-centrosymmetry requirement, in the case where we want to cancel imaginary speckles}:
\begin{align}
   & \Psi_{\bm{\alpha}}(\bm{u}) = \Psi_{\bm{\alpha}}(\bm{u'}) \quad \text{if} \quad \bm{\alpha} \cdot \bm{u} = \bm{\alpha} \cdot \bm{u'} \label{enum:sym1}, \\
   & \Psi_{\bm{\alpha}}(\bm{u}) = -\Psi_{\bm{\alpha}}(-\bm{u}) \label{enum:sym2}.
\end{align}

We also assume that the apodization function is small, i.e. that $\Psi_{\bm{\alpha}}(\bm{u}) \ll 1 $. This is consistent with the 
requirement that $\eta_\rmplanet^{\rm inj}$ should be as close to 1 as possible. This can be seen by using a linear development of Eq.~\eqref{eq:apod_planet} to the second order: 
\begin{equation}
   \eta_\rmplanet^{\rm inj} = \left|\scalar{e^{i\Psi_{\bm{\alpha}}(\bm{u})}}{1}\right|^2 \simeq{} \left|\scalar{1}{1} - \frac{1}{2} \scalar{\Psi_{\bm{\alpha}}(\bm{u})}{\Psi_{\bm{\alpha}}(\bm{u})}\right|^2\,.
\end{equation}
The first order has canceled out due to the central symmetry of $\Psi_{\bm{\alpha}}$: $\scalar{\Psi_{\bm{\alpha}}}{1} =0 $. Also, 
we made use the fact that $\Psi$ is a real function, which implies $\scalar{\Psi_{\bm{\alpha}}^2}{1} = \scalar{\Psi_{\bm{\alpha}}}{\Psi_{\bm{\alpha}}}$ from the definition of $\scalar{\cdot}{\cdot}$ given in Eq.~\eqref{eq:scalar}. With such a form, $\eta_\rmplanet^{\rm inj}$ can be expected to be close to $\scalar{1}{1} = 0.81$ if $\Psi_{\bm{\alpha}}$ is small.

The first order development of Eq.~\eqref{eq:apod_star} for the starlight apodization, however, gives:
\begin{align}
   \eta_{\rmstar}^{\rm inj} & =\left|\scalar{e^{i\Psi(\bm{u})}}{e^{-ik\bm{\alpha}\cdot\bm{u}}}\right|^2 \\
   &\simeq{} \left|\scalar{1}{e^{-ik\bm{\alpha}\cdot{}\bm{u}}} + \scalar{i\Psi_{\bm{\alpha}}(\bm{u})}{e^{-ik\bm{\alpha}\cdot{}\bm{u}}}\right|^2 \\
   &\simeq{} \left|\scalar{1}{\cos\left(k\bm{\alpha}\cdot{}\bm{u}\right)} - \scalar{\Psi_{\bm{\alpha}}(\bm{u})}{\sin\left(k\bm{\alpha}\cdot{}\bm{u}\right)}\right|^2,
\end{align}
\noindent{}where the last line makes use of the fact that $\scalar{1}{\sin\left(k\bm{\alpha}\cdot{}\bm{u}\right)}$ and $\scalar{\Psi_{\bm{\alpha}}(\bm{u})}{\cos\left(k\bm{\alpha}\cdot{}\bm{u}\right)} = 0 $ due to the definition of $\scalar{\cdot}{\cdot}$ and the symmetries of the functions involved. This equation shows that the injected starlight can be minimised by setting:
\begin{equation}
   \label{eq:orthogonal}
   \scalar{\Psi_{\bm{\alpha}}(\bm{u})}{\sin\left(k\bm{\alpha}\cdot{}\bm{u}\right)} = \scalar{1}{\cos\left(k\bm{\alpha}\cdot{}\bm{u}\right)}.
\end{equation}

This guarantees that the injected starlight is 0 at the particular wavevector $k$, but does not guarantee that the injected light is minimum over the entire waveband. If the waveband is small enough, minimisation over the entire band can be ensured by setting the successive derivatives of $\eta_\rmstar^{\rm inj}$ with respect to $k$ to 0. The first and second derivates are given by:
\begin{small}
\begin{align}
   \frac{\mathrm{d}\eta_\rmstar^{\rm inj}}{\mathrm{d}k} & =  2\scalar{e^{i\Psi_{\bm{\alpha}}(\bm{u})}}{e^{-ik\bm{\alpha}\cdot{}\bm{u}}} \Re{\left(\scalar{e^{i\Psi_{\bm{\alpha}}(u)}}{\left[-i\bm{\alpha}\cdot{}\bm{u}\right]e^{-ik\bm{\alpha}\cdot{}\bm{u}}}\right)}, \\
   \nonumber \frac{\mathrm{d^2}\eta_\rmstar^{\rm inj}}{\mathrm{d}k^2} & = 2\scalar{e^{i\Psi_{\bm{\alpha}}(\bm{u})}}{\left[-i\bm{\alpha}\cdot{}\bm{u}\right]e^{-ik\bm{\alpha}\cdot{}\bm{u}}}\Re{\left(\scalar{e^{i\Psi_{\bm{\alpha}}(u)}}{\left[-i\bm{\alpha}\cdot{}\bm{u}\right]e^{-ik\bm{\alpha}\cdot{}\bm{u}}}\right)} \\ 
   & \quad + 2\scalar{e^{i\Psi_{\bm{\alpha}}(\bm{u})}}{e^{-ik\bm{\alpha}\cdot{}\bm{u}}} \frac{\mathrm{d}\Re{\left(\scalar{e^{i\Psi_{\bm{\alpha}}(u)}}{\left[-i\bm{\alpha}\cdot{}\bm{u}\right]e^{-ik\bm{\alpha}\cdot{}\bm{u}}}\right)}}{\mathrm{d}k}.
   \label{eq:d2}
\end{align}
\end{small}
\noindent{}Given our previous Eq.~\eqref{eq:orthogonal}, which guarantees that $\eta_\rmstar^{\rm inj} = \left|\scalar{e^{i\Psi_{\bm{\alpha}}(\bm{u})}}{e^{-ik\bm{\alpha}\cdot{}\bm{u}}}\right|^2 = 0$, the first derivative is necessary 0, and the second derivative can be simplified to:
\begin{small}
\begin{equation}
   \frac{\mathrm{d^2}\eta_\rmstar^{\rm inj}}{\mathrm{d}k^2} = 2\scalar{e^{i\Psi_{\bm{\alpha}}(\bm{u})}}{\left[i\bm{\alpha}\cdot{}\bm{u}\right]e^{-ik\bm{\alpha}\cdot{}\bm{u}}}\Re{\left(\scalar{e^{i\Psi_{\bm{\alpha}}(u)}}{\left[i\bm{\alpha}\cdot{}\bm{u}\right]e^{-ik\bm{\alpha}\cdot{}\bm{u}}}\right)}.
\end{equation}
\end{small}
\noindent{}This last expression shows that canceling the second derivative of $\eta_\rmstar^{\rm inj}$ with respect to $k$ requires canceling out the real part of $\scalar{e^{i\Psi_{\bm{\alpha}}(u)}}{\left[ik\bm{\alpha}\cdot{}\bm{u}\right]e^{-ik\bm{\alpha}\cdot{}\bm{u}}}$, where we took advantage of the linearity of the scalar product to add a factor $k$ in the second term and keep all elements dimensionless. 

By using again the first-order development of $e^{i\Psi_{\bm{\alpha}}}$ and the different symmetries, we find:
\begin{small}
\begin{align}
   \Re{\left(\scalar{e^{i\Psi_{\bm{\alpha}}(u)}}{\left[ik\bm{\alpha}\cdot{}\bm{u}\right]e^{-ik\bm{\alpha}\cdot{}\bm{u}}}\right)} & \simeq{} 
   \scalar{1}{\left[k\bm{\alpha}\cdot{}\bm{u}\right]\sin(k\bm{\alpha}\cdot{}\bm{u})} \nonumber \\
   & \quad + \scalar{\Psi_{\bm{\alpha}}(\bm{u})}{\left[k\bm{\alpha}\cdot{}\bm{u}\right]\cos(k\bm{\alpha}\cdot{}\bm{u})},
\end{align} 
\end{small}
which shows that in order to cancel out this real part, we need:
\begin{align}
   \scalar{\Psi_{\bm{\alpha}}(\bm{u})}{\left[k\bm{\alpha}\cdot{}\bm{u}\right]\cos(k\bm{\alpha}\cdot{}\bm{u})} = -\scalar{1}{\left[k\bm{\alpha}\cdot{}\bm{u}\right]\sin(k\bm{\alpha}\cdot{}\bm{u})}.
\end{align}

Putting everything together, to solve our apodization problem, we are therefore looking for a function $\Psi_{\bm{\alpha}}$ such that:
\begin{align}
   & \scalar{\Psi_\alpha(\bm{u})}{\Psi_\alpha(\bm{u})} = \norm{\Psi_{\bm{\alpha}}}^2 \text{ is as small as possible,} \label{eq:cond1}\\
   & \scalar{\Psi_{\bm{\alpha}}(\bm{u})}{H_1(u)} = z_1, \label{eq:cond2}\\
   & \scalar{\Psi_{\bm{\alpha}}(\bm{u})}{H_2(u)} = z_2, \label{eq:cond3}
\end{align}
\noindent{}where:
\begin{align}
   & H_1(u) = \sin\left(k\bm{\alpha}\cdot{}\bm{u}\right), \\
   & z_1 = \scalar{1}{\cos\left(k\bm{\alpha}\cdot{}\bm{u}\right)}, \\
   & H_2(u) = \left[k\bm{\alpha}\cdot{}\bm{u}\right]\cos\left(k\bm{\alpha}\cdot{}\bm{u}\right), \\
   & z_2 = - \scalar{1}{\left[k\bm{\alpha}\cdot{}\bm{u}\right]\sin\left(k\bm{\alpha}\cdot{}\bm{u}\right)}.
\end{align}

This system is readily solved by noting that in order to minimise its norm in Eq.~\eqref{eq:cond1}, while fulfilling the two conditions given by Eq.~\eqref{eq:cond2} and \eqref{eq:cond3}, $\Psi_{\bm{\alpha}}$ needs to be in the plane defined by $(H_1, H_2)$:
\begin{equation}
   \Psi_{\bm{\alpha}} = a H_1 + b H_2 \,.
\end{equation}   
Then, solving the linear system resulting from Eq.~\eqref{eq:cond2} and \eqref{eq:cond3} gives the following closed-form solution for $\Psi_{\bm{\alpha}}$:
\begin{small}
\begin{equation}
   \Psi_{\bm{\alpha}} = \frac{z_1 \norm{H_2}^2 - z_2 \scalar{H_1}{H_2}}{\norm{H_1}^2 \norm{H_2}^2 - {\scalar{H_1}{H_2}}^2} H_1 + \frac{z_2 \norm{H_1}^2 - z_1 \scalar{H_1}{H_2}}{\norm{H_1}^2 \norm{H_2}^2 - {\scalar{H_1}{H_2}}^2} H_2.
   \label{eq:solution}
\end{equation}
\end{small}

This closed-form solution for the apodization is valid under the conditions that $\Psi_{\bm{\alpha}}$ remains small and in the absence of atmospheric fluctuations. It also ignores the true dependency of the injected flux as a function of the wavelength, as we have only tried to cancel the flux at a particular wavenumber in Eqs.~\eqref{eq:cond2} and \eqref{eq:cond3}. Moreover, in practice, 
the optical path delays is applied by a deformable mirror and it is therefore in units of length. So our strategy is to use the solution given in Eq.~\eqref{eq:solution} as an initial guess in an optimisation process.

To do so, we keep $\Psi_{\bm{\alpha}}$ as a linear combination of two functions $H_1$ and $H_2$ defined for a central wavenumber $k_0$, and we directly integrate the square root of the scalar product over all wavenumbers $k$:
\begin{equation} 
   \eta_\rmstar^{\rm inj}(a, b) = \int_{k_0-\Delta k/2}^{k_0+\Delta k/2} \left|\scalar{e^{ i \frac{k}{k_0} \left[ a \sin(k_0 \bm{\alpha} \cdot \bm{u}) + b k_0 \bm{\alpha} \cdot \bm{u} \cos(k_0 \bm{\alpha} \cdot \bm{u}) \right]}}{e^{-ik\bm{\alpha}\cdot{}\bm{u}}}\right|^2 \frac{\mathrm{d}k}{\Delta k}\,.
\end{equation}
This apodization coefficient can then be numerically minimised to find the coefficients $a$ and $b$:
\begin{equation}
   (a, b) = \mathrm{argmin}\left\{\eta_\rmstar^{\rm inj}(a, b)\right\}\,,
   \label{eq:opt}
\end{equation}
which will give a set of values $(a, b)$ for any value of the angular separation $\bm{\alpha}$. 

\begin{figure}
   \centering
   \includegraphics[width=9cm]{"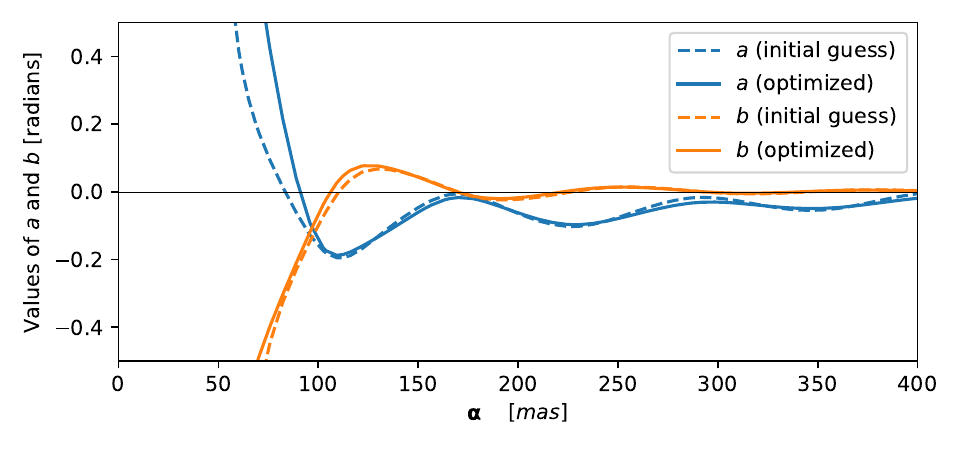"}
   \includegraphics[width=9cm]{"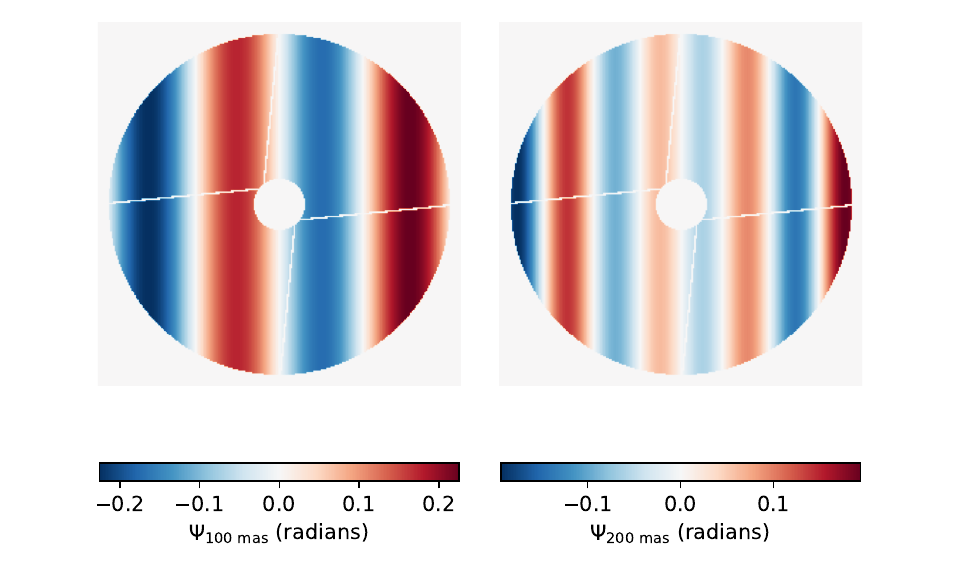"}
   \caption{Phase maps for pupil plane apodization. \textit{Upper panel:} Coefficients $(a, b)$ used to compute the phase apodizer as $\Psi_{\bm{\alpha}} = a H_1 + b H_2$. The solid lines show the optimised coefficient used in the final definition of the apodizer, i.e. the solution of Eq.~\eqref{eq:opt}. The dashed lines show the initial guess calculated from Eq.~\eqref{eq:solution}. The simulation has been made for K band observations ($k_0=2\pi/2.2\,\mu$m) and the shape of the VLT pupil. \textit{Lower panels :} The apodizer function, $\Psi_{\bm{\alpha}}(\bm u)$, for a target at an angular separation of $100\,$mas (left panel) and $200\,$mas (right panel).}
   \label{fig:kApod}
\end{figure}

In Figure~\ref{fig:kApod}, we show the values of the coefficients $a$ and $b$ obtained both by the closed-form solution of the approximated apodization problem from Eq.~\eqref{eq:solution}, and after the optimisation process from Eq.~\eqref{eq:opt}, as a function of the separation $\bm{\alpha}$. Also shown are the apodization functions $a H_1 + b H_2$ for the two specific cases of $\bm{\alpha} = \lambda/D$ and $\bm{\alpha} = 2\lambda/D$.

\subsubsection{Photon noise after pupil apodization}
\label{sec:photon_noise}

\begin{figure}
   \centering
   \includegraphics[width=9cm]{"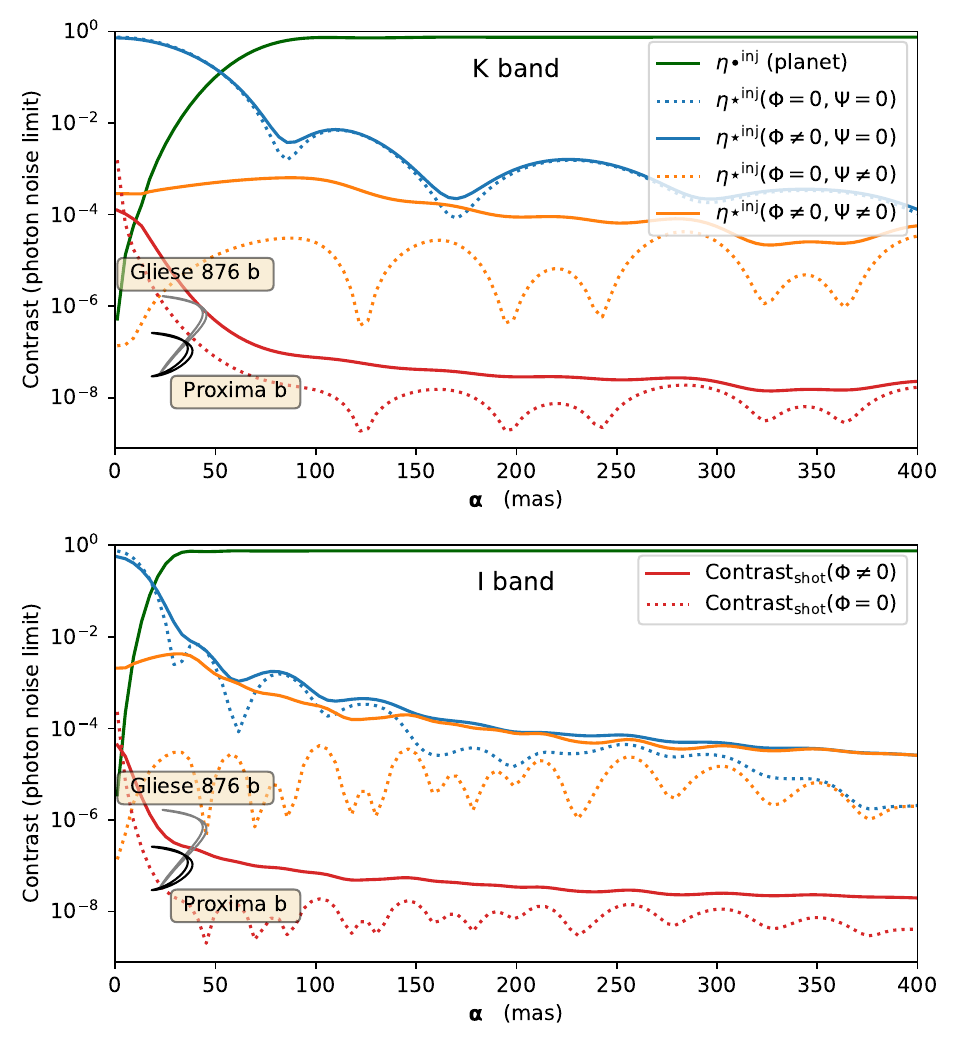"}
   \caption{ Apodization due to the spatial filtering caused by single-mode fibres as a function of the angular separation $\bm{\alpha}$ between the star and the exoplanet, in the K band (upper panel) and in the I band (lower panel). 
   The green curves show $\eta_\rmplanet^{\rm inj}$, the injection of the exoplanet flux into interferometer 2, while the blue and orange curves represent the injection of the stellar flux ($\eta_\rmstar^{\rm inj}$).
   The blue curves are without a phase apodization ($\Psi_{\bm \alpha}=0$), while the orange curves are with phase apodization ($\Psi_{\bm \alpha}=a H_1 + b H_2$). 
   The solid curves include the residuals of atmospheric correction by an extreme AO system, giving a Strehl ratio of 98\% in the K band and 72\% in the I band. The dotted curves are obtained assuming no atmospheric residuals: $\Phi(\bm u)=0$, hence a Strehl ratio of 100\%.
   The red curves correspond to the contrast limit, $C_{\rm shot}$ in Eq~\eqref{eq:C_shot}, that would be achieved when limited by photon noise, for an observation with parameters as stipulated in Table~\ref{tb:simulation}.
   The phase apodizer is efficient at decreasing the flux injected into the fibre in the K band. However, at shorter wavelengths, the lower Strehl ratio makes it less effective, with the injected flux mostly due to the atmospheric halo.}
   \label{fig:Astar}
\end{figure}

To calculate the amount of photon noise introduced by the starlight, we have to account for the atmospheric perturbations.
 To do so, we simulated the atmospheric perturbations using HCIPy, a Python framework for high-contrast imaging simulation \citep{Por2018}.
 The atmospheric residuals are obtained using a closed-loop adaptive optics system similar to the one used in the SPHERE instrument, which employs a Shack-Hartmann Wavefront Sensor (WFS) with 40 by 40 lenslets. The atmospheric conditions suppose a seeing of 0.6'' at 500\,nm, a coherence time of 5\,ms, and an outer scale parameter of 40 meters. 
 The simulations include WFS Photon Noise and the Time Lag of the AO system. The aberrations have been adjusted to the lab measurements of the SAXO optimal performance, giving 98\% in K band, and 72\% in I band \citep{fuscoSAXOSPHEREExtreme2016}.
 
 The phase masks $\Phi(u)$ generated by the atmospheric residuals are then added to the phase mask $\Psi(u)$  of the apodizer to calculate the two broad band attenuations coefficients:
\begin{equation} 
   \eta_\rmstar^{\rm inj} = \int_{k_0-\Delta k/2}^{k_0+\Delta k/2} \left|\scalar{e^{i\Phi(\bm{u})+i\Psi(\bm{u})}}{e^{-ik\bm{\alpha}\cdot\bm{u}}}\right|^2 \frac{\mathrm{d}k}{\Delta k}
\end{equation}
and
\begin{equation} 
   \eta_\rmplanet^{\rm inj} = \int_{k_0-\Delta k/2}^{k_0+\Delta k/2} \left|\scalar{e^{i\Phi(\bm{u})+i\Psi(\bm{u})}}{1}\right|^2 \frac{\mathrm{d}k}{\Delta k}\,.
\end{equation}

The two panels of Figure~\ref{fig:Astar} display simulations under these atmospheric conditions in the two bands. For reference, the dotted curves in both panels represent a system without atmosphere (Strehl ratio of 100\%). These simulations use parameters representative of the VLTI, as listed in Table~\ref{tb:simulation}.

Figure~\ref{fig:Astar} also shows the contrast limit obtained assuming a threshold of 3 on the S/N, using $S_{\rm interfero}$ for the signal as defined in Eq.~\eqref{eq:s_coh_2}, and $\noise_{\rm shot}$ for the noise as defined by Eq.~\eqref{eq:nshot}. According to this threshold, the contrast is defined as follows:
\begin{equation}
   C_{\rm shot} = 3 \frac{1}{\sqrt{\gamma F_\rmstar}} \frac{\sqrt{\eta_{\rmstar}^{\rm inj}}}{\eta_{\rmplanet}^{\rm inj}}\,.
   \label{eq:C_shot}
\end{equation}

These curves indicate that the photon noise contrast limit is just sufficient to detect the exoplanet Gliese 876b in the K band (if the albedo is 0.3). Proxima Cen b, on the other hand, is only observable at smaller angular separations and can only be detected at shorter wavelengths. 

Achieving a stellar extinction ratio of at least $1/\eta_\rmstar^{\rm inj} \geq 10^{3}$ is essential for detecting exoplanets with contrast ratios around $10^{-8}$. This can be accomplished either at longer wavelengths with a very high Strehl ratio and excellent apodization, or by leveraging the smaller diffraction pattern at shorter wavelengths, which somewhat relaxes the requirements on Strehl and apodization. Because this balance, which results in giving a raw contrast relatively independent of wavelength, may seem counterintuitive, it is discussed in more detail in Appendix~\ref{sec:appA}.

\subsection{Coherent speckle removal}
\label{sec:postprocessing}

\subsubsection{Principle of post-processing}

Due to the chromatic behaviour of stellar light and due to residual optical aberrations, an apodizer alone cannot completely eliminate the coherent stellar light. However, the residual coherent speckles can be removed by a dedicated post-processing algorithm. This is similar to what happens in direct imaging, where the use of a coronagraph is always complemented by a post-processing technique (for example ADI or SDI) to reach the highest levels of contrast.

In dual-field interferometry, the residual coherent speckles can be removed by taking advantage of the simultaneous observation of the central-star by the first interferometer. A coherent flux $\visibility{1}{\rmstar}$ is measured on the first interferometer (see Figure~\ref{fig:schema}), which can be used to phase-reference the stellar coherent speckles on the second interferometer, thanks to the internal monitoring of the non-common optical path by the metrology. 

In practice, the two coherent fluxes $\visibility{1}{\rmstar}$ and $\visibility{2}{\rmstar}$ are not equal, as they can have vastly different amplitudes. But they remain similar in the sense that neither their phase, nor their spectral shape are significantly different. Experience \citep[e.g.,][]{gravitycollaborationPeeringFormationHistory2020}, as well as numerical simulations \citep{2024arXiv240604003P}, shows that these differences are well captured by a low-order polynomial in $k$, the wavenumber:
\begin{equation}
   \frac{\visibility{2}{\rmstar}}{\visibility{1}{\rmstar}} = \sum_{n=0}^{n_{\rm poly}} a_n k^n,
\end{equation}
where the coefficients $a_n$ are complex-coefficients which represent the chromatic difference between the on-axis stellar light $\visibility{1}{\rmstar}$ and the off-axis stellar light $\visibility{2}{\rmstar}$. The order $n_{\rm poly}$ is typically 5 to 15 for a good approximation\footnote{We also tested spline-based linear models, similar to those used by \citet{2019AJ....158..200R,2023AJ....165..113R} to model the continuum, but did not observe any improvement for an identical number of degrees of freedom.}.

Another advantage of using a metrology system to reference one interferometer to the other is the ability to phase reference the planet light as well. It guarantees that the phase of the coherent flux observed in the second interferometer will follow the analytical formula for a binary companion at a given separation ${\bm{\alpha}}$:
\begin{equation} 
   \frac{\visibility{2}{\rmplanet}}{\visibility{1}{\rmstar}} = \left(\frac{\mathsf F_{\rmplanet}}{\mathsf F_{\rmstar}}\right)e^{-i k \left(\bm \alpha \cdot \bm B\right)} = C e^{-i k \left(\bm \alpha \cdot \bm B\right)} \,,
   \label{eq:visplanet_ref}
\end{equation}
ignoring the spectral differences between the planet and the star by reducing the contrast $F_{\rmplanet}/F_{\rmstar}$ to a single parameter $C$.

Hence, there is a structural difference between the phase of the low-order polynomial-like off-axis coherent flux, $\arg(\visibility{2}{\rmstar})$, and the phase of planet coherent flux, $\arg(\visibility{2}{\rmstar})$.
It is this difference which is used to filter out $\visibility{2}{\rmstar}$ from the combined coherent signal $\Gamma_2 = \visibility{2}{\rmstar} + \visibility{2}{\rmplanet}$ without significantly impacting the second. This idea is at the heart of the ExoGRAVITY Collaboration's post-processing technique, pioneered in \cite{gravitycollaborationFirstDirectDetection2019} and detailed in \cite{gravitycollaborationPeeringFormationHistory2020}. In the rest of this Section, we use the algorithms detailed in these publications to estimate the residual stellar contamination $\noise_{\rm speckle}$ in Eq.~\eqref{eq:snr}.

\subsubsection{Mathematical formulation}

As outlined by \cite{gravitycollaborationPeeringFormationHistory2020} and \cite{nowakDirectConfirmationRadialvelocity2020}, the formalism of the ExoGRAVITY method involves considering all wavelength-dependant quantities as column-vectors. In this formalism, the coherent flux $\Gamma$ is a column-vector $\left[\Gamma(k_1), \Gamma(k_2), \dots{}, \Gamma(k_n)\right]^T$. Retrieving the value of $C$ involves two steps: (1) projecting the coherent flux vector $\Gamma_2$ onto a subspace orthogonal to that of the coherent speckle model of $\visibility{2}{\rmstar}$; (2) projecting the theoretical planet coherent flux from Eq.~\eqref{eq:visplanet_ref} onto this same subspace to retrieve the contrast $C$.

In the first step, the subspace orthogonal to the coherent speckle is calculated from the series of vectors $J_n$ and $Q_n$ defined as:
\begin{align}
   J_n &= k^n \, \visibility{1}{\rmstar} \\
   Q_n &= i k^n \, \visibility{1}{\rmstar}.
\end{align}
The projector $P_\perp$ on the subspace orthogonal to all the $J_n$ and $Q_n$, $n\in \{0, \dots{}, n_{\rm poly}\}$ is the matrix $P_\perp$ defined by:
\begin{align}
   P_\perp = I_{n_{\rm poly}} - \sum_{n=0}^{n_{\rm poly}} \left[\frac{J_n^T}{J_n^T\cdot{}J_n}  + \frac{Q_n^T}{Q_n^T\cdot{}Q_n}\right].
\end{align}
The first step therefore consists in multiplying the coherent flux by $P_\perp$:
\begin{equation}
   P_\perp \cdot \Gamma_2 = P_\perp \cdot \visibility{2}{\rmstar} + P_\perp \cdot \visibility{2}{\rmplanet},
\end{equation}
with the hope that the starlight residual, $P_\perp \cdot{} \visibility{2}{\rmstar}$, will be close to 0. 

In the second step, we calculate the vector:
\begin{equation}
   K_{\bm{\alpha}} = P_\perp \cdot \left[ \visibility{1}{\rmstar} e^{-ik\left(\bm{\alpha}\cdot\bm{B}\right)} \right],
\end{equation}
which serves as a model for the coherent flux of the exoplanet $\visibility{2}{\rmplanet}$, as expected according to Eqs.~\eqref{eq:visplanet_ref}, projected in the same subspace as the data. Assuming $P_\perp \cdot \visibility{2}{\rmstar}=0$, we now have:
\begin{equation}
   P_\perp \cdot \Gamma_2 = C K_{\bm{\alpha}}, 
\end{equation}
which gives the solution:
\begin{equation}
   C = \frac{K_{\bm{\alpha}}^T \cdot P_\perp}{K_{\bm{\alpha}}^T \cdot K_{\bm{\alpha}}}\cdot \Gamma_2.
\end{equation}

This product post-processing is required to extract the planet signal from the stellar speckles, but it also attenuates the signal by reducing $\visibility{2}{\rmplanet}$.

\begin{figure}
   \centering
      \includegraphics[width=9cm]{"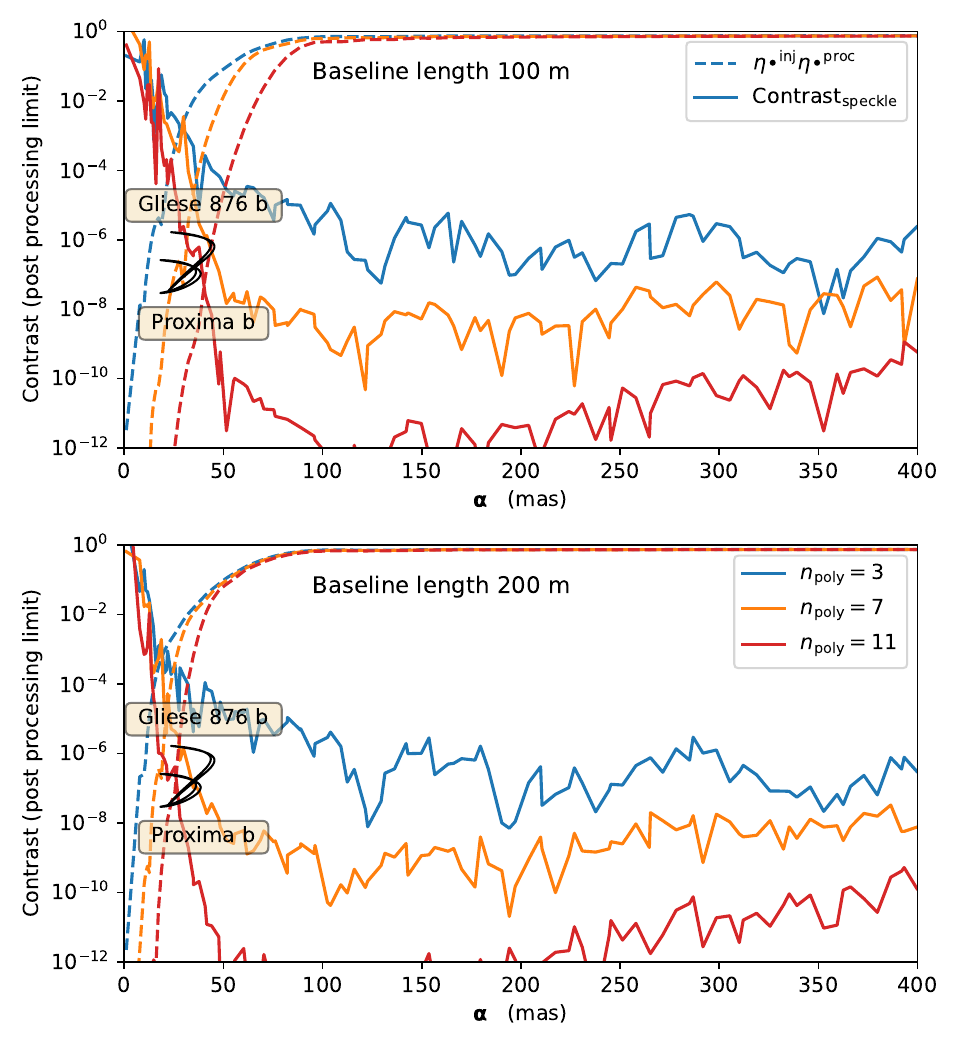"}
      \caption{Attenuation of the exoplanet signal ($\eta_{\rmplanet}^{\rm inj} \eta_{\rmplanet}^{\rm proc}$, in dashed lines) and the corresponding contrast limit ($C_{\rm speckle}=3\eta_{\rmstar}^{\rm inj} \eta_{\rmstar}^{\rm proc}/\eta_{\rmplanet}^{\rm inj} \eta_{\rmplanet}^{\rm proc}$, in solid lines) after post-processing of the stellar speckle, as a function of angular separation. The three colour curves represent three levels of polynomial fitting strength ($n_{\rm poly}$ of 3, 7, and 11) used for filtering the coherent speckles. The dotted curves show the attenuation of the exoplanet signal, illustrating that a lower polynomial order yield a smaller inner working angle. The solid curves indicate the contrast limit after post-processing of the stellar speckle, demonstrating that higher polynomial orders result in a better contrast limit. The upper plot represents observations with a 100\,m baseline, while the lower plot corresponds to a 200\,m interferometer baseline.}
      \label{fig:Aprocess}
\end{figure}

\subsubsection{Simulation of post-processing attenuations}

To calculate the level of attenuation of the coherent star and exoplanetary light, we must account for atmospheric perturbations. We again used HCIPy to simulate the wavefront residuals on two telescopes, each equipped with a SPHERE-like AO system with similar performance. As in Section~\ref{sec:photon_noise}, we then added a phase apodizer to obtain two sets of phase masks $\Phi_{\mathbb A}(\bm u)$ for telescope ${\mathbb A}$ and $\Phi_{\mathbb B}(\bm u)$ for telescope ${\mathbb B}$. Both $\Phi_{\mathbb A}$ and $\Phi_{\mathbb B}$ depend on $k$ and time.

The total attenuation can be calculated from the electric field expressed in Table~\ref{tb:elec}:
\begin{equation}
   \eta_\rmplanet^{\rm inj}\eta_\rmplanet^{\rm proc} = \frac{K_{\bm{\alpha}}^T \cdot P_\perp}{K_{\bm{\alpha}}^T \cdot K_{\bm{\alpha}}} 
   \cdot \begin{bmatrix} \scalar{e^{i\Phi_{\mathbb A}(\bm u)}}{1}\scalar{e^{i\Phi_{\mathbb B}(\bm u)}}{1}^\ast e^{-i k_1 \bm{\alpha}\cdot{}\bm{B}}\\ \vdots \\  \scalar{e^{i\Phi_{\mathbb A}(\bm u)}}{1}\scalar{e^{i\Phi_{\mathbb B}(\bm u)}}{1}^\ast e^{-i k_{n_\lambda} \bm{\alpha}\cdot{}\bm{B}}\end{bmatrix}.
\end{equation}

Similarly, the attenuation of the coherent stellar speckle can be written as:
\begin{equation}
   \eta_\rmstar^{\rm inj}\eta_\rmstar^{\rm proc} = \frac{K_{\bm{\alpha}}^T \cdot P_\perp}{K_{\bm{\alpha}}^T \cdot K_{\bm{\alpha}}} 
   \cdot \begin{bmatrix} \scalar{e^{i\Phi_{\mathbb A}(\bm u)}}{e^{-ik_1\bm{\alpha}\cdot{}\bm{u}}}\scalar{e^{i\Phi_{\mathbb B}(\bm u)}}{e^{-ik_1\bm{\alpha}\cdot{}\bm{u}}}^\ast \\ \vdots \\ \scalar{e^{i\Phi_{\mathbb A}(\bm u)}}{e^{-ik_{n_\lambda}\bm{\alpha}\cdot{}\bm{u}}}\scalar{e^{i\Phi_{\mathbb B}(\bm u)}}{e^{-ik_{n_\lambda}\bm{\alpha}\cdot{}\bm{u}}}^\ast\end{bmatrix}.
\end{equation}
The attenuations due to apodization and post-processing are combined in our analysis because it is difficult to calculate the attenuation of post-processing without including the spatial filtering of the single-mode fibre.

This residual starlight, which is neither rejected by the fibre and the apodization nor by the post-processing algorithm, corresponds to the speckle noise term in Eq.~\eqref{eq:snr}:
\begin{equation}
   \noise_{\rm speckle} = \gamma F_{\rmstar} \eta_{\rmstar}^{\rm inj} \eta_{\rmstar}^{\rm proc}.
   \label{eq:N_speck}
\end{equation}

In Figure~\ref{fig:Aprocess}, we show the evolution of $\eta_\rmplanet^{\rm inj}\eta_\rmplanet^{\rm proc}$ as a function of the angular separation $\bm{\alpha}$, for three different levels of aggressiveness in deconvolution, corresponding to three different orders for the polynomial model of the coherent residual starlight ($n_{\rm poly}=3$, 7, and 11), and two baseline lengths: $B=100~\mathrm{m}$ and 200~m. This graph shows how, as the planet gets closer to the central star ($\alpha \to 0$), its signal becomes too similar to the coherent flux of the central star and gets removed by the post-processing step ($\eta_\rmplanet^{\rm inj}\eta_\rmplanet^{\rm proc} \to 0 $).

We also plot a contrast limit assuming a threshold of 3 on the S/N, using $S_{\rm processed}$ for the signal as defined in Eq.~\eqref{eq:s_proc}, and $\noise_{\rm speckle}$ for the noise as defined in Eq.~\eqref{eq:N_speck}. According to this threshold, the contrast is defined as follows:
\begin{equation}
   C_{\rm speckle} = 3\ \frac{\eta_\rmstar^{\rm inj}\eta_\rmstar^{\rm proc} }{\eta_\rmplanet^{\rm inj}\eta_\rmplanet^{\rm proc}}\,.
   \label{eq:C_speckle}
\end{equation}

\subsection{Final contrast curve}
\label{sec:contrastcurve}

\begin{figure} 
   \centering
      \includegraphics[width=9cm]{"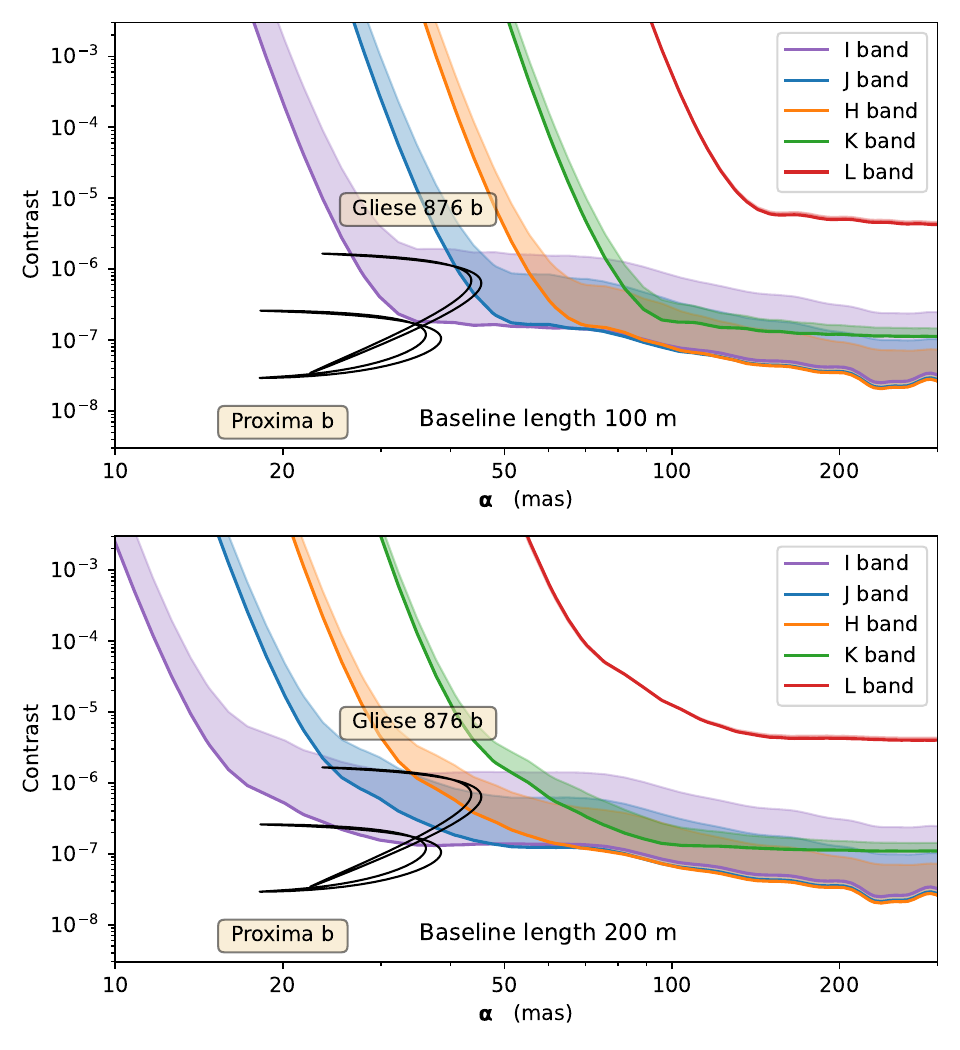"}
      \caption{Contrast limit as a function of angular separation, with simulation parameters detailed in Table~\ref{tb:simulation}. The upper panel illustrates the contrast dependence on waveband for a 100\,m baseline. The lower panel displays the contrast limits for a 200\,m baseline. The shaded area represents a range of K band Strehl ratios, from 88\% (worse contrast limit) to 98\% (best contrast limit). At short wavelength, the contrast range is limited by the Strehl. In the K and L bands, the contrast range is predominantly limited by thermal background noise. The inner working angles are constrained by both the telescope's point spread function and the interferometric resolution, both linear function of wavelength.}
      \label{fig:Cfull}
\end{figure}

Putting everything together, the contrast limit of a dual-field interferometer on the VLTI can be calculated using the formula presented in Eq.~\eqref{eq:snr}, replacing $F_\rmplanet$ by $C\,F_\rmstar$ to convert to contrast, and using an arbitrary threshold of $\mathrm{S/N} = 3$ to define a detection limit:
\begin{equation}
   C = 3\ \frac{\sqrt{\noise_{\rm readout}^2 + \noise_{\rm background}^2 + \gamma F_{\rmstar} \eta_{\rmstar}^{\rm inj} + ( \gamma  F_{\rmstar} \eta_{\rmstar}^{\rm inj} \eta_{\rmstar}^{\rm proc})^2}}{ \gamma  F_{\rmstar} \eta_{\rmplanet}^{\rm inj} \eta_{\rmplanet}^{\rm proc}}.
\end{equation}
The resulting contrast curve is given in Figure~\ref{fig:Cfull}.

This accounts for the cumulative effects of readout noise, thermal background noise, stellar photon noise, and residual speckle noise after post-processing, and provides critical insights into the system's performance. As shown in the upper panel of Fig.~\ref{fig:Cfull}, a contrast of $10^{-6}$ can be achieved at 25 mas in the I band, but this increases to 70 mas in the K band. Reducing the baseline length also results in an inversely proportional increase of the detection limit at low separations, highlighting the importance of baseline extension for reducing this angle beyond the current capabilities of the VLTI.

Another important observation relates to the Strehl ratio, which strongly influences the contrast that can be achieved, impacting both the signal injected into the fibre and the effectiveness of stellar flux apodization, as illustrated in Fig.~\ref{fig:Astar}. As anticipated, the Strehl ratio - and consequently, the contrast - is more adversely affected at shorter wavelengths. Two distinct regimes are observed: in a regime of good correction, above a Strehl ratio of $\simeq{}50\,\%$ our simulations show a preference for shorter wavelengths due to the beneficial increase in angular resolution that compensates for the slight decrease in Strehl ratio. In the regime where the Strehl ratio is low, the correction quality is insufficient for proper light injection into the single-mode fibre, and the advantages of a narrower point spread function no longer offset the losses.

Furthermore, the figure underscores the difficulty of detecting reflected light from exoplanets beyond 2 microns, attributed to the escalation of thermal background noise. At these wavelength, exploring thermal emissions from exoplanets emerges as a viable alternative, albeit likely requiring the colder operational conditions provided by space-based observatories.


\section{An additional Unit Telescope in the context of exoplanet detection}
\label{sec:NewUT5}
\subsection{The effect of the 2D geometry of the VLTI}

Figure~\ref{fig:Cfull} shows how the contrast limit changes as a function of separation for baseline lengths of 100 and 200\,m. However, this does not account for the geometry of the VLTI, which has six baselines. In practice, the geometry of the array is crucial, as the relevant quantity when determining the "baseline length" is the projected baseline vector, \(\bm{\alpha} \cdot \bm{B} / |\bm{\alpha}|\), projected both in the plane of the sky and along the position angle of the exoplanet.

The projection along the plane of the sky depends on the position of the star. In the upper panel of Fig.~\ref{fig:uv_iwa}, the UV coverage is shown for Proxima Cen when it passes the meridian, representing almost optimal conditions. The frequency domain covered by the VLTI (red points) is mostly constrained in the North-East direction due to the intrinsically elongated geometry of the VLTI along the UT1-UT4 baseline.

The second projection is along the position angle (PA) of the exoplanet (the argument of $\bm{\alpha}$): if the exoplanet PA aligns with the direction of the maximal baseline, the contrast will be optimal. However, if a baseline is orthogonal to the planet's position, the baseline becomes useless. Thus, the simultaneous usage of all baselines is crucial to cover all possible PAs of the exoplanet.

To illustrate this effect, Figure~\ref{fig:contrast} shows the contrast limit using all different baselines of the array as a function of the 2D vector $\bm{\alpha}$ shown in RA-DEC. To compute the contrast limit, we calculated the $S_{ij}/N_{ij}$ on each projected baseline and summed it as: 
\begin{equation}
S/N = \sqrt{ \sum_{ij} (S_{ij}/N_{ij})^2 } \,.
\end{equation}
Then, we estimated the final contrast using a requirement of 3 on the S/N. This clearly shows that the performance of the VLTI is fundamentally under-performing in the North-West direction.

\subsection{Limitations of the inner working angle}

\begin{figure}
   \centering
   \includegraphics[width=\linewidth]{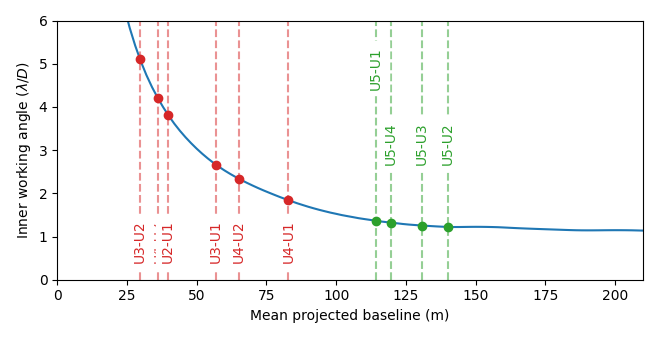}
   \caption{
      Inner working angle (IWA) as a function of the PA-averaged projected baseline length \(\bm{\alpha} \cdot \bm{B} / |\bm{\alpha}|=|\bm B|/\sqrt2\). For very long baseline lengths, the IWA is limited by telescope diffraction, approaching one \(\lambda/D\). However, for most of the VLTI baselines, the IWA is typically limited to several \(\lambda/D\). 
      The addition of a new telescope would create four new baselines with an IWA only limited by the 8-metre telescope diameter.}
   \label{fig:uv_iwa}
\end{figure}

\begin{figure*}
   \begin{center}
      \includegraphics[width=0.75\linewidth]{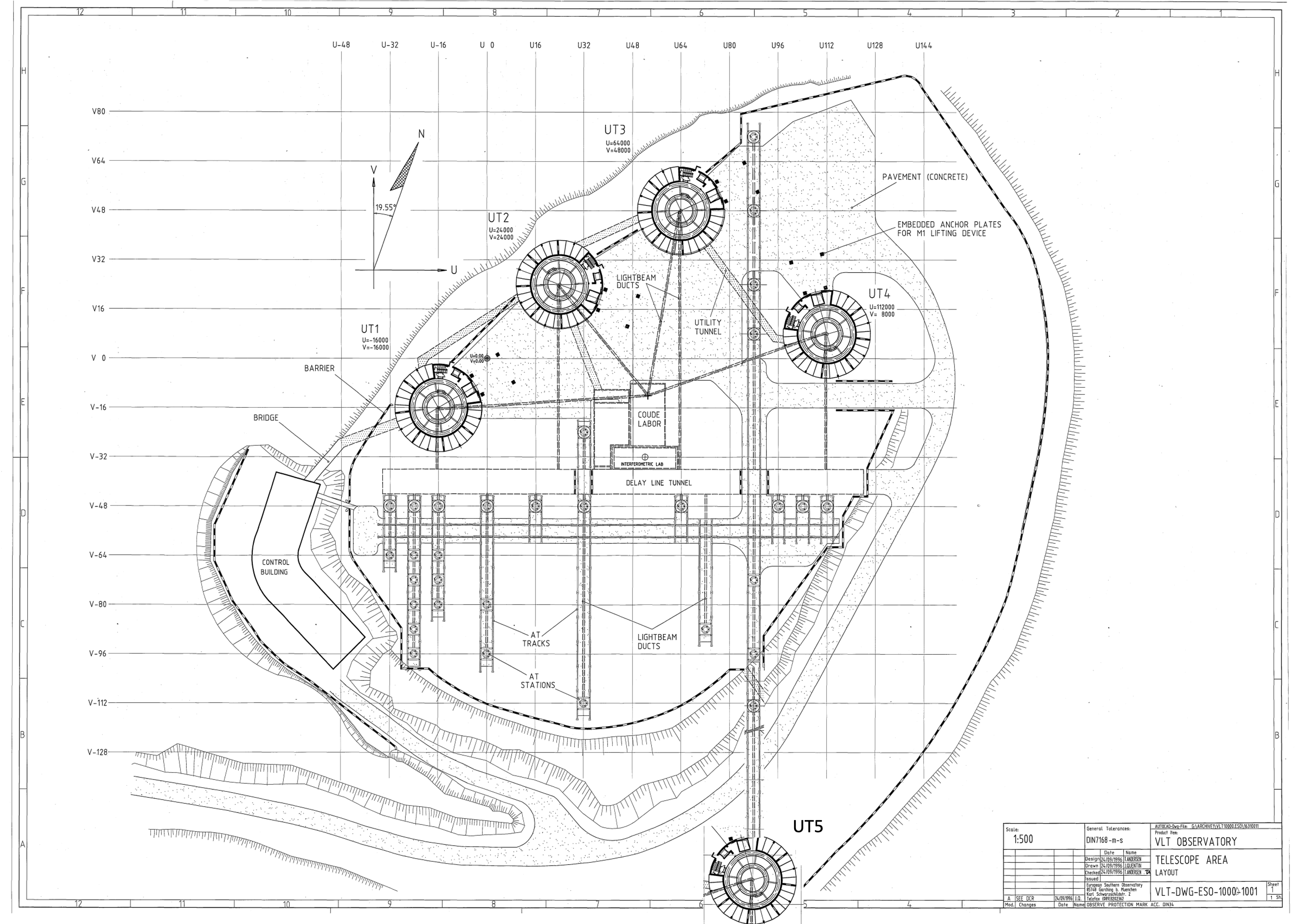}
      \includegraphics[width=0.8\linewidth]{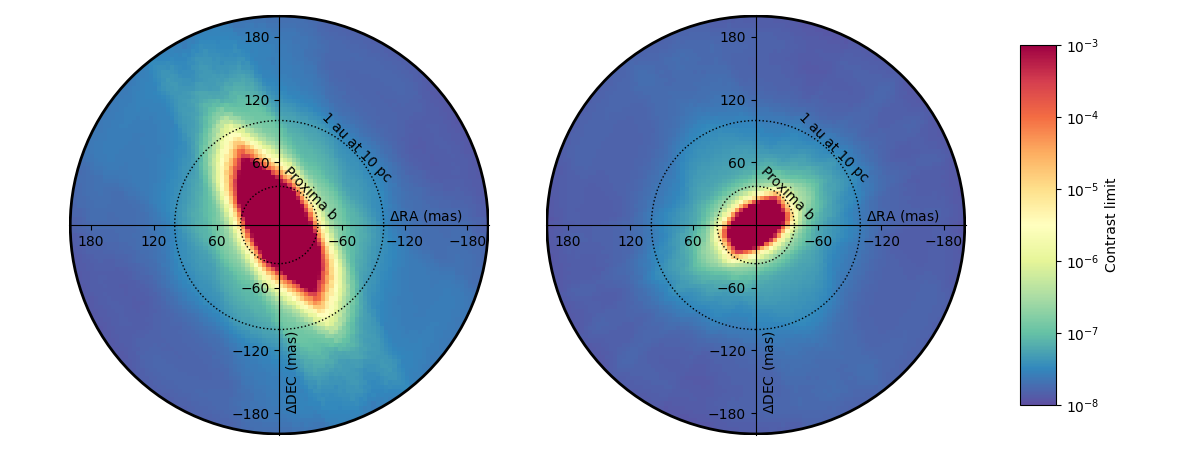}
      \caption{Aerial view of the proposed extension of the VLTI. The new UT5 telescope would be positioned to the southeast of the platform, on the opposite side of the access road. The beam would be propagated to the VLTI tunnel through the J arm of the existing VLTI/AT interferometer, creating four new baselines with north-south orientations, each approximately 200 meters in length. With a length of 220 meters and an orientation at 154 degrees to the South-East, the UT3-UT5 baseline would be almost orthogonal to the existing long UT1-UT4 baseline giving an optimal UV coverage. The long delay line could be obtained by serialising the two remaining, presently unused, VLTI delay lines.
      \textit{Lower pannels:}
      Contrast limit of a dual-field interferometer using the current four-UT configuration of the VLTI (left panel), compared to the proposed five-UT configuration (right panel).}
      \label{fig:contrast}
   \end{center}
\end{figure*}

As detailed in Eq.~\eqref{eq:contrast} of Section~\ref{sec:detect_exo}, the reflected light flux from an exoplanet decreases with the square of its distance from the host star. Consequently, the detectability of exoplanets heavily depends on the instrument's ability to observe companions close to their star. This capability is quantified by the inner working angle (IWA), a critical metric for assessing the effectiveness of astronomical instruments in detecting companions at short separation. 

The IWA showed in Fig.~\ref{fig:contrast} is influenced by two factors which both reduces the exoplanet signal: $\eta_{\rmplanet}^{\mathrm{inj}}$ and $\eta_{\rmplanet}^{\mathrm{proc}}$. The first parameter, $\eta_{\rmplanet}^{\mathrm{inj}}$, arises from the phase apodization used to reduce the stellar flux. Its impact gets bigger as $\bm \alpha$ becomes close to $\lambda/D$ (as shown in Figure~\ref{fig:Astar}), meaning that the extent of its effect is governed by the diameter $D$ of the individual telescopes of the array.

Conversely, the influence of $\eta_{\rmplanet}^{\mathrm{proc}}$, depicted in Figure~\ref{fig:Aprocess}, becomes significant when the separation is too small for the oscillatory term $e^{ik\bm{\alpha}\cdot{}\bm{B}}$ to effectively separate the exoplanet signal from the residual starlight. This starts to occur when the $\bm \alpha$ becomes close to a few $\lambda/\Delta\lambda$ of $\lambda/\bm B$, where $\bm B$ is the projected baseline.

 Hence, the IWA depends both on the diameter of the telescopes and the baseline length, whichever limit comes first. This is illustrated in 
 Fig.~\ref{fig:uv_iwa}, which plots the IWA as defined by the angular separation at which the quantity $\eta_{\rmplanet}^{\mathrm{inj}}\eta_{\rmplanet}^{\mathrm{proc}}$ equals 1/2. This figure shows the IWA's noticeable increase at shorter baselines (baseline-limited regime) and its stabilisation at longer baselines (telescope diameter-limited regime). The red vertical lines represent the lengths of the six existing baselines of the VLTI averaged over all possible orientations. They highlight that the current VLTI IWA is limited by its baselines, thus not fully capitalising on the individual telescopes' size. Four additional baselines with projected lengths above 100 meters would enable the observation of exoplanets with a minimal IWA.

\subsection{Physical implementation of the fifth UT}

Given the need to both enhance the UV coverage in the North-West direction and increase the baseline lengths for maximising the inner working angle, we have devised a proposal for integrating a fifth UT that addresses these challenges without extensive change to the VLTI architecture.

This new telescope could be constructed to the south of UT4, across the road. It would be aligned with the AT J band arm, facilitating the propagation of the beam towards the delay line tunnels. The required delay line for a baseline of approximately 200 meters could be achieved by serially utilising the two remaining unused delay lines of the VLTI. An aerial view of this proposed setup is presented in Figure~\ref{fig:UT5}.

In terms of UV coverage, the four new baselines would ideally complement the existing frequency coverage. With these four new baselines, the IWA would also reach optimal levels, approaching the theoretical limit of $\lambda/D$. Additionally, this configuration would enhance the contrast limit at many position angles, improving the detectability of exoplanets. As shown in the right panel of Fig.~\ref{fig:contrast}, this setup would enable the detection of an exoplanet at a 1 AU distance from a star located 10 parsecs away across all position angles. This setup would also allow for the detection of Proxima Centauri b at its most favourable position angle. The enhancements provided by these additional capabilities are quantified in the subsequent section.


\section{Detection rate of reflected light exoplanets} 
\label{sec:detect_exo}
 
\begin{figure*}
	\centering
	\includegraphics[width=8cm]{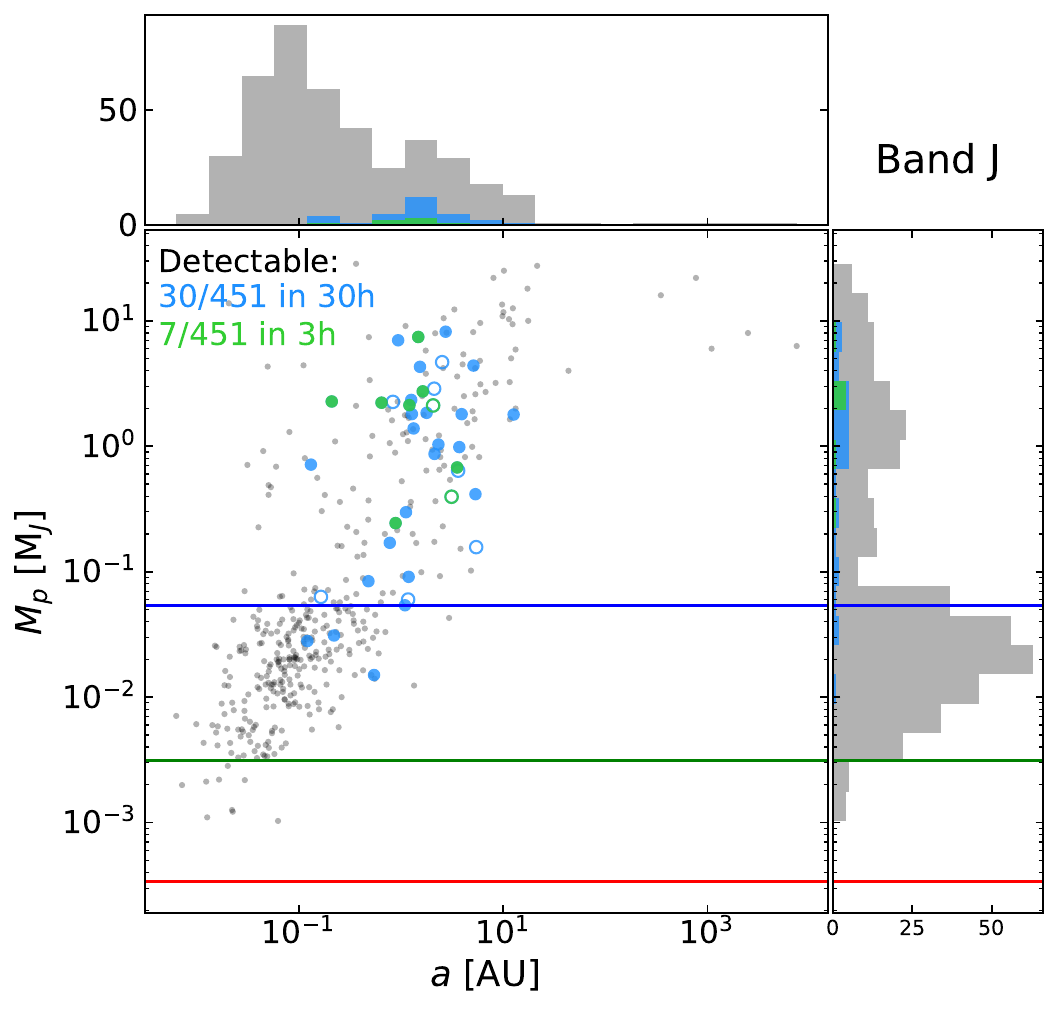} 
    \includegraphics[width=8cm]{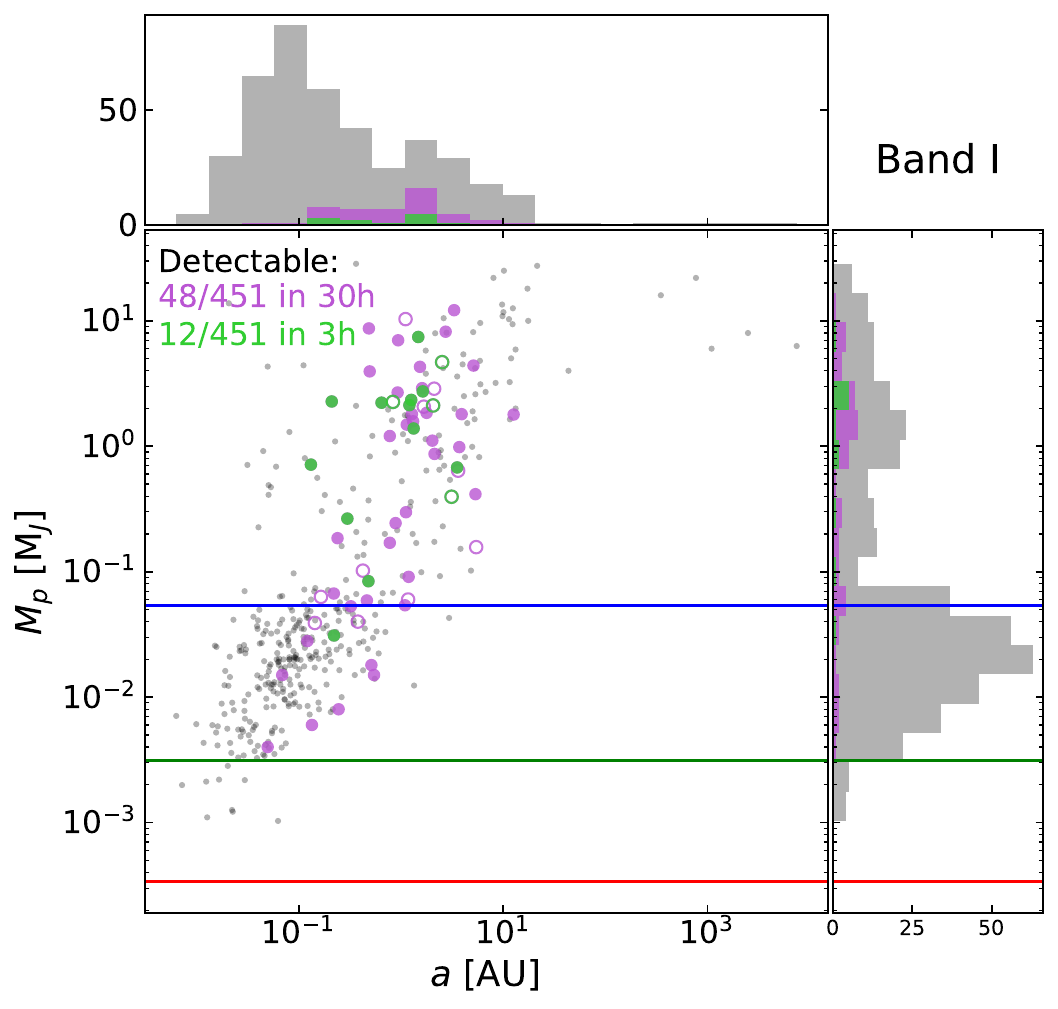}
    \\
	\caption{Detectable exoplanets with the current four-UT configuration with a 100~m baseline (left) at the J band and the proposed five-UT with 200~m baselines (right) at the I band. Blue and purple circles and histograms show the detectable planets in 30~h of integration time in each case, and green ones correspond to planets detectable in 3~h.
    Empty circles indicate planets that could be detectable by such an instrument but are out of the $\rm{DEC}\in\{-90^\circ, +30^\circ\}$ interval observable from Paranal, and thus not included in the tally of detectable planets.
    Grey dots and histograms show the ensemble of known exoplanets within 30~pc.
    For non-detectable exoplanets, we plot either $M_p$ or $M_p\,sin\,i$ as reported in the NASA Archive. Horizontal lines indicate the masses of Neptune (blue), Earth (green) and Mars (red).}
\label{fig:results_MpVSd}
\end{figure*} 

\subsection{Simulated orbital realisations}  \label{subsec:methods_orbits}

\begin{table*}
    \tiny
    \centering
    \caption{Detectability at each spectral band of the exoplanets within 30~pc that are potential targets ($P_{detect}$>25\%) in at least one band with 3~h of integration time. Both the configurations assuming 4 or 5 telescopes are shown. In all cases, we adopted a value of $A_g$=0.3 for bands I and J, and $A_g$=0.1 for H and K. The corresponding targets in 30~h of integration time are displayed in Tables \ref{table:detectability_30pc_30h_100m} and \ref{table:detectability_30pc_30h_200m}.}
    \label{table:detectability_30pc_3h} 
    \begin{tabular}{c l c c c c c c c c c c c }
    \hline 
    \hline
      &  & & &    &  \multicolumn{2}{c}{I} & \multicolumn{2}{c}{J} & \multicolumn{2}{c}{H} & \multicolumn{2}{c}{K} \\ 
      & Planet & $d$ & $a$ & $M_p$   & $P_{detect}$ & $\alpha_{obs}$ & $P_{detect}$ & $\alpha_{obs}$ & $P_{detect}$ & $\alpha_{obs}$ & $P_{detect}$ & $\alpha_{obs}$ \\
      &   & [pc] & [AU] & [$M_{J}$] & [\%] & [deg] &[\%] & [deg] & [\%] & [deg] & [\%] & [deg] \\ 
    \hline
\multirow{8}{*}{\rotatebox{90}{VLTI}}  & HD 62509 b		&	10.34	&	1.62	&	2.74	&100.00	&	[31$^{+ 26 }_{- 18 }$,127$^{+ 2 }_{- 5 }$]		&	100.00	&	[31$^{+ 26 }_{- 14 }$,123$^{+ 1 }_{- 2 }$]&99.90	&	[32$^{+ 25 }_{- 2 }$,101$^{+ 2 }_{- 2 }$]		&	97.80	&	[44$^{+ 11 }_{- 2 }$,86$^{+ 2 }_{- 3 }$]	\\
 & eps Eri b		&	3.20	&	3.53	&	0.68	&100.00	&	[14$^{+ 12 }_{- 5 }$,115$^{+ 1 }_{- 1 }$]		&	100.00	&	[14$^{+ 12 }_{- 6 }$,100$^{+ 6 }_{- 2 }$]	&100.00	&	[24$^{+ 14 }_{- 3 }$,71$^{+ 9 }_{- 5 }$]		&	0.00	&	$-$	\\
 & alf Tau b		&	20.44	&	1.47	&	7.44	&100.00	&	[28$^{+ 29 }_{- 6 }$,118$^{+ 1 }_{- 1 }$]		&	100.00	&	[37$^{+ 21 }_{- 6 }$,115$^{+ 1 }_{- 1 }$]	&49.70	&	[61$^{+ 6 }_{- 4 }$,81$^{+ 3 }_{- 3 }$]		&	0.00	&	$-$	\\
 & alf Ari b		&	20.21	&	1.20	&	2.13	&100.00	&	[31$^{+ 28 }_{- 2 }$,105$^{+ 1 }_{- 1 }$]		&	100.00	&	[39$^{+ 20 }_{- 1 }$,98$^{+ 1 }_{- 2 }$]	&0.00	&	$-$		&	0.00	&	$-$	\\
 & GJ 876 b		&	4.68	&	0.21	&	2.28	&100.00	&	[41$^{+ 1 }_{- 1 }$,97$^{+ 1 }_{- 1 }$]		&	100.00	&	[53$^{+ 0 }_{- 0 }$,94$^{+ 1 }_{- 1 }$]		&	0.00	&	$-$		&	0.00	&	$-$	\\
 & GJ 896 A b		&	6.26	&	0.64	&	2.23	&71.60	&	[53$^{+ 4 }_{- 3 }$,73$^{+ 3 }_{- 5 }$]		&	99.90	&	[47$^{+ 2 }_{- 2 }$,80$^{+ 2 }_{- 3 }$]		&	0.00	&	$-$		&	0.00	&	$-$	\\
 & GJ 876 c		&	4.68	&	0.13	&	0.71	&100.00	&	[46$^{+ 0 }_{- 0 }$,92$^{+ 1 }_{- 1 }$]		&	0.00	&	$-$		&	0.00	&	$-$		&0.00	&	$-$	\\
 & GJ 9066 c		&	4.47	&	0.88	&	0.24	&12.70	&	[56$^{+ 5 }_{- 5 }$,66$^{+ 5 }_{- 4 }$]		&	47.60	&	[54$^{+ 15 }_{- 10 }$,87$^{+ 2 }_{- 8 }$]		&0.00	&	$-$		&	0.00	&	$-$	\\

    \hline
\multirow{14}{*}{\rotatebox{90}{Extended VLTI}}  & HD 62509 b		&	10.34	&	1.62	&	2.74	&100.00	&	[31$^{+ 26 }_{- 21 }$,127$^{+ 2 }_{- 5 }$]		&	100.00	&	[31$^{+ 26 }_{- 17 }$,123$^{+ 2 }_{- 2 }$]&99.90	&	[31$^{+ 26 }_{- 8 }$,102$^{+ 3 }_{- 3 }$]		&	99.30	&	[33$^{+ 23 }_{- 2 }$,91$^{+ 3 }_{- 2 }$]	\\
 & alf Tau b		&	20.44	&	1.47	&	7.44	&100.00	&	[28$^{+ 29 }_{- 10 }$,119$^{+ 1 }_{- 2 }$]		&	100.00	&	[29$^{+ 28 }_{- 5 }$,119$^{+ 1 }_{- 1 }$]	&99.70	&	[44$^{+ 13 }_{- 5 }$,96$^{+ 2 }_{- 3 }$]		&	43.50	&	[62$^{+ 6 }_{- 4 }$,82$^{+ 2 }_{- 3 }$]	\\
 & eps Eri b		&	3.20	&	3.53	&	0.68	&100.00	&	[14$^{+ 12 }_{- 5 }$,113$^{+ 1 }_{- 2 }$]		&	100.00	&	[14$^{+ 12 }_{- 6 }$,108$^{+ 5 }_{- 5 }$]	&100.00	&	[24$^{+ 7 }_{- 3 }$,71$^{+ 6 }_{- 4 }$]		&	0.00	&	$-$	\\
 & alf Ari b		&	20.21	&	1.20	&	2.13	&100.00	&	[31$^{+ 28 }_{- 10 }$,109$^{+ 1 }_{- 1 }$]		&	100.00	&	[31$^{+ 28 }_{- 4 }$,110$^{+ 1 }_{- 1 }$]	&0.00	&	$-$		&	0.00	&	$-$	\\
 & GJ 876 c		&	4.68	&	0.13	&	0.71	&100.00	&	[34$^{+ 0 }_{- 0 }$,110$^{+ 0 }_{- 1 }$]		&	100.00	&	[47$^{+ 0 }_{- 0 }$,100$^{+ 1 }_{- 0 }$]	&0.00	&	$-$		&	0.00	&	$-$	\\
 & GJ 876 b		&	4.68	&	0.21	&	2.28	&100.00	&	[31$^{+ 0 }_{- 0 }$,103$^{+ 1 }_{- 1 }$]		&	100.00	&	[38$^{+ 0 }_{- 0 }$,111$^{+ 1 }_{- 1 }$]	&0.00	&	$-$		&	0.00	&	$-$	\\
 & GJ 896 A b		&	6.26	&	0.64	&	2.23	&67.00	&	[45$^{+ 4 }_{- 3 }$,79$^{+ 4 }_{- 9 }$]		&	100.00	&	[39$^{+ 2 }_{- 4 }$,88$^{+ 3 }_{- 3 }$]		&	0.00	&	$-$		&	0.00	&	$-$	\\
 & HD 27442 b		&	18.27	&	1.27	&	1.81	&7.20	&	[56$^{+ 2 }_{- 1 }$,63$^{+ 2 }_{- 2 }$]		&	71.30	&	[46$^{+ 3 }_{- 2 }$,66$^{+ 4 }_{- 5 }$]		&	0.00	&	$-$		&	0.00	&	$-$	\\
 & nu Oct A b		&	21.15	&	1.25	&	2.34	&30.60	&	[55$^{+ 5 }_{- 3 }$,70$^{+ 3 }_{- 4 }$]		&	47.10	&	[53$^{+ 5 }_{- 4 }$,68$^{+ 4 }_{- 4 }$]		&	0.00	&	$-$		&	0.00	&	$-$	\\
 & 61 Vir d		&	8.50	&	0.48	&	0.08	&41.20	&	[55$^{+ 15 }_{- 8 }$,84$^{+ 16 }_{- 14 }$]		&	26.60	&	[62$^{+ 12 }_{- 7 }$,86$^{+ 17 }_{- 14 }$]&0.00	&	$-$		&	0.00	&	$-$	\\
 & GJ 9066 c		&	4.47	&	0.88	&	0.24	&4.90	&	[55$^{+ 4 }_{- 8 }$,67$^{+ 5 }_{- 4 }$]		&	54.50	&	[51$^{+ 17 }_{- 10 }$,87$^{+ 3 }_{- 11 }$]		&0.00	&	$-$		&	0.00	&	$-$	\\
 & HD 26965 b		&	5.04	&	0.22	&	0.03	&29.70	&	[61$^{+ 13 }_{- 11 }$,89$^{+ 13 }_{- 19 }$]		&	21.80	&	[64$^{+ 13 }_{- 6 }$,94$^{+ 8 }_{- 16 }$]	&1.20	&	[84$^{+ 2 }_{- 2 }$,94$^{+ 1 }_{- 3 }$]		&	0.00	&	$-$	\\
 & HD 147513 b		&	12.90	&	1.32	&	1.39	&50.30	&	[54$^{+ 3 }_{- 2 }$,63$^{+ 3 }_{- 3 }$]		&	0.00	&	$-$		&	0.00	&	$-$		&0.00	&	$-$	\\
 & HD 3651 b		&	11.13	&	0.30	&	0.27	&42.70	&	[86$^{+ 4 }_{- 5 }$,100$^{+ 2 }_{- 3 }$]		&	0.00	&	$-$		&	0.00	&	$-$&	0.00	&	$-$	\\

    \hline
    \end{tabular}
\end{table*}

We used the Planetary System database of the NASA Exoplanet  Archive\footnote{\url{https://exoplanetarchive.ipac.caltech.edu}} \citep{2013PASP..125..989A} as our main source of information on the confirmed exoplanets.
As of \textcolor{black}{January 25, 2024}, it contained a total of \textcolor{black}{5572} exoplanets.
We focused in this study on the \textcolor{black}{451} known exoplanets within \textcolor{black}{30} pc.

We applied the methodology of \citet{carriongonzalezetal2021a} and computed 1000 orbital realisations for each of these planets (we refer to Sects. 3 and 4 therein for a more detailed description).
These 1000 simulated orbits account for the fact that the planet location will only be known to within a given confidence interval, related to the uncertainties in the orbital parameters reported in the NASA Archive.
Each orbital realisation was discretised into 360 orbital positions, at which we computed the planet-star angular separation ($\Delta \theta$) and the brightness of the planet in reflected starlight, which is given by the planet-to-star contrast ratio ($F_\rmplanet / F_\rmstar$).
For each orbital position, $F_\rmplanet / F_\rmstar$ is computed at a given wavelength ($\lambda$) as:
\begin{equation}
\label{eq:contrast}
\frac{F_\rmplanet}{F_{\star}}= \left(\frac{R_\rmplanet}{r(t)} \right)^2  A_g(\lambda) \, \Phi(\alpha, \lambda)
,\end{equation}
with $R_\rmplanet$ being the planetary radius, $r$ the distance between the planet and the star, and $\alpha$ the corresponding star-planet-observer phase angle.
We assumed for all planets a Lambertian scattering phase law ($\Phi$), and carried out our analysis for geometric albedo ($A_g$) values of 0.3 and 0.1 (see Sect. \ref{sec:results}).

When a planet lacked a radius measurement, we computed it using either the mass-radius relationships from \citet{otegietal2020} for rocky and for volatile-rich exoplanets, or the one from \citet{hatzes-rauer2015} for giant planets.
In case of a missing value for the orbital inclination ($i$) or the argument of periastron of the planet ($\omega_p$), we assumed an isotropic distribution of orbital orientations and drew random values from $\cos(i)\in[-1,1]$ and $\omega_p\in[0,2\pi]$ at each orbital realisation.
Similarly, we drew values for the orbital eccentricity ($e$) from $e \in[0,1)$ if it had not been measured.
We assumed that the NASA Exoplanet Archive reports for each planet the argument of periastron of the host star ($\omega_\rmstar$), following the radial-velocity convention.
Sect. 4.1.1 in \citet{carriongonzalezetal2021a} provides details on the lack of homogeneous criteria to report the planetary argument of periastron on catalogues, and its impact for exoplanet detectability studies.

If no value was available for the stellar mass, radius or temperature ($M_\rmstar$, $R_\rmstar$, $T_\rmstar$), we used those in the Planetary Systems Composite Data database of the NASA Exoplanet Archive.
Also, if the stellar magnitude at bands I, J, H or K was not reported in the NASA Exoplanet Archive, we obtained it from the SIMBAD database \citep{wengeretal2000}.

This affects in particular the I magnitude, which is unavailable for \textcolor{black}{198} of the planets within \textcolor{black}{30}~pc.
If also unavailable in SIMBAD, we adopted the value of the J magnitude as an approximation -- this however happened for none of the detectable-exoplanet hosts in Table \ref{table:detectability_30pc_3h}.

With the S/N analytical model from Sect. \ref{sec:simulation}, assuming the VLTI parameters used to make Fig.~\ref{fig:contrast} with a projection along the best possible orientation, and adapting in each case for the stellar luminosity, we computed the detectability of the orbital realisations simulated for each exoplanet.
A planet is considered detectable at a given orbital position of a given orbital realisation if an $\mathrm{S/N}=3$ is achievable in less than 3~h of integration time.
The probability of a planet to be detectable ($P_{\rm detect}$) is given by the number of detectable orbits -- those with at least one orbital position detectable -- divided by the total number of orbits simulated.
As described below, we also explored the number of planets detectable if the integration time is increased to 30~h.

\subsection{Detectable exoplanets} \label{sec:results}

We applied our methods to the \textcolor{black}{451} confirmed exoplanets within \textcolor{black}{30}~pc, and obtained a list of detectable planets at each spectral band.
For each planet, the value of $A_g$ will strongly depend on the structure and composition of its atmosphere.
In the case of the exoplanets under study here, this information will be a priori completely unknown.
For reference, Solar-System bodies have geometrical albedos at spectral bands I, J, H and K ranging from about 0.0 to 0.5, significantly impacted by the presence of clouds and by methane absorption bands in the case of gaseous planets \citep[e.g.][]{fink-larson1979, madden-kaltenegger2018}.
In agreement with values from Solar System giants, we adopted in Table \ref{table:detectability_30pc_3h} an albedo of 0.3 at bands I and J, and $A_g =0.1$ for bands H and K.

In order to build a target list that is statistically robust, we only considered as suitable targets those planets meeting $P_{\rm detect}>25\%$.
Similarly, we discarded any planet with declination out of the interval $\rm{DEC}\in\{-90^\circ, +30^\circ\}$ and thus not observable from Paranal.

Table \ref{table:detectability_30pc_3h} contains the list of targets and their detectability conditions, and Fig. \ref{fig:orbits_and_histograms} shows the simulated orbits for each planet compared to the sensitivity curves of the instrument.
For integration times of 3~h, we find \textcolor{black}{8} exoplanets that are detectable in at least one of the four bands under study with the current VLTI configuration. 
For the extended VLTI configuration (ie., the current VLTI plus UT5), this count rises to \textcolor{black}{14} detectable planets.
We find that the majority of the detectable targets are long-period planets in the mass range of Jupiter, although the new UT5 can allow to detect planets in the Neptune-mass regime (Fig. \ref{fig:results_MpVSd}).
These targets, and in particular the lower-mass ones, are detectable mostly at the I and J spectral bands (Table \ref{table:detectability_30pc_3h}), while the expected low-albedos at H and K bands makes them undetectable at such longer wavelengths.
Extending the integration time to 30~h increases the tally of detectable planets to 40 with the VLTI and 50 with the extended VLTI (Tables \ref{table:detectability_30pc_30h_100m} and \ref{table:detectability_30pc_30h_200m}).
With the new UTs, low-mass planets down to the Earth-like range (e.g. Proxima Cen b) become detectable (see Sect. \ref{subsubsec:proximab}).

We find wide ranges of observable phase angles ($\alpha_{obs}$) for several of the exoplanets detectable even in 3~h of integration (Table \ref{table:detectability_30pc_3h}), reaching phases both before and after quadrature ($\alpha=90^\circ$).
These are prime targets for atmospheric characterisation, as combining the information of multi-phase observations has been found to significantly improve the atmospheric retrievals in direct imaging \citep{damianoetal2020, carriongonzalezetal2021b}.
For 30~h of integration, this list of targets detectable at multiple phase angles increases significantly, including several low-mass planets (Tables \ref{table:detectability_30pc_30h_100m} and \ref{table:detectability_30pc_30h_200m}).
In addition, several multi-planet systems have more than one detectable planet offering a good opportunity to study the architecture of planetary systems, and the evolution of the planets therein and their atmospheres.

Notably, two promising targets for this instrument in 3~h of integration (HD 62509 b and $\epsilon$ Eri b) are potentially detectable as well with the Nancy Grace Roman Space Telescope in reflected starlight \citep{carriongonzalezetal2021a}.
Roman will be the first space-based facility capable of directly imaging cold and temperate exoplanets in reflected starlight.
It will be launched in late 2026 or 2027, with a mission duration of five years and a potential extension of five more.
With the spectral range of Roman's coronagraph instrument limited to within 575 and 825~nm \citep{akesonetal2019, zellemetal2022}, this showcases synergies between both facilities for complementary --and potentially simultaneous-- observations in a wider spectral range.

Moreover, this proposed instrument will be able to search for new exoplanets in direct imaging with sensitivity levels never reached before. Such a search phase is not planned for the Roman's coronagraph with its current time allocations \citep{baileyetal2023}. This will reveal planets that eluded detection in RV or astrometry due to e.g. their long period or their host-star properties, and will also help confirm planet candidates from these techniques. These new planets will be in range of mass and orbital separations as those detectable in Fig. \ref{fig:results_MpVSd}.

\begin{figure}
   \centering
    \includegraphics[width=8.5cm]{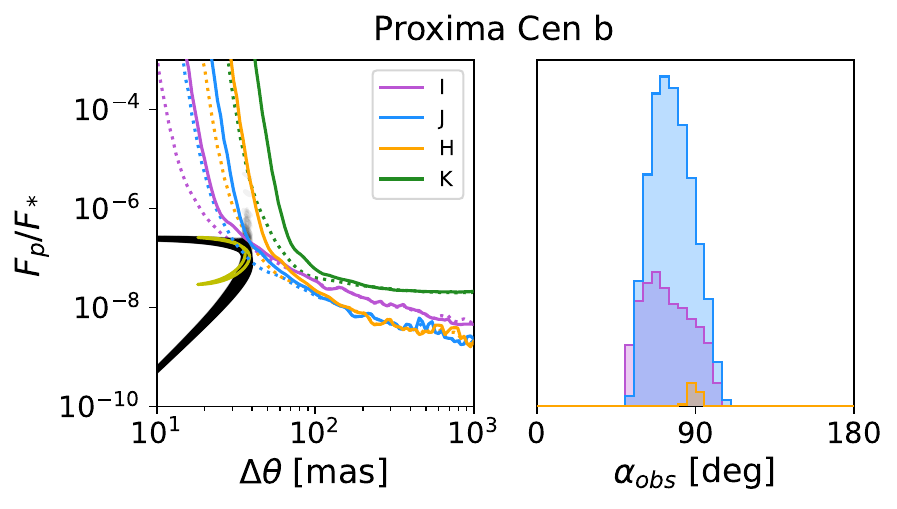}
   \caption{Detectability conditions for Proxima Cen b with the extended VLTI configuration. Left panel show the simulated orbits  (black lines) assuming $A_g =0.3$, as well as the mean orbital realisation (as shown in Figs. \ref{fig:Astar}-\ref{fig:Cfull}) in yellow. Coloured lines show the detectability limits of the interferometer at each spectral band with the VLTI (solid lines) and the extended VLIT (Fig.~\ref{fig:UT5}, dotted lines) configurations. In this case, we show the detectability limits for a $\rm{S/N} = 3$ in 30~h of integration time, instead of 3~h.
  The histograms on the right panel show the range of $\alpha_{obs}$ at each band.}
\label{fig:proximab}
\end{figure} 

\subsection{Pushing the instrumental limits: The case of Proxima Cen b}\label{subsubsec:proximab}

The adopted detectability criteria ($\mathrm{S/N}=3$ in less than 3~h of integration time, $P_{\rm detect}$>25\%) ensure the robustness of the target list in Table \ref{table:detectability_30pc_3h}.
However, tens of additional known exoplanets that are not included in Table \ref{table:detectability_30pc_3h} are at the limits of our detectability criteria.
These planets could be considered as potential targets for high-risk high-gain observations by increasing the integration time beyond 3~h.
Tables \ref{table:detectability_30pc_30h_100m} and \ref{table:detectability_30pc_30h_200m} showcase the detectable exoplanets ($P_{detect}$>25\%) for an integration time of 30~h.

Several temperate low-mass exoplanets are among these, including the habitable-zone super Earths $\tau$~Cet~e and f, as well as Proxima Cen b, a potentially rocky planet at the habitable zone of our nearest stellar neighbour.
We find that for a detection threshold of $\mathrm{S/N}=3$, 30~h of integration would be enough to detect Proxima Cen b with the upgraded VLTI configuration at bands I and J (Fig. \ref{fig:proximab}).
The a priori unknown atmospheric properties of the planet will also modify the detectability, and we note that if its albedo is for instance similar to that of Venus \citep[0.45 at band I, 0.4 at band J,][]{madden-kaltenegger2018} the prospects for detecting the planet will be more favourable than reported here.

\begin{figure*}
   \centering
   \includegraphics[width=15.5cm]{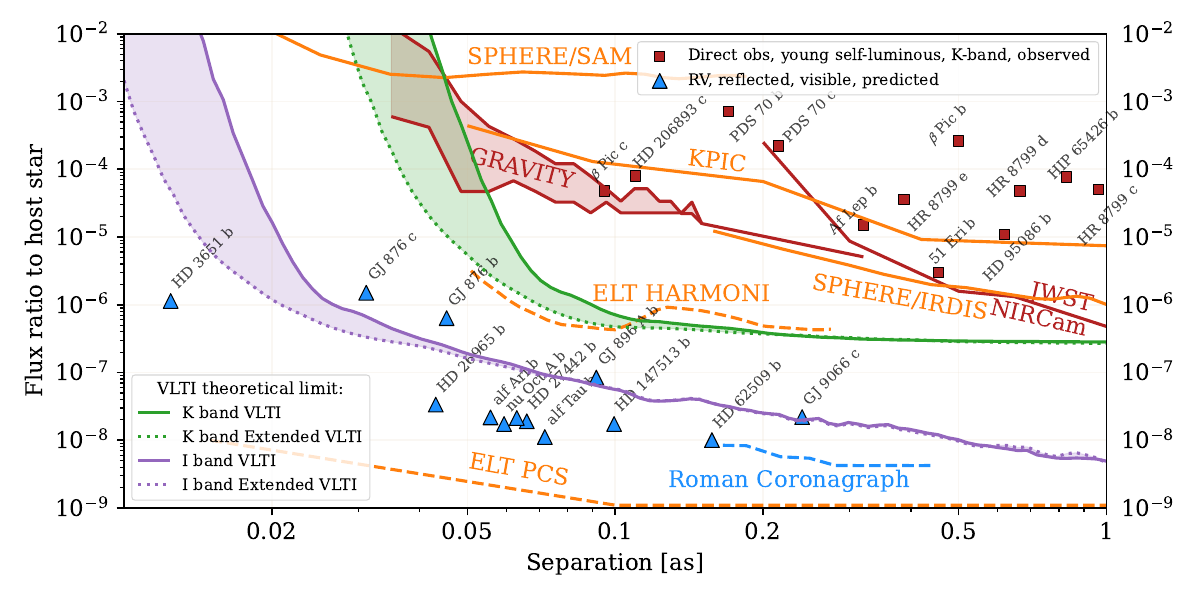}
   \caption{An overview of current and predicted contrast limits for present and future instruments. The red squares represent directly detected exoplanets, while the blue triangles show contrast predictions for exoplanets detected via radial velocities (targets found in Table~\ref{table:detectability_30pc_3h}, assuming an albedo of 0.3). The solid orange lines correspond to three techniques using single-dish telescopes: aperture masking \citep{2024A&A...682A.101S}, high-contrast imaging combined with high-resolution spectroscopy \citep{2024SPIE13096E..1XW}, and high-contrast imaging \citep{2023A&A...675A.205C}. The red lines illustrate the actual capabilities of JWST/NIRCAM \citep{2024ApJ...974L..11F} and GRAVITY \citep[two lines, one for the best and one for the worst orientation of the baselines,][]{2024arXiv240604003P}. The dashed lines show predictions for the ELT/HARMONI instrument \citep{2021A&A...652A..67H}, ELT/PCS \citep{kasperetal2021} and the Roman Space Telescope coronagraph \citep{baileyetal2023}. The green and purple solid lines represent the theoretical $3\sigma$ contrast limits that could be achieved with the VLTI in 3 hours of observation for the K and I bands, respectively. The shaded regions down to the dotted lines represent the benefits of an extended VLTI configuration, in addition to provide a more homogeneous position angle coverage. In theory, the VLTI's contrast should reach the contrast range predicted for HARMONI and could also observe targets accessible to the Roman Space Telescope coronagraph. The plotted contrast range of GRAVITY is limited by the capabilities of the old MACAO AO system and systematics, which are expected to be overcome with GRAVITY+.}
\label{fig:final_contrast}
\end{figure*}

\section{Summary and perspectives}
\label{sec:summary}

\subsection{Summary}

In this paper, we have demonstrated that:
\begin{enumerate}
   \item Dual-field interferometry shows significant potential for detecting exoplanets in reflected light, positioning it as a competitive technique alongside cross-correlation methods used with the ELT or space-based coronagraph.
   \item The primary limitation of this technique is shot noise from stellar light, which can be effectively mitigated through the combined use of phase apodization and a high-performance AO system.
   \item Sensitivity is enhanced at shorter wavelengths, provided the AO system performs well (Strehl ratio above 50\%). Furthermore, planets are expected to exhibit higher geometrical albedo at these wavelengths, further improving detectability.
   \item The inner working angle is constrained by both the telescope diameter and the baseline length, with optimal performance achieved when the projected baseline length is \(\geq 10\) times the telescope diameter.
\end{enumerate}

\subsection{Comparison with the current GRAVITY Instrument}

Fig.~\ref{fig:final_contrast} illustrates the potential of the VLTI compared to other current and future projects\footnote{Figure adapted from Vanessa Bailey's GitHub \url{https://github.com/nasavbailey/DI-flux-ratio-plot}.}. A comparison with the experimental GRAVITY contrast curve published by \citet{2024arXiv240604003P} reveals an inner working angle similar to our predictions. In some cases, when the position angle of the exoplanet aligns with the VLTI, the inner working angle may even be smaller due to their use of a lower-order polynomial fit. However, their contrast ratio at separations of several \(\lambda/D\) is more limited: at 100\,mas, they achieved only $2\times10^{-5}$, whereas our simulations suggest that limits as low as $10^{-6}$ could theoretically be achieved. We attribute this discrepancy to a combination of factors: (i) the outdated MACAO adaptive optics system, (ii) non-common-path aberrations, and (iii) systematic noise.

The limitations of the MACAO system are well known and motivated the GRAVITY+ upgrade. It suffers from a low Strehl ratio (below 30\%) and poor raw contrast. Non-common-path aberrations (arising because the AO system is located in the coudé room) also play a significant role, as distortions in the point spread function inject a considerable amount of photons into the off-axis fibre. Indeed, typical GRAVITY observations (pre-GRAVITY+) \citep[Figure 2 in][]{2024CRPhy..24S.144L} show a raw contrast of about 1/10 at 100\,mas, whereas our simulations suggest a raw contrast of just a few $10^{-4}$ could be achievable.

Therefore, the discrepancy between \citet{2024arXiv240604003P} and the theoretical results presented in this paper could, in part, be explained by low throughput and photon noise. However, additional limitations not included in our simulations can also be contributors. One such issue, highlighted by \citet{2024arXiv240604003P}, is the presence of systematic noise, colloquially referred to as "wiggles." These are residual fringes that remain after polynomial fitting and manifest as fluctuating coherent flux across the wavelength range. This phenomenon is a known issue. Possible causes include optical ghosts, polarisation interference, dispersion effects, or artefacts in the data reduction pipeline.

We look forward to investigating these issues further in the context of GRAVITY+, particularly after the installation and commissioning of the new GRAVITY+ AO system (GPAO). This system will include non-common-path corrections and a dark-hole approach. What Fig.~\ref{fig:final_contrast} demonstrates is that the VLTI has not yet reached its full potential for high-contrast observations, and additional instrumental investigations are still required to fully realise this capability.

\subsection{Comparison with other present and future projects}

Regarding other existing instruments: at angular separations below 150\,mas, 10-metre class single-dish telescopes are currently not competitive. Indeed, the primary limitation for single-dish telescopes is the inner working angle, which is constrained by the coronagraph or angular differential techniques. Even the James Webb Space Telescope (JWST), despite its exceptional optical system, faces challenges in detecting exoplanets close to their host star due to this constraint. For example, during observations of AF\,Lep\,b with NIRCam, \citet{2024ApJ...974L..11F} detected the planet with a flux reduced by the coronagraph by 93\%, illustrating both the observatory's strengths and its limitation to observe at small angular separations.

Because of its sheer size, the ELT holds great promise. For HARMONI, \citet{2021A&A...652A..67H} predict contrast limits on the order of $10^{-6}$ at 100\,mas using high-contrast imaging combined with high-resolution spectroscopy. This creates a strong synergy with optical interferometry, as both will explore the same parameter space using different deconvolution techniques. The ANDES instrument (not featured in Fig.~\ref{fig:final_contrast}) is expected to further enhance this approach, possibly allowing the detection of terrestrial exoplanets in reflected light through Doppler spectroscopy. The most promising candidates, represented by blue triangles in the figure, could be targets for both the VLTI, ANDES, and PCS, enabling detection via three different deconvolution techniques, each providing complementary data (astrometry, molecular maps, and continuum, respectively).

The Roman Space Telescope will also be in perfect synergy with the VLTI, targeting the same exoplanets but at even shorter wavelengths, without relying on cross-correlation techniques. However, its 2.4-metre aperture limits its ability to observe exoplanets at angular separations below 100\,mas, reducing the number of exoplanets that both observatories can detect in common.

\subsection{Conclusion}

The current phased implementation of a new extreme AO system for the VLTI is a major step forward\footnote{This is part of the GRAVITY+ upgrade, expected to be completed by 2025 with the installation of a laser source on each UT \citep{2022Msngr.189...17A}.}. We should soon see whether it can increase GRAVITY's contrast limits by an order of magnitude, as our simulations suggest. The next step should be to upgrade the VLTI with dual-field interferometry at shorter wavelengths, which would improve both contrast limits and the inner working angle.

The proposed addition of a new Unit Telescope to the south of the VLTI platform would significantly enhance its efficiency. This upgrade would increase the number of baselines over 100 meters by a factor five, double the angular resolution, and extend the UV coverage to include the northwest direction. Such improvements would greatly boost the VLTI’s capability, offering the best opportunity to detect signals from temperate, Earth-mass exoplanets, such as Proxima Centauri b, in reflected light.

Currently, no direct imaging observations of exoplanets in reflected starlight have been successful, as they require sensitivity levels beyond the reach of existing facilities. The proposed evolution of the VLTI will join a handful of other missions and concepts capable of achieving this sensitivity in the near- and mid-term future, such as the Planetary Camera and Spectrograph (PCS) instrument envisioned for the ELT \citep{kasperetal2021}, the Roman Space Telescope's Coronagraph Instrument technology demonstrator \citep{baileyetal2023}, and the future Habitable Worlds Observatory.

\begin{acknowledgements}
   This research made use of HCIPy, an open-source object-oriented framework written in Python for performing end-to-end simulations of high-contrast imaging instruments \citep{Por2018}. SL acknowledges the support of the French "Agence Nationale de la Recherche" (ANR), under grants ANR-21-CE31-0017 (project ExoVLTI) and ANR-22-EXOR-0005 (PEPR Origins). The authors thank Dr. A. Kellerer and Dr. G. Bourdarot for their thorough proofreading and extensive review of the mathematical formulations. A warm thank you to Dr. N. Pourré and Dr. V. Bailey for providing their script to make Fig.~\ref{fig:final_contrast}.
\end{acknowledgements}

%
%

 \bibliographystyle{aa}
 \bibliography{Full}
  
\begin{appendix} 

   \onecolumn

   \section{Beam waist for maximum coupling into a single-mode fibre behind a telescope without central obstruction}
   \label{sec:optimalI}
   
   The coupling into a single-mode fibre has been extensively described in the literature \citep[e.g.,][]{1988ApOpt..27.2334S,1997A&AS..121..379C}. However, for completeness, we provide here the calculation we performed to determine the optimal beam waist in the case of a circular aperture of diameter $D$. The injection efficiency is given by:
   \begin{equation}
      \eta^{\rm inj} = \frac{ \left| \iint_D  \mathsf{M}(\bm{u}) \ d\bm{u} \right|^2 }{\iint_\infty \left| \mathsf{M}(\bm{u}) \right|^2\ d\bm{u} \ \iint_D 1 d\bm{u}}\,.
   \end{equation}
   Assuming a Gaussian beam mode for the fibre:
   \begin{equation}
      \mathsf{M}(\bm{u}) \propto \exp\left(-\frac{|\bm{u}|^2}{2 w^2}\right),
   \end{equation}
   where $w$ is the beam waist as well as assuming a circular pupil without obstruction, $\iint_D 1 d\bm{u} = \pi D^2/4$. The optimal coupling is obtained by finding $w$ that maximise the injection efficiency:
   \begin{equation}
      \eta^{\rm inj} = \left| \iint_D \frac{2}{\pi D w} \exp\left(-\frac{|\bm{u}|^2}{2 w^2}\right) \ d\bm{u} \right|^2.
   \end{equation}
   Although this maximisation cannot be solved analytically, it can be solved numerically. The optimal beam waist is then found to be approximately $D/1.3715$, which we further approximate in the text as $\sqrt{0.1}D$. In this case, $\eta$ is slightly above 81\%.

   \section{Raw contrast variations with wavelength: the role of strehl and spatial filtering}
   \label{sec:appA}


\begin{figure*}[h!]
   \centering
   \includegraphics[width=9.3cm]{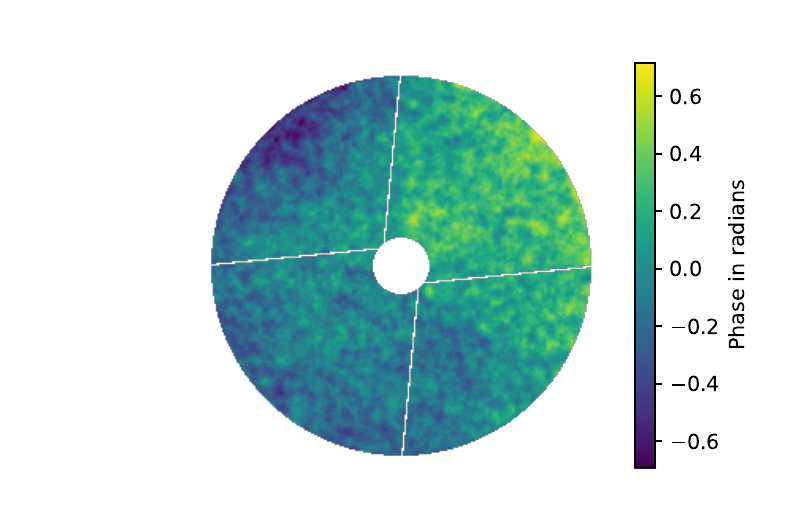}
   \includegraphics[width=9.cm]{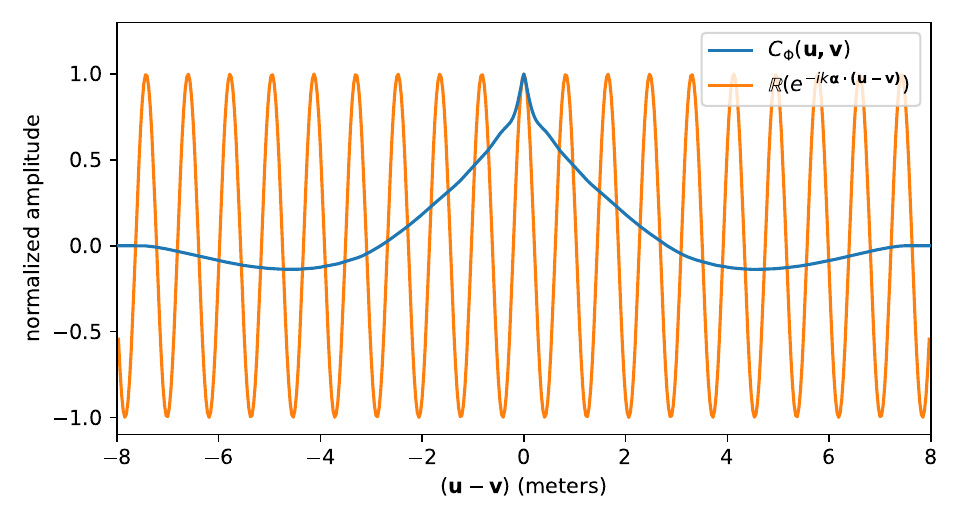}
   \includegraphics[width=8.5cm]{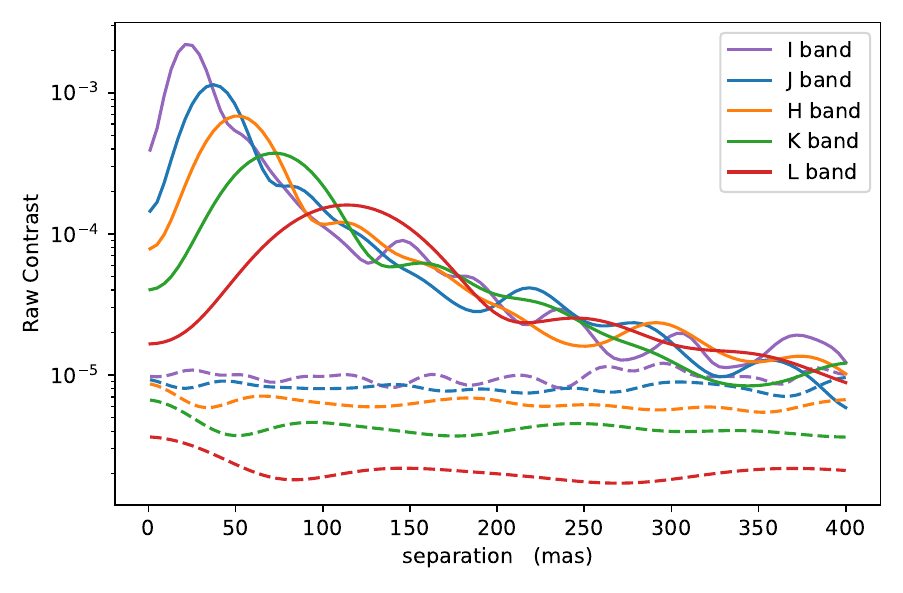}
   \caption{Raw contrast of atmospheric residual as a function of wavelength. \textit{Upper left panel:} Instantaneous phase map obtained using HCIPy simulations (parameters for the AO system are detailed in Section~\ref{sec:photon_noise}), yielding a Strehl ratio of 97.8\% in the K band. \textit{Upper right panel:} Normalised autocorrelation function of the phase map, $C_\Phi ( \bm u, \bm v )$ (blue curve), overlaid with the real part of $e^{-ik\bm{\alpha}\cdot(\bm{u-v})}$ for an object at a separation of 200\,mas in the I band. The scalar product of these two functions gives the residual flux injected into the fibre, $\EX[\eta_{\rmstar}^{\rm inj}]$. \textit{Lower panel:} Raw contrast as a function of wavelength and angular separation, assuming two cases: (i) solid lines represent $\EX[\eta_{\rmstar}^{\rm inj}]$ assuming a correlated phase noise (as in our simulation), and (ii) dashed lines represent the raw contrast if we would have uncorrelated  phase noise ($C_\Phi ( \bm u, \bm v ) = cte$).}
\label{fig:raw_contrast}
\end{figure*} 

The behaviour of raw contrast with wavelength at a given angular separation depends on two opposing phenomena.
 At longer wavelengths, phase residuals are smaller, leading to higher Strehl ratios and less stray light. Conversely, at shorter wavelengths, the angular resolution is higher, making the single-mode fibre filtering via the $e^{-ik\bm{\alpha}\cdot\bm{u}}$ scalar product more effective at removing stellar light.
   
   The effectiveness of these two phenomena is captured in the equation for the off-axis stellar flux injected into the fibre, as given in Eq.~\eqref{eq:apod_star}:
   \begin{equation}
    \eta_{\rmstar}^{\rm inj} = \left|\scalar{e^{i\Phi(\bm{u}) + i\Psi(\bm{u})}}{e^{-ik\bm{\alpha}\cdot\bm{u}}}\right|^2 \,.
   \end{equation}
   
   To simplify, let's assume high Strehl conditions where the phase perturbation $\Phi(\bm{u})$ is small. A first-order Taylor expansion of $e^{i\Phi(\bm{u})}$ yields:
   \begin{equation}
    \eta_{\rmstar}^{\rm inj} \approx  \left|  \scalar{ e^{i\Psi(\bm{u})} }{e^{-ik\bm{\alpha}\cdot\bm{u}}} + \scalar{ i\Phi(\bm{u}) e^{i\Psi(\bm{u})}}{e^{-ik\bm{\alpha}\cdot\bm{u}}}    \right|^2 \,.
   \end{equation}
   
   Assuming an ideal phase apodizer and small $\Psi(\bm{u})$, the residual flux inside the fibre can be approximated as:
   \begin{equation}
    \eta_{\rmstar}^{\rm inj} 
    \approx  \left|\scalar{ \Phi(\bm{u}) }{e^{-ik\bm{\alpha}\cdot\bm{u}}}\right|^2 \,.
   \end{equation}
   
   The expected value of $\eta_{\rmstar}^{\rm inj}$ after multiple iterations converges to:
   \begin{equation}
      \EX[\eta_{\rmstar}^{\rm inj}] = \sigma_\Phi^2  \dscalar{ \EX[\Phi(\bm{u}) \Phi(\bm{v})] }{e^{-ik\bm{\alpha}\cdot(\bm{u-v})}} \,.
   \end{equation}
   
   The challenge lies in computing the autocorrelation function of the phase perturbation, $C_\Phi(\bm u, \bm v) = \EX[\Phi(\bm{u}) \Phi(\bm{v})]$, which is the Fourier transform of the phase perturbation's power spectrum. For uncorrelated phase maps (i.e., white noise), the autocorrelation function is a Dirac delta function:
   \begin{equation}
      \EX[\eta_{\rmstar}^{\rm inj}]_{\Phi \, \text{uncorrelated}} = \sigma_\Phi^2,
   \end{equation}
   which increases as the wavelength decreases.
   
   However, in realistic conditions, the phase map is spatially correlated, and $C_\Phi(\bm u, \bm v) \neq \sigma_\Phi^2 \delta(\bm u - \bm v)$. This correlation depends on both atmospheric perturbations and the adaptive optics system. Although no simple formula exists for this, numerical simulations can test it. Using HCIPy, we simulated realistic phase residuals and obtained the correlation function shown in the middle panel of Fig.~\ref{fig:raw_contrast}.
   
   In the lower panel of Fig.~\ref{fig:raw_contrast}, we plot the expected value of $\eta_{\rmstar}^{\rm inj}$ as a function of angular separation. For uncorrelated Gaussian noise, the raw contrast remains constant with angular separation but decreases with wavenumber due to $\sigma_\Phi$ being proportional to it, as shown by the dashed lines. For correlated phase maps, however, the raw contrast decreases with angular separation, as the frequency of the multiplicative term $e^{-ik\bm{\alpha}\cdot(\bm{u-v})}$ increases.
   
   Interestingly, our simulations show that far from the star, the amplitude of atmospheric residuals remains roughly independent of wavelength: the increase in phase residuals with wavenumber is compensated by the modulation caused by the off-axis tilt of the wavefront.

   \begin{figure*}[h!]
\section{Simulated orbits and detectability of known exoplanets}

	\centering
	\includegraphics[width=6.0cm]{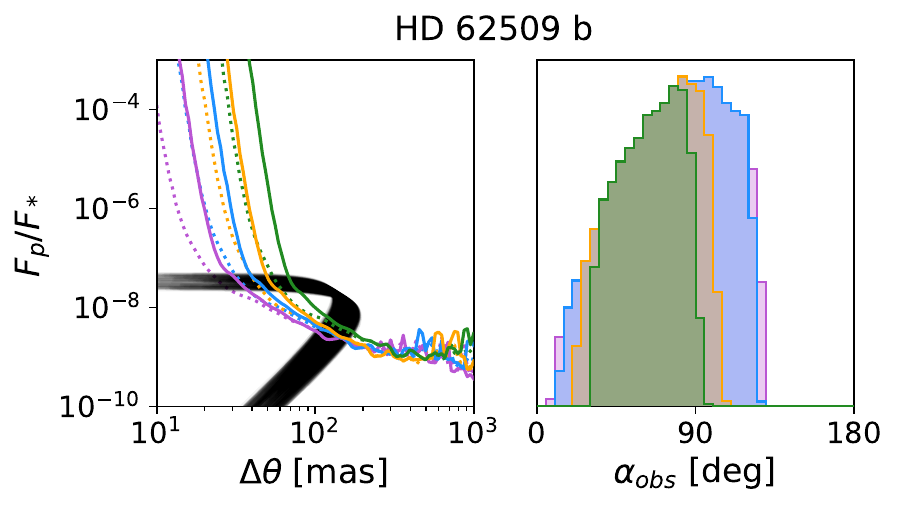}
	\hfill
	\includegraphics[width=6.0cm]{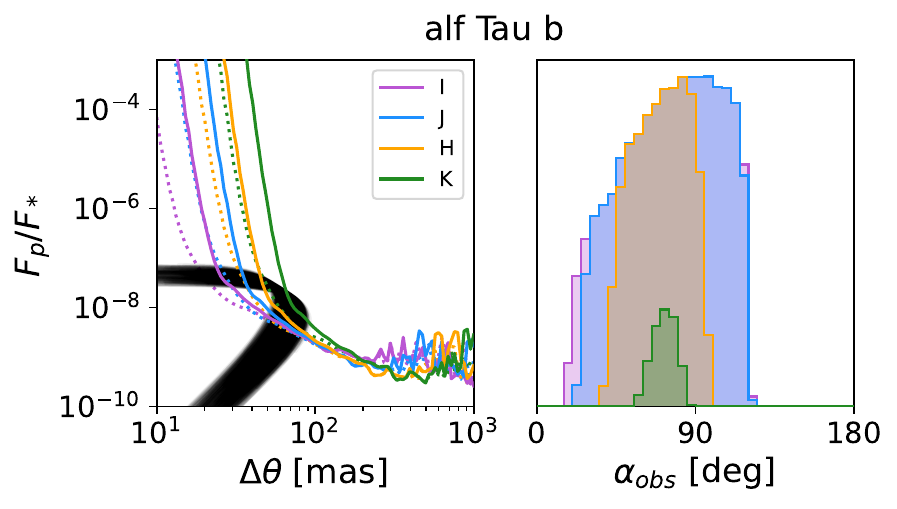}
	\hfill
	\includegraphics[width=6.0cm]{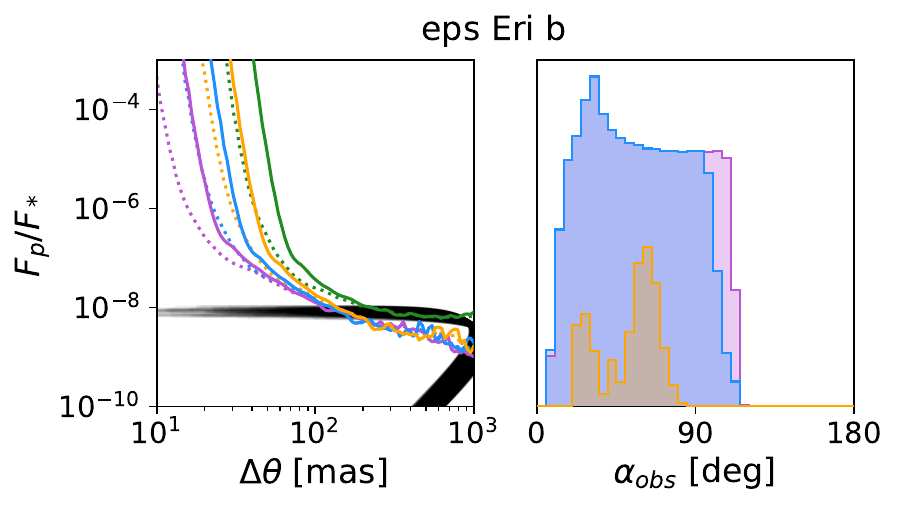}
\vspace{-0.05cm}
	\\
	\includegraphics[width=6.0cm]{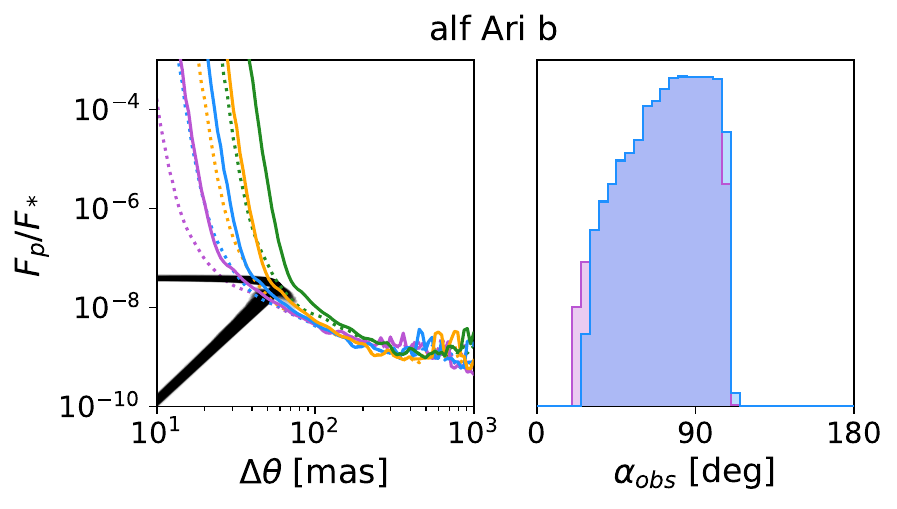}
	\hfill
	\includegraphics[width=6.0cm]{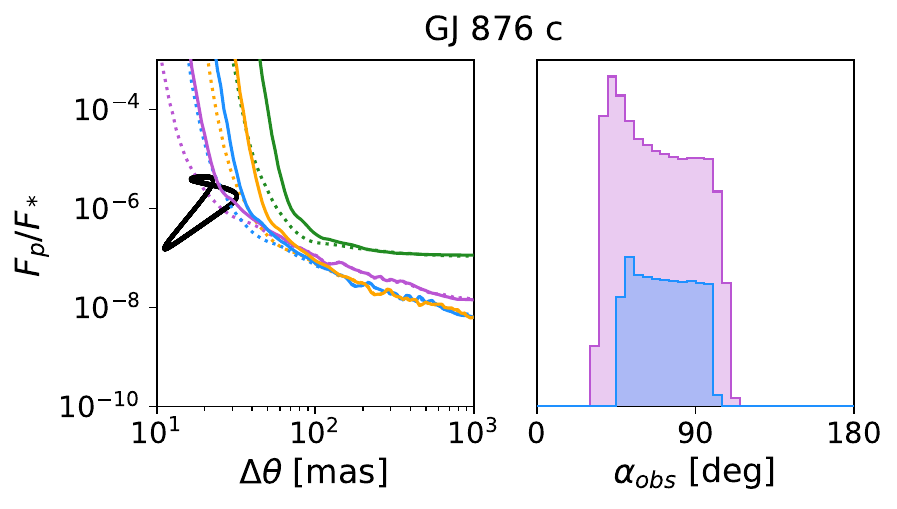}
\hfill
	\includegraphics[width=6.0cm]{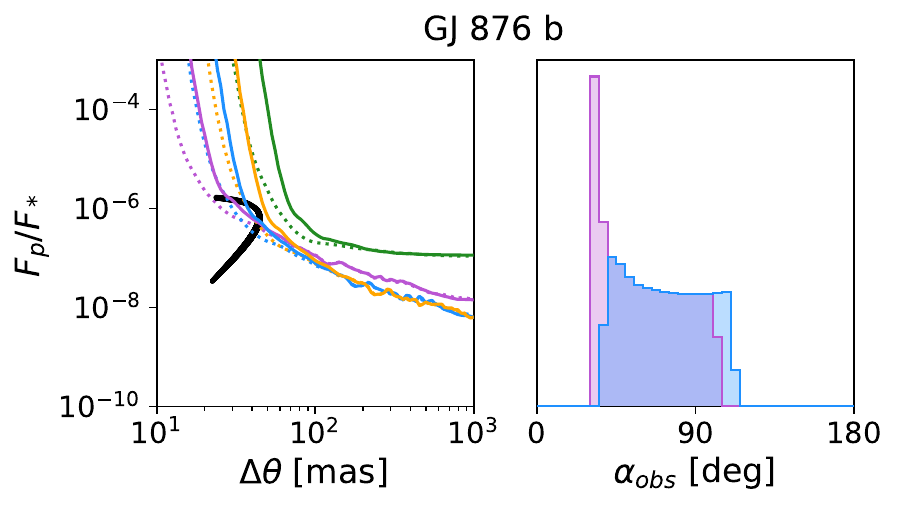}
\vspace{-0.05cm}
	\\
	\includegraphics[width=6.0cm]{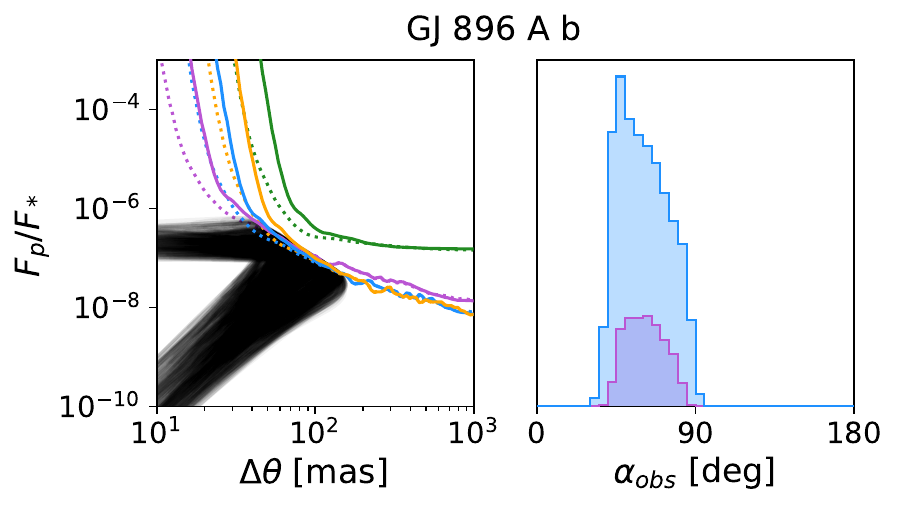}
	\hfill
	\includegraphics[width=6.0cm]{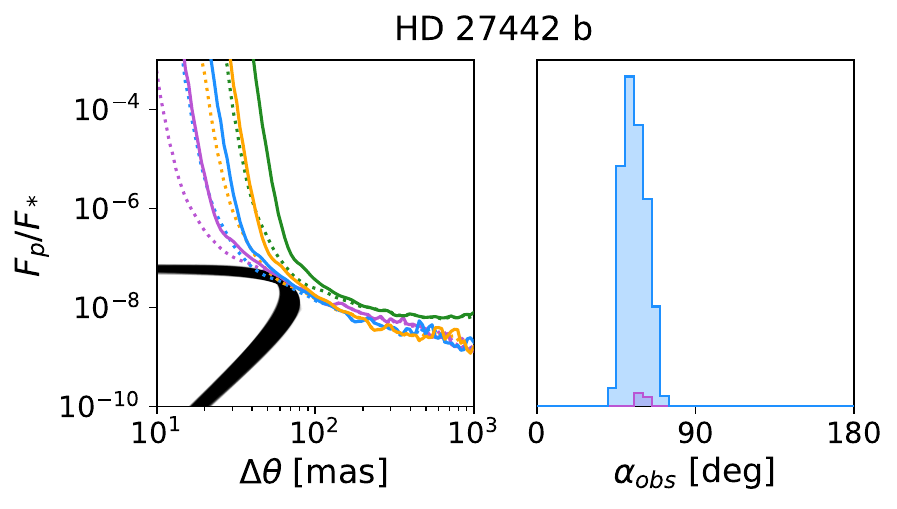}
	\hfill
	\includegraphics[width=6.0cm]{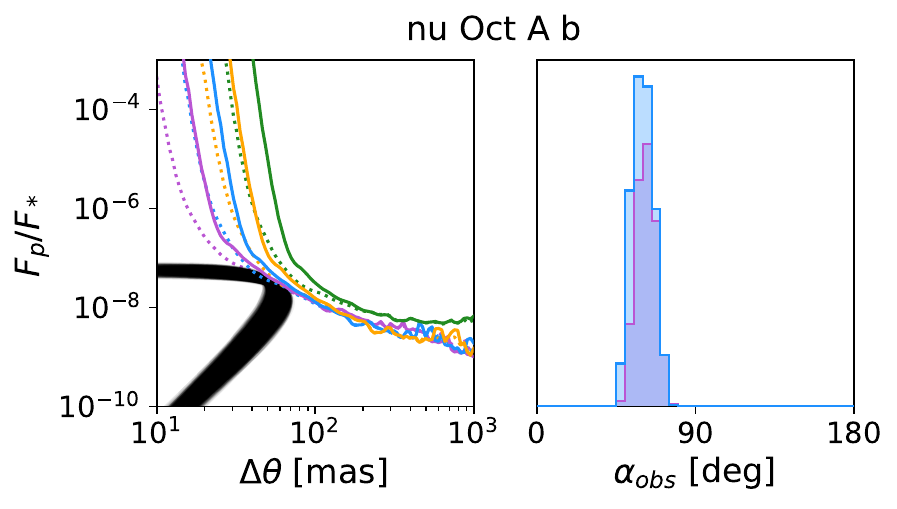}
\vspace{-0.05cm}
	\\
	\includegraphics[width=6.0cm]{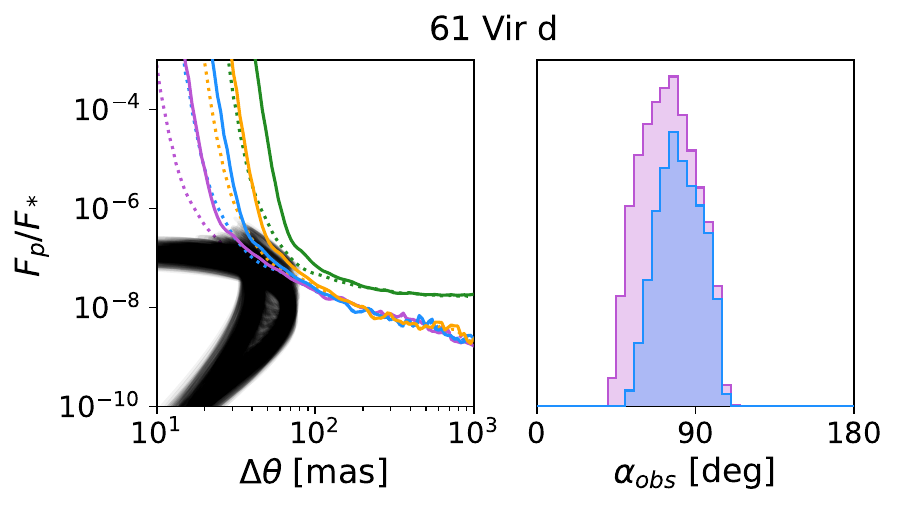}
	\hfill
	\includegraphics[width=6.0cm]{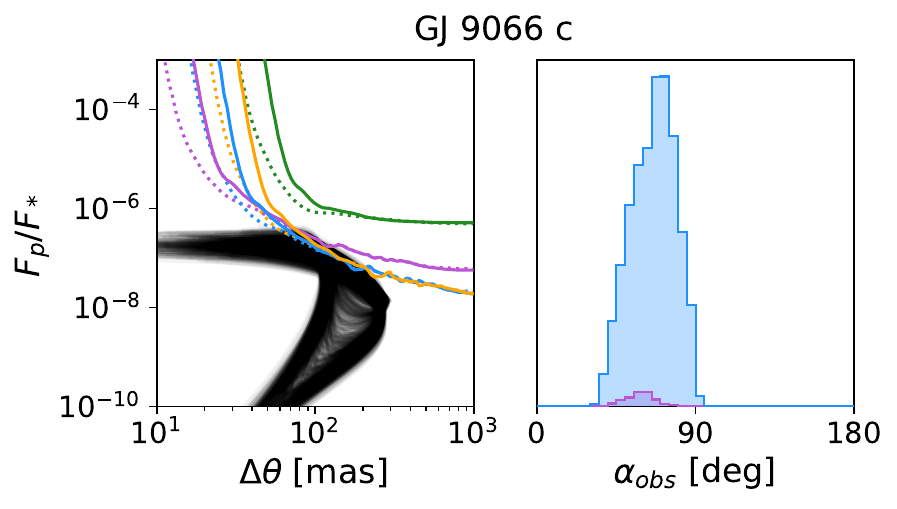}
	\hfill
	\includegraphics[width=6.0cm]{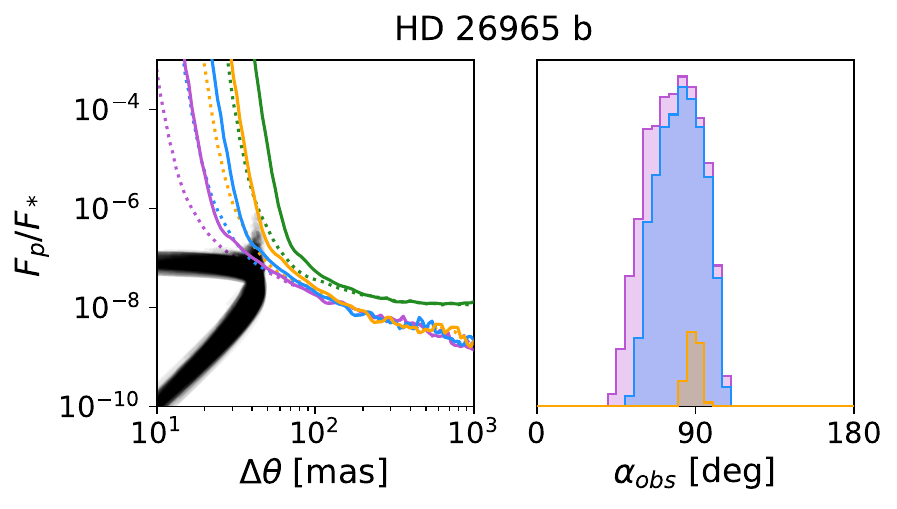}
\vspace{-0.05cm}
	\\
	\includegraphics[width=6.0cm]{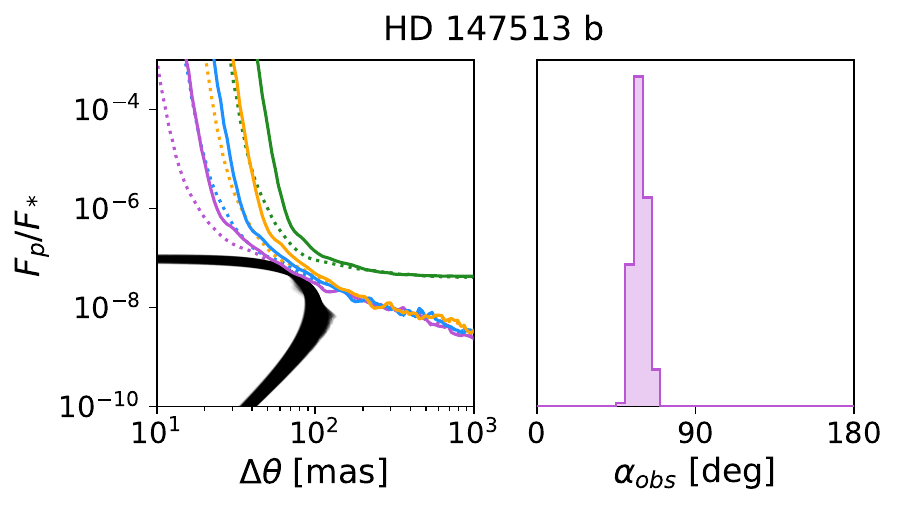}
	\includegraphics[width=6.0cm]{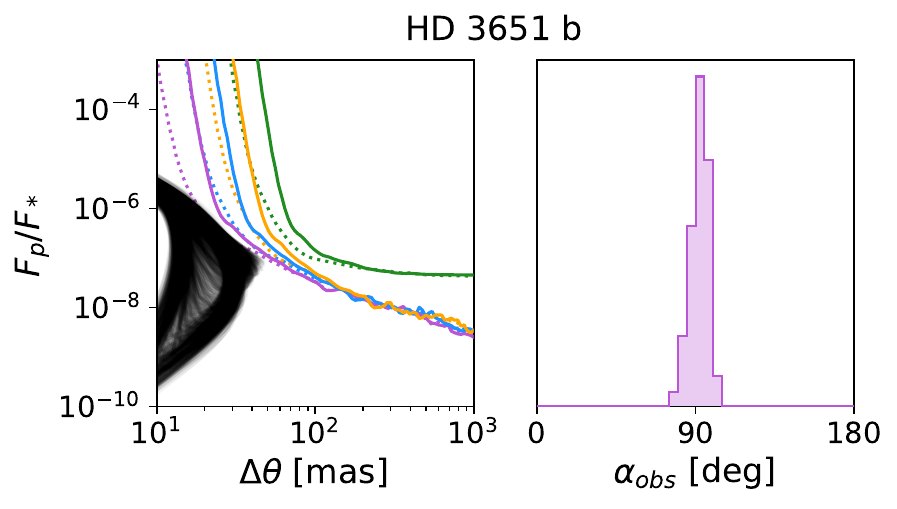}
\vspace{-0.05cm}
	\\
	\caption{Detectability conditions for the targets with $P_{detect}$>25\% in 3~h of integration. Left panels show the simulated orbits of each planet (black lines) assuming $A_g$=0.3. Coloured lines show the detectability limits of the interferometer at each spectral band with the current VLTI (solid lines) and the extended VLTI (dotted lines) baselines, for a $S/N$=3 in 3~h of integration.
	The histograms on the right panels show the range of $\alpha_{obs}$ at each band with the extended VLTI configuration.}
	\label{fig:orbits_and_histograms}
	\end{figure*}

\vspace{6cm}

\ 
\newline
\ 
\newline
\ 
\newline
\ 
\newline
\ 
\newline
\ 
\newline
\ 

\begin{table*}[h!]
\section{Detactable targets in 30~h of integration time}

    \tiny
    \centering
    \caption{List of detectable targets as in Table \ref{table:detectability_30pc_3h}, but for an integration time of 30~h with the current VLTI configuration.}
    \label{table:detectability_30pc_30h_100m} 
    \begin{tabular}{c l c c c c c c c c c c c }
    \hline 
    \hline
      &  & & &    &  \multicolumn{2}{c}{I} & \multicolumn{2}{c}{J} & \multicolumn{2}{c}{H} & \multicolumn{2}{c}{K} \\ 
      & Planet & $d$ & $a$ & $M_p$   & $P_{detect}$ & $\alpha_{obs}$ & $P_{detect}$ & $\alpha_{obs}$ & $P_{detect}$ & $\alpha_{obs}$ & $P_{detect}$ & $\alpha_{obs}$ \\
      &   & [pc] & [AU] & [$M_{J}$] & [\%] & [deg] &[\%] & [deg] & [\%] & [deg] & [\%] & [deg] \\ 
    \hline
\multirow{39}{*}{\rotatebox{90}{VLTI}} & HD 62509 b		&	10.34	&	1.62	&	2.74	&100.00	&	[31$^{+ 26 }_{- 20 }$,140$^{+ 1 }_{- 17 }$]		&	100.00	&	[31$^{+ 26 }_{- 17 }$,137$^{+ 1 }_{- 14 }$]&	100.00	&	[31$^{+ 26 }_{- 11 }$,123$^{+ 1 }_{- 1 }$]		&	100.00	&	[31$^{+ 26 }_{- 5 }$,115$^{+ 1 }_{- 1 }$]	\\
 & eps Eri b		&	3.20	&	3.53	&	0.68	&100.00	&	[14$^{+ 12 }_{- 9 }$,130$^{+ 2 }_{- 2 }$]		&	100.00	&	[14$^{+ 12 }_{- 9 }$,128$^{+ 2 }_{- 9 }$]	&100.00	&	[14$^{+ 12 }_{- 5 }$,92$^{+ 16 }_{- 3 }$]		&	97.30	&	[16$^{+ 9 }_{- 2 }$,47$^{+ 5 }_{- 11 }$]	\\
 & alf Tau b		&	20.44	&	1.47	&	7.44	&100.00	&	[28$^{+ 29 }_{- 8 }$,133$^{+ 1 }_{- 10 }$]		&	100.00	&	[34$^{+ 24 }_{- 6 }$,130$^{+ 1 }_{- 8 }$]	&100.00	&	[47$^{+ 11 }_{- 7 }$,112$^{+ 2 }_{- 4 }$]		&	62.70	&	[64$^{+ 10 }_{- 7 }$,96$^{+ 2 }_{- 3 }$]	\\
 & alf Ari b		&	20.21	&	1.20	&	2.13	&100.00	&	[31$^{+ 28 }_{- 11 }$,124$^{+ 1 }_{- 3 }$]		&	100.00	&	[31$^{+ 28 }_{- 2 }$,117$^{+ 1 }_{- 1 }$]	&99.50	&	[50$^{+ 9 }_{- 2 }$,84$^{+ 2 }_{- 2 }$]		&	0.00	&	$-$	\\
 & GJ 896 A b		&	6.26	&	0.64	&	2.23	&100.00	&	[24$^{+ 14 }_{- 1 }$,111$^{+ 1 }_{- 1 }$]		&	100.00	&	[26$^{+ 12 }_{- 3 }$,113$^{+ 1 }_{- 1 }$]	&87.10	&	[54$^{+ 3 }_{- 3 }$,71$^{+ 4 }_{- 5 }$]		&	0.00	&	$-$	\\
 & GJ 9066 c		&	4.47	&	0.88	&	0.24	&98.30	&	[38$^{+ 22 }_{- 9 }$,99$^{+ 7 }_{- 10 }$]		&	100.00	&	[35$^{+ 24 }_{- 8 }$,106$^{+ 8 }_{- 8 }$]	&40.70	&	[57$^{+ 12 }_{- 10 }$,84$^{+ 2 }_{- 4 }$]		&	0.00	&	$-$	\\
 & nu Oct A b		&	21.15	&	1.25	&	2.34	&100.00	&	[34$^{+ 22 }_{- 3 }$,106$^{+ 2 }_{- 2 }$]		&	100.00	&	[43$^{+ 14 }_{- 4 }$,101$^{+ 2 }_{- 3 }$]	&0.00	&	$-$		&	0.00	&	$-$	\\
 & bet Pic c		&	19.74	&	2.72	&	8.20	&100.00	&	[30$^{+ 2 }_{- 2 }$,98$^{+ 3 }_{- 3 }$]		&	100.00	&	[43$^{+ 4 }_{- 3 }$,84$^{+ 5 }_{- 6 }$]		&	0.00	&	$-$		&	0.00	&	$-$	\\
 & HD 27442 b		&	18.27	&	1.27	&	1.81	&100.00	&	[31$^{+ 25 }_{- 1 }$,106$^{+ 1 }_{- 1 }$]		&	100.00	&	[36$^{+ 19 }_{- 1 }$,105$^{+ 1 }_{- 1 }$]	&0.00	&	$-$		&	0.00	&	$-$	\\
 & HD 160691 b		&	15.60	&	1.52	&	4.30	&100.00	&	[30$^{+ 1 }_{- 0 }$,99$^{+ 2 }_{- 2 }$]		&	100.00	&	[40$^{+ 2 }_{- 1 }$,88$^{+ 2 }_{- 5 }$]		&	0.00	&	$-$		&	0.00	&	$-$	\\
 & HD 147513 b		&	12.90	&	1.32	&	1.39	&100.00	&	[29$^{+ 28 }_{- 3 }$,106$^{+ 1 }_{- 1 }$]		&	100.00	&	[38$^{+ 19 }_{- 1 }$,100$^{+ 1 }_{- 1 }$]	&0.00	&	$-$		&	0.00	&	$-$	\\
 & GJ 876 c		&	4.68	&	0.13	&	0.71	&100.00	&	[40$^{+ 0 }_{- 0 }$,114$^{+ 0 }_{- 0 }$]		&	100.00	&	[68$^{+ 1 }_{- 1 }$,81$^{+ 1 }_{- 1 }$]		&0.00	&	$-$		&	0.00	&	$-$	\\
 & GJ 876 b		&	4.68	&	0.21	&	2.28	&100.00	&	[31$^{+ 0 }_{- 0 }$,120$^{+ 0 }_{- 0 }$]		&	100.00	&	[45$^{+ 0 }_{- 0 }$,117$^{+ 0 }_{- 0 }$]	&0.00	&	$-$		&	0.00	&	$-$	\\
 & GJ 832 b		&	4.96	&	3.70	&	0.98	&100.00	&	[39$^{+ 2 }_{- 2 }$,82$^{+ 2 }_{- 2 }$]		&	100.00	&	[39$^{+ 2 }_{- 2 }$,93$^{+ 1 }_{- 3 }$]		&	0.00	&	$-$		&	0.00	&	$-$	\\
 & 7 CMa c		&	19.81	&	2.12	&	0.87	&100.00	&	[25$^{+ 0 }_{- 0 }$,106$^{+ 1 }_{- 1 }$]		&	100.00	&	[30$^{+ 1 }_{- 1 }$,105$^{+ 1 }_{- 1 }$]	&0.00	&	$-$		&	0.00	&	$-$	\\
 & 7 CMa b		&	19.80	&	1.77	&	1.85	&100.00	&	[26$^{+ 1 }_{- 1 }$,105$^{+ 1 }_{- 1 }$]		&	100.00	&	[32$^{+ 1 }_{- 1 }$,104$^{+ 1 }_{- 1 }$]	&0.00	&	$-$		&	0.00	&	$-$	\\
 & HD 160691 e		&	15.60	&	0.93	&	7.00	&99.70	&	[38$^{+ 13 }_{- 3 }$,97$^{+ 2 }_{- 2 }$]		&	93.90	&	[54$^{+ 10 }_{- 2 }$,86$^{+ 2 }_{- 2 }$]	&0.00	&	$-$		&	0.00	&	$-$	\\
 & tau Cet e		&	3.60	&	0.54	&	0.01	&94.70	&	[41$^{+ 18 }_{- 9 }$,95$^{+ 13 }_{- 17 }$]		&	87.00	&	[44$^{+ 16 }_{- 8 }$,94$^{+ 12 }_{- 13 }$]&8.00	&	[74$^{+ 8 }_{- 7 }$,94$^{+ 6 }_{- 15 }$]		&	2.80	&	[82$^{+ 3 }_{- 4 }$,93$^{+ 4 }_{- 9 }$]	\\
 & HD 114613 b		&	20.28	&	5.33	&	0.41	&87.20	&	[37$^{+ 12 }_{- 2 }$,79$^{+ 9 }_{- 12 }$]		&	96.90	&	[45$^{+ 11 }_{- 2 }$,77$^{+ 3 }_{- 5 }$]	&0.00	&	$-$		&	0.00	&	$-$	\\
 & 61 Vir d		&	8.50	&	0.48	&	0.08	&98.40	&	[46$^{+ 13 }_{- 6 }$,98$^{+ 10 }_{- 12 }$]		&	78.10	&	[59$^{+ 9 }_{- 6 }$,96$^{+ 11 }_{- 12 }$]	&3.50	&	[83$^{+ 4 }_{- 4 }$,100$^{+ 1 }_{- 4 }$]		&	0.00	&	$-$	\\
 & HD 190360 b		&	16.01	&	3.90	&	1.80	&100.00	&	[34$^{+ 1 }_{- 1 }$,76$^{+ 2 }_{- 3 }$]		&	56.30	&	[61$^{+ 4 }_{- 8 }$,75$^{+ 4 }_{- 11 }$]		&0.00	&	$-$		&	0.00	&	$-$	\\
 & HD 26965 b		&	5.04	&	0.22	&	0.03	&97.40	&	[48$^{+ 12 }_{- 7 }$,96$^{+ 11 }_{- 11 }$]		&	45.20	&	[62$^{+ 9 }_{- 5 }$,91$^{+ 13 }_{- 13 }$]	&0.70	&	[86$^{+ 3 }_{- 1 }$,93$^{+ 1 }_{- 3 }$]		&	0.00	&	$-$	\\
 & GJ 649 b		&	10.38	&	1.11	&	0.30	&62.20	&	[45$^{+ 15 }_{- 5 }$,82$^{+ 3 }_{- 4 }$]		&	72.20	&	[45$^{+ 15 }_{- 6 }$,87$^{+ 2 }_{- 5 }$]	&0.00	&	$-$		&	0.00	&	$-$	\\
 & 55 Cnc f		&	12.59	&	0.77	&	0.17	&88.00	&	[47$^{+ 12 }_{- 7 }$,92$^{+ 14 }_{- 12 }$]		&	34.00	&	[60$^{+ 8 }_{- 6 }$,91$^{+ 9 }_{- 13 }$]	&0.00	&	$-$		&	0.00	&	$-$	\\
 & GJ 887 c		&	3.29	&	0.12	&	0.03	&76.90	&	[53$^{+ 14 }_{- 9 }$,93$^{+ 11 }_{- 12 }$]		&	42.60	&	[63$^{+ 15 }_{- 11 }$,90$^{+ 12 }_{- 13 }$]&	1.70	&	[83$^{+ 8 }_{- 6 }$,90$^{+ 5 }_{- 7 }$]		&	0.10	&	[89$^{+ 0 }_{- 0 }$,89$^{+ 0 }_{- 0 }$]	\\
 & HD 10647 b		&	17.34	&	2.02	&	1.11	&99.00	&	[37$^{+ 22 }_{- 1 }$,88$^{+ 1 }_{- 2 }$]		&	19.70	&	[58$^{+ 3 }_{- 3 }$,66$^{+ 2 }_{- 3 }$]		&0.00	&	$-$		&	0.00	&	$-$	\\
 & HD 192310 c		&	8.80	&	1.18	&	0.09	&61.30	&	[51$^{+ 14 }_{- 10 }$,90$^{+ 14 }_{- 14 }$]		&	51.00	&	[53$^{+ 14 }_{- 10 }$,90$^{+ 15 }_{- 14 }$]&	5.00	&	[77$^{+ 4 }_{- 6 }$,92$^{+ 2 }_{- 11 }$]		&	0.00	&	$-$	\\
 & HD 39091 b		&	18.27	&	3.29	&	12.16	&99.20	&	[49$^{+ 3 }_{- 2 }$,75$^{+ 3 }_{- 4 }$]		&	0.00	&	$-$		&	0.00	&	$-$		&0.00	&	$-$	\\
 & HR 810 b		&	17.32	&	0.92	&	2.69	&99.10	&	[43$^{+ 14 }_{- 2 }$,91$^{+ 2 }_{- 2 }$]		&	0.00	&	$-$		&	0.00	&	$-$&	0.00	&	$-$	\\
 & 70 Vir b		&	17.90	&	0.48	&	8.70	&98.90	&	[54$^{+ 8 }_{- 2 }$,94$^{+ 1 }_{- 2 }$]		&	0.00	&	$-$		&	0.00	&	$-$		&0.00	&	$-$	\\
 & HD 19994 b		&	22.52	&	1.31	&	1.58	&93.10	&	[52$^{+ 3 }_{- 2 }$,77$^{+ 2 }_{- 3 }$]		&	0.00	&	$-$		&	0.00	&	$-$		&0.00	&	$-$	\\
 & HD 140901 c		&	15.25	&	12.61	&	1.79	&35.20	&	[43$^{+ 10 }_{- 4 }$,62$^{+ 11 }_{- 12 }$]		&	56.40	&	[57$^{+ 5 }_{- 5 }$,65$^{+ 5 }_{- 6 }$]		&0.00	&	$-$		&	0.00	&	$-$	\\
 & tau Cet h		&	3.60	&	0.24	&	0.01	&60.30	&	[55$^{+ 8 }_{- 6 }$,78$^{+ 21 }_{- 8 }$]		&	19.10	&	[64$^{+ 11 }_{- 8 }$,93$^{+ 9 }_{- 14 }$]	&1.40	&	[83$^{+ 2 }_{- 1 }$,92$^{+ 3 }_{- 5 }$]		&	0.20	&	[88$^{+ 0 }_{- 0 }$,92$^{+ 0 }_{- 0 }$]	\\
 & GJ 849 b		&	8.80	&	2.31	&	1.03	&0.00	&	$-$		&	75.70	&	[38$^{+ 1 }_{- 1 }$,51$^{+ 4 }_{- 4 }$]		&	0.00	&	$-$		&0.00	&	$-$	\\
 & HD 60532 c		&	25.98	&	1.60	&	2.91	&70.10	&	[55$^{+ 3 }_{- 2 }$,69$^{+ 3 }_{- 3 }$]		&	0.00	&	$-$		&	0.00	&	$-$		&0.00	&	$-$	\\
 & HD 160691 c		&	15.60	&	5.10	&	4.40	&34.20	&	[47$^{+ 9 }_{- 15 }$,97$^{+ 3 }_{- 19 }$]		&	35.90	&	[46$^{+ 14 }_{- 5 }$,85$^{+ 4 }_{- 7 }$]	&0.00	&	$-$		&	0.00	&	$-$	\\
 & GJ 229 b		&	5.76	&	1.09	&	0.05	&30.10	&	[59$^{+ 13 }_{- 10 }$,82$^{+ 14 }_{- 13 }$]		&	32.70	&	[57$^{+ 13 }_{- 9 }$,86$^{+ 15 }_{- 15 }$]&4.20	&	[77$^{+ 3 }_{- 5 }$,90$^{+ 8 }_{- 10 }$]		&	0.00	&	$-$	\\
 & HD 102365 b		&	9.29	&	0.46	&	0.06	&42.60	&	[56$^{+ 15 }_{- 9 }$,89$^{+ 13 }_{- 13 }$]		&	17.30	&	[67$^{+ 10 }_{- 8 }$,88$^{+ 7 }_{- 13 }$]	&0.00	&	$-$		&	0.00	&	$-$	\\
 & HD 3651 b		&	11.13	&	0.30	&	0.27	&52.80	&	[92$^{+ 4 }_{- 3 }$,110$^{+ 4 }_{- 7 }$]		&	1.10	&	[94$^{+ 4 }_{- 2 }$,96$^{+ 3 }_{- 3 }$]		&0.00	&	$-$		&	0.00	&	$-$	\\
 & 61 Vir c		&	8.50	&	0.22	&	0.07	&40.00	&	[67$^{+ 6 }_{- 5 }$,92$^{+ 12 }_{- 12 }$]		&	0.00	&	$-$		&	0.00	&	$-$&	0.00	&	$-$	\\

    \hline
    \end{tabular}
\end{table*}

\begin{table*}[h!]
    \tiny
    \centering
    \caption{List of detectable targets as in Table \ref{table:detectability_30pc_3h}, but for an integration time of 30~h with the extended VLTI configuration.}
    \label{table:detectability_30pc_30h_200m} 
    \begin{tabular}{c l c c c c c c c c c c c }
    \hline 
    \hline
      &  & & &    &  \multicolumn{2}{c}{I} & \multicolumn{2}{c}{J} & \multicolumn{2}{c}{H} & \multicolumn{2}{c}{K} \\ 
      & Planet & $d$ & $a$ & $M_p$   & $P_{detect}$ & $\alpha_{obs}$ & $P_{detect}$ & $\alpha_{obs}$ & $P_{detect}$ & $\alpha_{obs}$ & $P_{detect}$ & $\alpha_{obs}$ \\
      &   & [pc] & [AU] & [$M_{J}$] & [\%] & [deg] &[\%] & [deg] & [\%] & [deg] & [\%] & [deg] \\ 
    \hline
\multirow{49}{*}{\rotatebox{90}{Extended VLTI}}  & alf Tau b		&	20.44	&	1.47	&	7.44	&100.00	&	[28$^{+ 29 }_{- 13 }$,134$^{+ 1 }_{- 12 }$]		&	100.00	&	[28$^{+ 29 }_{- 8 }$,134$^{+ 1 }_{- 11 }$]&100.00	&	[36$^{+ 22 }_{- 5 }$,120$^{+ 1 }_{- 2 }$]		&	100.00	&	[49$^{+ 10 }_{- 7 }$,111$^{+ 2 }_{- 4 }$]	\\
 & HD 62509 b		&	10.34	&	1.62	&	2.74	&100.00	&	[31$^{+ 26 }_{- 21 }$,140$^{+ 1 }_{- 17 }$]		&	100.00	&	[31$^{+ 26 }_{- 20 }$,138$^{+ 1 }_{- 15 }$]&	100.00	&	[31$^{+ 26 }_{- 15 }$,124$^{+ 1 }_{- 2 }$]		&	100.00	&	[31$^{+ 26 }_{- 10 }$,118$^{+ 1 }_{- 2 }$]	\\
 & eps Eri b		&	3.20	&	3.53	&	0.68	&100.00	&	[14$^{+ 12 }_{- 9 }$,126$^{+ 5 }_{- 2 }$]		&	100.00	&	[14$^{+ 12 }_{- 9 }$,126$^{+ 2 }_{- 2 }$]	&100.00	&	[14$^{+ 12 }_{- 5 }$,100$^{+ 4 }_{- 2 }$]		&	99.00	&	[17$^{+ 9 }_{- 2 }$,49$^{+ 5 }_{- 13 }$]	\\
 & alf Ari b		&	20.21	&	1.20	&	2.13	&100.00	&	[31$^{+ 28 }_{- 16 }$,128$^{+ 1 }_{- 7 }$]		&	100.00	&	[31$^{+ 28 }_{- 10 }$,126$^{+ 1 }_{- 5 }$]&100.00	&	[36$^{+ 23 }_{- 1 }$,104$^{+ 1 }_{- 2 }$]		&	52.90	&	[62$^{+ 5 }_{- 4 }$,75$^{+ 3 }_{- 4 }$]	\\
 & GJ 896 A b		&	6.26	&	0.64	&	2.23	&100.00	&	[21$^{+ 16 }_{- 3 }$,113$^{+ 1 }_{- 2 }$]		&	100.00	&	[22$^{+ 16 }_{- 4 }$,116$^{+ 2 }_{- 1 }$]	&100.00	&	[39$^{+ 4 }_{- 2 }$,86$^{+ 5 }_{- 3 }$]		&	0.00	&	$-$	\\
 & GJ 876 b		&	4.68	&	0.21	&	2.28	&100.00	&	[31$^{+ 0 }_{- 0 }$,125$^{+ 1 }_{- 1 }$]		&	100.00	&	[31$^{+ 0 }_{- 0 }$,126$^{+ 0 }_{- 0 }$]	&100.00	&	[52$^{+ 0 }_{- 0 }$,101$^{+ 1 }_{- 1 }$]		&	0.00	&	$-$	\\
 & HD 27442 b		&	18.27	&	1.27	&	1.81	&100.00	&	[30$^{+ 26 }_{- 8 }$,108$^{+ 1 }_{- 1 }$]		&	100.00	&	[30$^{+ 26 }_{- 3 }$,109$^{+ 1 }_{- 1 }$]	&85.00	&	[55$^{+ 3 }_{- 2 }$,72$^{+ 3 }_{- 4 }$]		&	0.00	&	$-$	\\
 & GJ 9066 c		&	4.47	&	0.88	&	0.24	&97.80	&	[37$^{+ 24 }_{- 10 }$,99$^{+ 7 }_{- 11 }$]		&	100.00	&	[33$^{+ 27 }_{- 11 }$,106$^{+ 7 }_{- 8 }$]&52.20	&	[53$^{+ 15 }_{- 12 }$,88$^{+ 2 }_{- 11 }$]		&	0.00	&	$-$	\\
 & 61 Vir d		&	8.50	&	0.48	&	0.08	&100.00	&	[36$^{+ 22 }_{- 5 }$,103$^{+ 8 }_{- 9 }$]		&	100.00	&	[46$^{+ 12 }_{- 5 }$,101$^{+ 10 }_{- 9 }$]&13.00	&	[74$^{+ 4 }_{- 6 }$,95$^{+ 7 }_{- 10 }$]		&	0.00	&	$-$	\\
 & HD 26965 b		&	5.04	&	0.22	&	0.03	&100.00	&	[35$^{+ 25 }_{- 4 }$,101$^{+ 10 }_{- 9 }$]		&	100.00	&	[48$^{+ 12 }_{- 5 }$,100$^{+ 9 }_{- 9 }$]	&8.20	&	[76$^{+ 7 }_{- 6 }$,95$^{+ 4 }_{- 11 }$]		&	0.10	&	[87$^{+ 0 }_{- 0 }$,89$^{+ 0 }_{- 0 }$]	\\
 & nu Oct A b		&	21.15	&	1.25	&	2.34	&100.00	&	[30$^{+ 26 }_{- 7 }$,109$^{+ 2 }_{- 2 }$]		&	100.00	&	[34$^{+ 23 }_{- 4 }$,109$^{+ 2 }_{- 2 }$]	&1.90	&	[61$^{+ 1 }_{- 2 }$,68$^{+ 2 }_{- 2 }$]		&	0.00	&	$-$	\\
 & bet Pic c		&	19.74	&	2.72	&	8.20	&100.00	&	[25$^{+ 3 }_{- 3 }$,97$^{+ 3 }_{- 3 }$]		&	100.00	&	[33$^{+ 5 }_{- 2 }$,87$^{+ 4 }_{- 5 }$]		&	0.00	&	$-$		&	0.00	&	$-$	\\
 & HD 160691 b		&	15.60	&	1.52	&	4.30	&100.00	&	[30$^{+ 0 }_{- 0 }$,103$^{+ 4 }_{- 6 }$]		&	100.00	&	[32$^{+ 3 }_{- 1 }$,94$^{+ 4 }_{- 3 }$]		&0.00	&	$-$		&	0.00	&	$-$	\\
 & HD 147513 b		&	12.90	&	1.32	&	1.39	&100.00	&	[29$^{+ 28 }_{- 10 }$,106$^{+ 1 }_{- 1 }$]		&	100.00	&	[29$^{+ 28 }_{- 2 }$,102$^{+ 1 }_{- 1 }$]	&0.00	&	$-$		&	0.00	&	$-$	\\
 & GJ 876 c		&	4.68	&	0.13	&	0.71	&100.00	&	[31$^{+ 0 }_{- 0 }$,124$^{+ 0 }_{- 0 }$]		&	100.00	&	[39$^{+ 0 }_{- 0 }$,113$^{+ 0 }_{- 0 }$]	&0.00	&	$-$		&	0.00	&	$-$	\\
 & GJ 832 b		&	4.96	&	3.70	&	0.98	&100.00	&	[39$^{+ 2 }_{- 2 }$,70$^{+ 8 }_{- 4 }$]		&	100.00	&	[39$^{+ 2 }_{- 2 }$,89$^{+ 2 }_{- 2 }$]		&	0.00	&	$-$		&	0.00	&	$-$	\\
 & 7 CMa c		&	19.81	&	2.12	&	0.87	&100.00	&	[20$^{+ 1 }_{- 1 }$,106$^{+ 1 }_{- 1 }$]		&	100.00	&	[22$^{+ 0 }_{- 0 }$,107$^{+ 1 }_{- 1 }$]	&0.00	&	$-$		&	0.00	&	$-$	\\
 & 7 CMa b		&	19.80	&	1.77	&	1.85	&100.00	&	[20$^{+ 1 }_{- 1 }$,105$^{+ 1 }_{- 1 }$]		&	100.00	&	[24$^{+ 0 }_{- 0 }$,107$^{+ 1 }_{- 1 }$]	&0.00	&	$-$		&	0.00	&	$-$	\\
 & HD 160691 e		&	15.60	&	0.93	&	7.00	&100.00	&	[30$^{+ 14 }_{- 0 }$,103$^{+ 4 }_{- 8 }$]		&	99.70	&	[40$^{+ 15 }_{- 4 }$,92$^{+ 6 }_{- 2 }$]	&0.00	&	$-$		&	0.00	&	$-$	\\
 & 70 Vir b		&	17.90	&	0.48	&	8.70	&100.00	&	[41$^{+ 16 }_{- 2 }$,105$^{+ 1 }_{- 2 }$]		&	97.00	&	[63$^{+ 7 }_{- 3 }$,92$^{+ 2 }_{- 2 }$]		&0.00	&	$-$		&	0.00	&	$-$	\\
 & HR 810 b		&	17.32	&	0.92	&	2.69	&99.60	&	[32$^{+ 25 }_{- 3 }$,96$^{+ 1 }_{- 1 }$]		&	96.90	&	[46$^{+ 10 }_{- 3 }$,86$^{+ 2 }_{- 2 }$]	&0.00	&	$-$		&	0.00	&	$-$	\\
 & tau Cet e		&	3.60	&	0.54	&	0.01	&94.40	&	[42$^{+ 18 }_{- 10 }$,94$^{+ 13 }_{- 17 }$]		&	88.00	&	[42$^{+ 18 }_{- 9 }$,94$^{+ 12 }_{- 14 }$]&9.70	&	[72$^{+ 9 }_{- 8 }$,93$^{+ 7 }_{- 16 }$]		&	3.80	&	[80$^{+ 4 }_{- 7 }$,95$^{+ 3 }_{- 7 }$]	\\
 & GJ 887 c		&	3.29	&	0.12	&	0.03	&94.10	&	[43$^{+ 20 }_{- 9 }$,101$^{+ 10 }_{- 10 }$]		&	87.00	&	[53$^{+ 17 }_{- 10 }$,99$^{+ 9 }_{- 12 }$]&8.60	&	[76$^{+ 12 }_{- 11 }$,90$^{+ 8 }_{- 13 }$]		&	0.70	&	[85$^{+ 4 }_{- 2 }$,88$^{+ 5 }_{- 1 }$]	\\
 & HD 10647 b		&	17.34	&	2.02	&	1.11	&98.50	&	[34$^{+ 25 }_{- 2 }$,88$^{+ 2 }_{- 2 }$]		&	86.80	&	[48$^{+ 4 }_{- 2 }$,74$^{+ 3 }_{- 3 }$]		&0.00	&	$-$		&	0.00	&	$-$	\\
 & HD 19994 b		&	22.52	&	1.31	&	1.58	&97.20	&	[37$^{+ 18 }_{- 2 }$,84$^{+ 2 }_{- 2 }$]		&	86.80	&	[51$^{+ 3 }_{- 2 }$,71$^{+ 3 }_{- 4 }$]		&0.00	&	$-$		&	0.00	&	$-$	\\
 & tau Cet h		&	3.60	&	0.24	&	0.01	&95.80	&	[47$^{+ 12 }_{- 6 }$,82$^{+ 13 }_{- 8 }$]		&	83.40	&	[51$^{+ 10 }_{- 5 }$,79$^{+ 16 }_{- 9 }$]	&3.50	&	[79$^{+ 3 }_{- 5 }$,94$^{+ 4 }_{- 14 }$]		&	0.80	&	[85$^{+ 2 }_{- 2 }$,94$^{+ 3 }_{- 5 }$]	\\
 & HD 114613 b		&	20.28	&	5.33	&	0.41	&86.70	&	[37$^{+ 11 }_{- 2 }$,79$^{+ 9 }_{- 12 }$]		&	96.00	&	[42$^{+ 14 }_{- 3 }$,76$^{+ 4 }_{- 6 }$]	&0.00	&	$-$		&	0.00	&	$-$	\\
 & 55 Cnc f		&	12.59	&	0.77	&	0.17	&96.50	&	[39$^{+ 18 }_{- 9 }$,95$^{+ 12 }_{- 13 }$]		&	75.60	&	[47$^{+ 12 }_{- 6 }$,86$^{+ 17 }_{- 15 }$]&0.00	&	$-$		&	0.00	&	$-$	\\
 & GJ 649 b		&	10.38	&	1.11	&	0.30	&61.00	&	[43$^{+ 16 }_{- 6 }$,81$^{+ 3 }_{- 4 }$]		&	79.80	&	[43$^{+ 15 }_{- 9 }$,91$^{+ 2 }_{- 9 }$]	&0.00	&	$-$		&	0.00	&	$-$	\\
 & HD 190360 b		&	16.01	&	3.90	&	1.80	&100.00	&	[34$^{+ 1 }_{- 1 }$,77$^{+ 2 }_{- 3 }$]		&	36.40	&	[62$^{+ 4 }_{- 8 }$,74$^{+ 4 }_{- 11 }$]		&0.00	&	$-$		&	0.00	&	$-$	\\
 & tau Cet g		&	3.60	&	0.13	&	0.01	&100.00	&	[45$^{+ 13 }_{- 3 }$,86$^{+ 9 }_{- 6 }$]		&	33.00	&	[65$^{+ 7 }_{- 5 }$,85$^{+ 15 }_{- 8 }$]	&0.80	&	[85$^{+ 1 }_{- 0 }$,93$^{+ 1 }_{- 1 }$]		&	0.10	&	[90$^{+ 0 }_{- 0 }$,91$^{+ 0 }_{- 0 }$]	\\
 & Proxima Cen b		&	1.30	&	0.05	&	0.00	&27.80	&	[59$^{+ 14 }_{- 6 }$,77$^{+ 15 }_{- 10 }$]		&	99.30	&	[61$^{+ 5 }_{- 5 }$,85$^{+ 9 }_{- 5 }$]		&	0.70	&	[85$^{+ 2 }_{- 1 }$,93$^{+ 1 }_{- 4 }$]		&	0.00	&	$-$	\\
 & HD 3651 b		&	11.13	&	0.30	&	0.27	&100.00	&	[71$^{+ 6 }_{- 5 }$,116$^{+ 5 }_{- 7 }$]		&	26.10	&	[98$^{+ 5 }_{- 4 }$,112$^{+ 3 }_{- 6 }$]	&0.00	&	$-$		&	0.00	&	$-$	\\
 & 61 Vir c		&	8.50	&	0.22	&	0.07	&99.30	&	[50$^{+ 10 }_{- 5 }$,101$^{+ 9 }_{- 10 }$]		&	19.90	&	[73$^{+ 5 }_{- 5 }$,98$^{+ 6 }_{- 8 }$]		&0.00	&	$-$		&	0.00	&	$-$	\\
 & HD 192310 c		&	8.80	&	1.18	&	0.09	&59.30	&	[52$^{+ 14 }_{- 10 }$,90$^{+ 14 }_{- 14 }$]		&	53.20	&	[51$^{+ 16 }_{- 11 }$,91$^{+ 15 }_{- 15 }$]&	6.10	&	[76$^{+ 5 }_{- 5 }$,97$^{+ 2 }_{- 16 }$]		&	0.00	&	$-$	\\
 & HD 102365 b		&	9.29	&	0.46	&	0.06	&60.80	&	[51$^{+ 16 }_{- 11 }$,92$^{+ 13 }_{- 15 }$]		&	40.80	&	[56$^{+ 15 }_{- 10 }$,89$^{+ 13 }_{- 13 }$]&	3.40	&	[79$^{+ 4 }_{- 4 }$,88$^{+ 5 }_{- 8 }$]		&	0.00	&	$-$	\\
 & HD 39091 b		&	18.27	&	3.29	&	12.16	&100.00	&	[42$^{+ 2 }_{- 1 }$,82$^{+ 2 }_{- 2 }$]		&	0.00	&	$-$		&	0.00	&	$-$		&0.00	&	$-$	\\
 & HD 60532 b		&	25.98	&	0.77	&	1.21	&99.90	&	[43$^{+ 15 }_{- 1 }$,89$^{+ 2 }_{- 2 }$]		&	0.00	&	$-$		&	0.00	&	$-$&	0.00	&	$-$	\\
 & HD 60532 c		&	25.98	&	1.60	&	2.91	&95.60	&	[46$^{+ 6 }_{- 2 }$,78$^{+ 2 }_{- 3 }$]		&	0.00	&	$-$		&	0.00	&	$-$		&0.00	&	$-$	\\
 & GJ 3021 b		&	17.56	&	0.49	&	3.95	&84.80	&	[73$^{+ 5 }_{- 3 }$,93$^{+ 1 }_{- 2 }$]		&	0.00	&	$-$		&	0.00	&	$-$		&0.00	&	$-$	\\
 & HD 140901 c		&	15.25	&	12.61	&	1.79	&34.50	&	[43$^{+ 10 }_{- 4 }$,62$^{+ 11 }_{- 12 }$]		&	40.40	&	[57$^{+ 5 }_{- 10 }$,64$^{+ 5 }_{- 5 }$]	&0.00	&	$-$		&	0.00	&	$-$	\\
 & GJ 849 b		&	8.80	&	2.31	&	1.03	&0.00	&	$-$		&	74.70	&	[39$^{+ 1 }_{- 1 }$,50$^{+ 3 }_{- 4 }$]		&	0.00	&	$-$		&0.00	&	$-$	\\
 & HD 160691 c		&	15.60	&	5.10	&	4.40	&36.50	&	[48$^{+ 9 }_{- 15 }$,97$^{+ 8 }_{- 20 }$]		&	36.00	&	[45$^{+ 11 }_{- 10 }$,85$^{+ 11 }_{- 7 }$]&0.00	&	$-$		&	0.00	&	$-$	\\
 & GJ 229 b		&	5.76	&	1.09	&	0.05	&28.60	&	[60$^{+ 13 }_{- 11 }$,82$^{+ 14 }_{- 13 }$]		&	37.30	&	[54$^{+ 16 }_{- 9 }$,83$^{+ 18 }_{- 15 }$]&5.30	&	[76$^{+ 4 }_{- 6 }$,89$^{+ 8 }_{- 10 }$]		&	0.50	&	[79$^{+ 2 }_{- 2 }$,81$^{+ 5 }_{- 2 }$]	\\
 & 55 Cnc c		&	12.59	&	0.24	&	0.18	&70.70	&	[62$^{+ 8 }_{- 5 }$,96$^{+ 10 }_{- 8 }$]		&	0.00	&	$-$		&	0.00	&	$-$&	0.00	&	$-$	\\
 & HD 192310 b		&	8.80	&	0.32	&	0.05	&43.10	&	[48$^{+ 9 }_{- 6 }$,78$^{+ 8 }_{- 8 }$]		&	7.30	&	[65$^{+ 3 }_{- 3 }$,77$^{+ 4 }_{- 4 }$]		&	0.00	&	$-$		&	0.00	&	$-$	\\
 & HD 20794 e		&	6.00	&	0.51	&	0.02	&27.10	&	[59$^{+ 12 }_{- 11 }$,87$^{+ 13 }_{- 14 }$]		&	20.60	&	[63$^{+ 11 }_{- 10 }$,87$^{+ 13 }_{- 16 }$]&	1.30	&	[82$^{+ 3 }_{- 4 }$,93$^{+ 2 }_{- 6 }$]		&	0.50	&	[86$^{+ 1 }_{- 0 }$,91$^{+ 4 }_{- 1 }$]	\\
 & HD 180617 b		&	5.91	&	0.34	&	0.05	&16.50	&	[67$^{+ 10 }_{- 9 }$,89$^{+ 11 }_{- 14 }$]		&	27.30	&	[60$^{+ 14 }_{- 8 }$,86$^{+ 14 }_{- 14 }$]&1.90	&	[82$^{+ 4 }_{- 4 }$,94$^{+ 3 }_{- 9 }$]		&	0.00	&	$-$	\\
 & HD 210277 b		&	21.30	&	1.13	&	1.49	&26.30	&	[59$^{+ 5 }_{- 4 }$,70$^{+ 2 }_{- 3 }$]		&	0.00	&	$-$		&	0.00	&	$-$		&0.00	&	$-$	\\
 & GJ 887 b		&	3.29	&	0.07	&	0.01	&25.20	&	[66$^{+ 10 }_{- 5 }$,96$^{+ 7 }_{- 11 }$]		&	0.80	&	[85$^{+ 2 }_{- 2 }$,93$^{+ 1 }_{- 2 }$]		&0.00	&	$-$		&	0.00	&	$-$	\\

    \hline
    \end{tabular}
\end{table*}
\end{appendix}

\end{document}